\documentclass[fleqn,usenatbib]{mnras}

\usepackage{newtxtext,newtxmath}

\usepackage[T1]{fontenc}
\usepackage{ae,aecompl}

\DeclareRobustCommand{\VAN}[3]{#2}
\let\VANthebibliography\thebibliography
\def\thebibliography{\DeclareRobustCommand{\VAN}[3]{##3}\VANthebibliography}

\usepackage{graphicx}	
\usepackage{amsmath}	
\usepackage{multicol}        
\usepackage{bm}		

\usepackage[normalem]{ulem}
\usepackage{color} 
\usepackage{hyperref} 

\usepackage{makecell} 

\newcommand{\fitbox}{0.9\columnwidth}

\newcommand*\trades{\texttt{TRADES}}

\newcommand*\kepler{\textit{Kepler}}
\newcommand*\gaia{\textit{Gaia}}

\newcommand*\pycheops{\texttt{pycheops}}
\newcommand*\emcee{\texttt{emcee}}

\newcommand*{\geff}{\ensuremath{\mathrm{G}_{\mathrm{EFF}}}}
\newcommand*{\cpreff}{\ensuremath{\mathrm{cpr}_{\mathrm{EFF}}}}

\newcommand{\dd}{\mathrm{d}}

\newcommand{\feh} {\mbox{$[\mathrm{Fe}/\mathrm{H}]$}}

\definecolor{darkcyan}{rgb}{0.0, 0.55, 0.55}
\definecolor{dartmouthgreen}{rgb}{0.05, 0.5, 0.06}
\definecolor{magenta}{rgb}{1., 0., 1.}
\definecolor{gray}{rgb}{0.5, 0.5, 0.5}

\title[CHEOPS-EXPLORE/TTV]{Exploiting timing capabilities of the CHEOPS mission with warm-Jupiter planets}

\author[L.~Borsato]{
L.~Borsato$^{1}$\thanks{E-mail: luca.borsato@inaf.it},
G.~Piotto$^{2,1}$,
D.~Gandolfi$^{3}$,
V.~Nascimbeni$^{1}$,
G.~Lacedelli$^{2,1}$,
F.~Marzari$^{2}$,\cr
N.~Billot$^{4}$,
P.F.L.~Maxted$^{5}$,
S.~Sousa$^{6}$,
A.C.~Cameron$^{7}$,
A.~Bonfanti$^{8}$,
T.G.~Wilson$^{7}$,\cr
L.M.~Serrano$^{3}$,
Z.~Garai$^{9,10,11}$,
Y.~Alibert$^{12}$,
R.~Alonso$^{13,14}$,
J.~Asquier$^{15}$,\cr
T.~B\'arczy$^{16}$,
T.~Bandy$^{17}$,
D.~Barrado$^{18}$,
S.C.C.~Barros$^{6,19}$,
W.~Baumjohann$^{8}$,\cr
M.~Beck$^{4}$,
T.~Beck$^{17}$,
W.~Benz$^{12,17}$,
X.~Bonfils$^{20}$,
A.~Brandeker$^{21}$,
C.~Broeg$^{12}$,\cr
J.~Cabrera$^{22}$,
S.~Charnoz$^{23}$,
S.~Csizmadia$^{22}$,
M.B.~Davies$^{24}$,
M.~Deleuil$^{25}$,\cr
L.~Delrez$^{26,27,4}$,
O.~Demangeon$^{6,19}$,
B.-O.~Demory$^{17}$,
A.L.~des~Etangs$^{28}$,
D.~Ehrenreich$^{4}$,\cr
A.~Erikson$^{22}$,
G.A.~Escud\'e$^{29,30}$,
A.~Fortier$^{12}$,
L.~Fossati$^{8}$,
M.~Fridlund$^{31,32}$,\cr
M.~Gillon$^{27}$,
M.~Guedel$^{33}$,
J.~Hasiba$^{8}$,
K.~Heng$^{17,34}$,
S.~Hoyer$^{25}$,
K.G.~Isaak$^{15}$,\cr
L.~Kiss$^{35}$,
E.~Kopp$^{22}$,
J.~Laskar$^{36}$,
M.~Lendl$^{4}$,
C.~Lovis$^{4}$,
D.~Magrin$^{1}$,\cr
M.~Munari$^{37}$,
G.~Olofsson$^{21}$,
R.~Ottensamer$^{33}$,
I.~Pagano$^{37}$,
E.~Pall\'e$^{13,14}$,
G.~Peter$^{22}$,\cr
D.~Pollacco$^{34}$,
D.~Queloz$^{4,38}$,
R.~Ragazzoni$^{2,1}$,
N.~Rando$^{15}$,
H.~Rauer$^{22,39,40}$,\cr
I.~Ribas$^{29,30}$,
D.~S\'egransan$^{4}$,
N.C.~Santos$^{6,19}$,
G.~Scandariato$^{37}$,
A.~Simon$^{12}$,\cr
A.M.S.~Smith$^{22}$,
M.~Steller$^{8}$,
G.~Szab\'o$^{9,10}$,
N.~Thomas$^{12}$,
S.~Udry$^{4}$,
V.~Van~Grootel$^{26}$,\cr
N.~Walton$^{38}$,

\\
\textit{(Affiliations can be found after the references)}
}

\date{Accepted 2021 June 18. Received 2021 March 30.}

\pubyear{2021}

\begin{document}
\label{firstpage}
\pagerange{\pageref{firstpage}--\pageref{lastpage}}
\maketitle

\begin{abstract}
We present 17 transit light curves of seven known warm-Jupiters observed with the CHaracterising ExOPlanet Satellite (CHEOPS). 
The light curves have been collected as part of the CHEOPS Guaranteed Time Observation (GTO) program
that searches for transit-timing variation (TTV) of warm-Jupiters
induced by a possible external perturber to shed light on
the evolution path of such planetary systems. 
We describe the CHEOPS observation process, from the planning to the data analysis. 
In this work we focused on the timing performance of CHEOPS,
the impact of the sampling of the transit phases,
and the improvement we can obtain combining multiple transits together. 
We reached the highest precision on the transit time of about 13--16~s for the brightest target
(WASP-38, $G = 9.2$) in our sample.
From the combined analysis of multiple transits of fainter targets with $G \geq 11$
we obtained a timing precision of $\sim 2$~min.
Additional observations with CHEOPS, covering a longer temporal baseline,
will further improve the precision on the transit times 
and will allow us to detect possible TTV signals induced by an external perturber.
\end{abstract}

\begin{keywords}
techniques: photometric -- planets and satellites: individual: (HAT-P-17 b, KELT-6 b, WASP-8 b, WASP-38 b, WASP-106 b, WASP-130 b, K2-287 b)
\end{keywords}



\section{Introduction}

The \kepler{} space-mission showed that hot-Jupiters are usually lone planets
that do not show transit time variation (TTV) signals \citep{Agol2005MNRAS.359..567A,HolmanMurray2005Sci...307.1288H,Steffen2012PNAS..109.7982S}.
The occurrence rate of companions to hot-Jupiters is currently uncertain and unreliable \citep{Huang2016ApJ...825...98H}.
On the other hand, almost $50\%$ of warm-Jupiters 
(gas giant planets with orbital periods between $\sim$8 and 200 days)
of the \kepler{} sample are found in multi-planet systems \citep{Huang2016ApJ...825...98H}.
These warm-Jupiters show a wide variety of orbital configurations
possibly resulting from different formation and migration mechanisms \citep{Wu2018AJ....156...96W,Kley2019},
i.e. disk migration \citep{Lin1996Natur.380..606L,Baruteau2016SSRv..205...77B}
or high-eccentricity migration \citep{Nagasawa2008ApJ...678..498N}.
The measurement of the sky-projected orientation of
the planet orbit with respect to the spin axis of the star
(the so called projected spin-orbit angle $\lambda$)
can help to discern between these two models. 
Misaligned warm-Jupiters could be formed by high-eccentricity migration,
while circular and aligned warm-Jupiters,
potentially in a mean motion resonance (MMR) with an outer companion,
are expected to be the result of disk-driven migration process \citep{Baruteau2016SSRv..205...77B}.
In this scenario, the outer companion is expected
to be less massive than the inner warm-Jupiter,
if produced by convergent migration \citep{Kley2019}.
Although the orbits should be nearly circular and well aligned,
a mild eccentricity of the outer planet is expected to build up because of the resonant perturbations
\citep{Baruteau2016SSRv..205...77B}.
The TTVs of a resonant pair of planets are particularly strong
and might be found even if the companion has a significantly lower mass
that cannot be easily detected using high-precision radial velocity (RV) measurements
\citep{Agol2005MNRAS.359..567A,HolmanMurray2005Sci...307.1288H,Steffen2012PNAS..109.7982S}.
\par

Observing an outer perturber on possibly eccentric and inclined orbit
in a system where an eccentric (and misaligned) warm-Jupiter is present
would be the hint for a high-eccentricity mechanism,
driven by planet-planet (P--P) scattering \citep{Marzari2019AA...625A.121M}
followed by tidal interactions with the host star.
\par

Finding planetary perturbers of known transiting exoplanets
can provide precious insights onto the architecture and the evolution of planetary systems
\citep{Malavolta2017,Teyssandier2020AA...643A..11T,MacDonald2020ApJ...891...20M,Kane2019AJ....157..171K,Masuda2020AJ....159...38M,Poon2020MNRAS.491.5595P}.
Detecting a TTV signal of a known transiting warm-Jupiter
induced by a perturber of planetary nature would help to understand their evolution path,
which is expected to be different from that of hot-Jupiters
\citep{Huang2016ApJ...825...98H,Frewen2016MNRAS.455.1538F}.
\par

The CHaracterising ExOPlanet Satellite \citep[CHEOPS,][]{Benz2020ExA...tmp...53B}
was launched on December 18, 2019, and
it started observations in April, 2020.
CHEOPS is a follow-up mission that aims at characterising exoplanets
known to transit their host star using high-precision photometry.
It already demonstrated its performances 
improving the precision on the planetary parameters
of KELT-11 b \citep{Benz2020ExA...tmp...53B}.
\citet{Lendl2020AA...643A..94L} used the transit and the occultation observed with CHEOPS
to characterise the atmosphere and the spin-orbit obliquity of the highly-irradiated WASP-189 b,
measuring the asymmetry of the transit shape due to the stellar gravity darkening.
Furthermore, CHEOPS has been already used to characterise
two multiple-planet systems,
improving the ephemerides and the orbital parameters of the system TOI-1233 \citep{Bonfanti2021AA...646A.157B}
and solving the orbital configuration of TOI-178 \citep{Leleu2021arXiv210109260L}.
\par

As part of the CHEOPS Guaranteed Time Observation (GTO) programme, 
we are currently searching for TTV signals in a selected sample of known transiting warm-Jupiters (Section~\ref{sec:warmJup}).
The purpose of this work is to demonstrate CHEOPS' capability to schedule multiple observations and
obtain transit time measurements with sufficient accuracy to allow detection of TTV signals.
In Section~\ref{sec:timing} we present the first 17 CHEOPS transit light curves of seven targets of our GTO program,
we describe the strategy and planning of our observations,
and the data analysis of single and multiple transits for each target.
We summarise and discuss the results in Section~\ref{sec:results}
and draw our conclusions in Section~\ref{sec:conclusions}.
\par

\section{Target selection of the sample}
\label{sec:warmJup}

The planets in our sample have significantly non-zero eccentricity
measured from Doppler observations and,
when possible, measured spin-orbit angle, $\lambda$, 
from observations of the Rossiter-McLaughlin (RM) effect \citep{Ohta2005ApJ...622.1118O} or
Doppler tomography \citep[e.g.,][]{Brown2012ApJ...760..139B}.
We based our initial sample selection on the TEPCAT catalogue \citep{Southworth2011MNRAS.417.2166S},
then we checked if each candidate target was observable with CHEOPS using
the Feasibility Checker (FC) provided by the Consortium
\footnote{Available through ESA website; for more information see \url{https://www.cosmos.esa.int/web/cheops-guest-observers-programme}.}.
\par

The possible high mutual inclination ($\Delta i$) of the perturber
expected by a P-P scattering event
implies an almost null probability
of transit and reduces the RV semi-amplitude ($K_\mathrm{RV}$).
Nevertheless, the mass of the perturber, coupled with the eccentricity and the inclination,
is expected to induce a detectable TTV signal of the known transiting warm-Jupiter.
The lack of a TTV signal in highly eccentric and misaligned transiting warm-Jupiters
would indicate that P--P scattering is not efficient in producing eccentric and
misaligned warm gas giant planets.
We expect to observe 15 transits per target during 3.5 years, the nominal duration of the CHEOPS mission.
After the first five transits we should be able to have hint or rule out the presence of a TTV, 
but only with the full 15 transits we will be able to sample the TTV period and amplitude 
and draw conclusions about the existence of a perturber and
on the formation path (P-P scattering or migration in disk).
\par

We estimated the expected amplitude of the TTV signal  ($A_\mathrm{TTV}$)
produced by an outer perturber on a transiting warm-Jupiter,
following a procedure similar to that used by \citet{Borsato2021arXiv210309239B}.
We used the parameters of the transiting warm-Jupiter from the literature and
we assumed the existence of a hypothetical outer planetary companion.
The main parameters of the perturber that influence the period and the amplitude of the TTV
are the mass ($M_\mathrm{perturber}$), the period ($P_\mathrm{perturber}$),
the eccentricity ($e_\mathrm{perturber}$), and the mutual inclination ($\Delta i_\mathrm{perturber}$)
of the perturber,
as widely demonstrated analytically and numerically by, e.g.,
\citet{Agol2005MNRAS.359..567A, HolmanMurray2005Sci...307.1288H} and \cite{Nesvorny2009ApJ...701.1116N}.
We created different TTV maps based on different initial values of this set of four parameters of the perturber.

We computed the orbits with \trades{}\footnote{Publicly available at \url{https://github.com/lucaborsato/trades}.}
\citep{Borsato2014AA...571A..38B,Nespral2017,Malavolta2017,Borsato2019MNRAS.484.3233B}
over a grid of mass and period values of the perturber
with 30 log-spaced values of masses,
ranging from $1\, M_{\earth}$ to $1\, M_\mathrm{Jup}$,
and 30 log-spaced values of different orbital periods. 
The period grid of the perturber 
ranged from slightly longer values than the period of the transiting planet to 100 days.
We used \trades{} to integrate the orbits for 3.5 years 
(i.e., the nominal duration of the CHEOPS mission) and computed transit times ($T_0$) and linear ephemerides.
We then selected 15 random transits (without replacement),
i.e., the expected maximum number of transits to be obtained for each target
during the CHEOPS nominal mission,
re-computed the linear ephemeris,
and calculated the $A_\mathrm{TTV}$ as the semi-amplitude of the $O-C$
(selected transit times, $O$, minus the newly computed linear ephemeris, $C$).
This was done for each simulation and repeated for 100 times.
The final $A_\mathrm{TTV}$ was computed as the median of the $A_\mathrm{TTV}$ of the 100 repetitions.
We obtained a map of the $A_\mathrm{TTV}$ as a function of mass ($M_\mathrm{perturber}$)
and period ($P_\mathrm{perturber}$) of the perturber.
It is well known that the eccentricity of the perturber ($e_\mathrm{perturber}$)
boosts the $A_\mathrm{TTV}$ \citep{Agol2005MNRAS.359..567A,HolmanMurray2005Sci...307.1288H}.
We also took into account the effect of mutual inclination ($\Delta i$).
We repeated the same analysis
with different sets of initial conditions of the perturber:
$e_\mathrm{perturber} = 0.0$ and 0.1,
$\Delta i_\mathrm{perturber} = 0\degr$ and $60\degr$
(see Figures from \ref{fig:grid_hatp17} to \ref{fig:grid_k2-287} in Appendix~\ref{apdx:ttv}
for a selection of simulation outcomes).
\par

We found that a perturber less massive than the transiting planet
on an external orbit can induce a TTV with amplitude of a few minutes,
detectable with about 15 transits.
Finally, combining information on planet characterisation,
target visibility with CHEOPS,
and dynamical simulations,
we selected a sample of eight warm-Jupiters to follow-up with CHEOPS and
measure their transit times with the purpose of detecting possible TTV signals.
In this work we present the analysis of the timing of
CHEOPS observations obtained so far 
within the context of TTV search of the warm-Jupiters.
\par

\section{Exploiting transit timing from CHEOPS data}
\label{sec:timing}

We present the analysis of 17 CHEOPS single visits
of the transits of seven targets 
(HAT-P-17 b, KELT-6 b, WASP-8 b, WASP-38 b, WASP-106 b, WASP-130 b, K2-287 b)
out of the eight targets of our sample,
with the purpose to investigate the performance of CHEOPS on the timing precision
of the first year of observations.
Currently, for five targets (HAT-P-17 b, WASP-8 b, WASP-38 b, WASP-130 b, K2-287 b)
we have multiple visits (from two to four visits),
that is we have multiple transit observations.
Four targets, HAT-P-17 b, WASP-8 b, WASP-130 b, and K2-287 b
have been observed with an exposure time of $60$~s, 
while we used an exposure of $55$~s for WASP-38.
We used the CHEOPS Exposure Time Calculator (ETC\footnote{Available at \url{https://cheops.unige.ch/pht2/exposure-time-calculator/}.}) to determine the exposure time of each target.
\par

\subsection{Observing strategy}
\label{sec:strategy}

The CHEOPS orbit \citep[with period of 98.77 minutes, for more details see][]{Benz2020ExA...tmp...53B}
affects the scheduling and
the strategy of the observations.
Each CHEOPS observation is called visit.
We aimed to collect CHEOPS data with visit duration ($\mathrm{dur_{vis}}$) that covers
the transit event with an out-of-transit baseline long enough
to sample astrophysical and instrumental noise sources (systematics).
Furthermore, to increase the chance to schedule a transit observation,
it is advisable 
to allow for some level of flexibility in the start time of the visit
including a start lag ($l$),
defined as 
the difference between an earliest and latest starting phase
($\phi_\mathrm{start,earliest}$ and $\phi_\mathrm{start,latest}$, respectively).
We defined the starting phase ($\phi_\mathrm{start}$)
at half visit duration with respect to the expected centre of the transit,
but the observation can start between
$\phi_\mathrm{start} - l/2$ and $\phi_\mathrm{start} + l/2$.
We used a start lag, $l$, of half transit duration, 
enough to take into account 
the uncertainties on the transit duration and the linear ephemeris,
the possible presence of a TTV,
and making more flexible the visit scheduling.
Our visit duration definition changed with time after the analysis of the collected data
and planetary parameters update.
We found that 
a possible good choice for the visit duration, especially in case of short transits,
is given by
$\mathrm{dur_{vis}} = \mathrm{max} ( T_{14} + l + n_c \times c_\mathrm{o}, 2.5 \times T_{14})$,
where
$T_{14}$ is the total transit duration \citep[elapsed time from first to fourth contact, eq. 30 of][]{Kipping2010MNRAS.407..301K},
$c_\mathrm{o} = 98.77$~min is CHEOPS orbit duration
and $n_c$ is the minimum number of CHEOPS orbits to cover the out-of-transit light curve.
We need at least one CHEOPS orbit before and one after the transit to sample the
possible systematics,
so we decided to set $n_c = 3$ to have a more robust analysis.
We remind the reader that this definition of the visit duration is indicative and specific
for our targets, 
and it must be computed carefully based on the characteristics of the transiting exoplanet
and on the purpose of the observation.
With the aim of precisely measuring the transit time ($T_0$)
we need high temporal sampling of the ingress and egress phases.
The global efficiency of a CHEOPS visit (\geff),
defined as the ratio between the time effectively spent on target and the total visit duration,
depends on the satellite pointing 
exclusion angles,
Earth occultations,
straylight conditions,
and passages through the South Atlantic Anomaly (SAA).
A low \geff{} translates into periodic gaps in the light curve that
for a minimum \geff{} of $50\%$
can be as long as
about half an orbit in duration each.
This impacts how well we can sample the ingress and egress phases of the transit 
(critical phase ranges efficiency, \cpreff),
and so it greatly affects the precision on the mid-time of transit $T_0$.
However, the \geff{} and \cpreff{} predicted by the feasibility checker
can be inaccurate as the CHEOPS orbit implemented in the FC
is an approximation to the satellite's true orbit on the date of the observation.
The uncertainty of CHEOPS exact position along its orbit makes the exact timing
of these gaps obsolete beyond a few weeks.
As the FC is not updated on a weekly basis to take the revised CHEOPS orbit into account,
we cannot predict \geff{} and \cpreff{} far in advance.
The precision and the accuracy on the transit linear ephemeris
and on the parameters of the exoplanet also impact the \cpreff.
We set as minimum value of the global efficiency \geff{} $\ge 50\%$.
When possible, we selected the \cpreff{} transit-by-transit,
favouring events with
at least one \cpreff{} (ingress or egress) $\ge 70\%$ and the other one at least $\ge 30\%$,
or both \cpreff{} $\ge 50\%$.
The selection of the visits evolved in time and with updated 
planetary parameters and FC version.
Furthermore, some of the predicted \cpreff{} from FC mismatched
the sampling of the ingress and egress phases of the transit observations,
as we will explain in Section~\ref{sec:results}.
It would also be advisable to have non-consecutive transits
to increase the temporal baseline for the TTV identification and analysis,
but, due to all constraints and to the automatic scheduling,
in a few cases CHEOPS observed consecutive transits of the same exoplanet.
\par

\subsection{Data analysis}
\label{sec:data_analysis}

For all the visits we used the light curve extracted by the CHEOPS Data Reduction Pipeline 
\citep[DRP version 12,][]{Hoyer2020A&A...635A..24H}
with the default aperture size of 25 pixels (which corresponds to $25\arcsec$).
We used the same aperture size for all the visits of all the targets for consistency.
The DRP extracts the flux,
the error on flux measurement,
the background, 
the centroid position (and the offset position in $x$ and $y$ pixel coordinates),
the contamination,
and the roll angle of the satellite 
\citep[for further details see][]{Hoyer2020A&A...635A..24H,Bonfanti2021AA...646A.157B,Leleu2021arXiv210109260L}.
We clipped out the outliers by filtering out values
5 times the mean absolute deviation away
from the median-smoothed\footnote{We used the scipy.signal.medfilt} light curve.

\subsubsection{Stellar parameters}

We obtained
the stellar effective temperature $T_{\mathrm{eff}}$,
surface gravity $\log g$,
and the metallicy \feh{}
from SWEET-Cat \citep[e.g.,][]{Santos2013AA...556A.150S,Sousa2018AA...620A..58S}.
For K2-287 the spectroscopic parameters were reviewed with more recent
spectroscopic data within the CHEOPS Stellar Characterization working group.
The parameters were derived with \texttt{ARES+MOOG} \citep{Sneden1973PhDT.......180S,Sousa2015A&A...577A..67S}
following the same procedure as for SWEET-Cat \citep[e.g.,][]{Sousa2014dapb.book..297S, Bonfanti2021AA...646A.157B}.
We use the infrared flux method \citep[IRFM;][]{Blackwell1977MNRAS.180..177B}
to determine the stellar radius $R_{\star}$ of targets in this study via comparison between optical and infrared broadband fluxes,
and synthetic photometry of stellar atmospheric models,
and using known relationships between stellar angular diameter,
effective temperature, and parallax (\gaia{} DR2).
This is conducted in a Markov chain Monte Carlo (MCMC) approach
by taking the spectral parameter values derived above
as priors on stellar spectral energy distribution selection to be used for synthetic photometry.
We retrieved broadband photometry for the following bandpasses from the most recent data releases, that are
\textit{Gaia} $G$, $G_\mathrm{BP}$, and $G_\mathrm{RP}$, 2MASS $J$, $H$, and $K$,
and \textit{WISE} W1 and W2 \citep{Gaia2016_A&A...595A...1G,GaiaDR2_2018yCat.1345....0G,Skrutskie2006AJ....131.1163S, Wright2010AJ....140.1868W},
and we used the \textit{Gaia} DR2 parallax and \textsc{atlas} Catalogues \citep{Castelli2003IAUS..210P.A20C} of models.
Stellar mass $M_{\star}$ and age 
values were determined
by combining the results coming from two different sets of stellar evolutionary models,
namely PARSEC v1.2S \citep[PAdova \& TRieste Stellar Evolutionary Code,][]{marigo17} and
CLES \citep[Code Li\`{e}geois d'\'{E}volution Stellaire][]{scuflaire08}. 
The adopted input parameters were
$T_{\mathrm{eff}}$, metallicity \feh, and $R_{\star}$.
In particular, the results from PARSEC were inferred employing the isochrone placement algorithm
described in \citet{bonfanti15,bonfanti16},
which interpolates within a pre-computed grid of models to retrieve the best-fit parameters.
Instead, the results from CLES are retrieved by directly modelling the star with CLES code
following a Levenberg-Marquardt minimisation \citep{Salmon2020arXiv201114932S}.
The final adopted values for $M_{\star}$ and age $t_{\star}$
derive from
a careful combination of the two pairs of outputs,
as described in details in \citet{Bonfanti2021AA...646A.157B}.
\par
Of all the stellar properties of all the targets,
we found that three values agree with literature within 3-$\sigma$,
four values are within 2-$\sigma$,
and all others agree within 1-$\sigma$.
\par

\subsubsection{Light curves analysis}

We analysed all single and multiple visits with 
\pycheops\footnote{Publicly available at \url{https://github.com/pmaxted/pycheops}. We used version 0.9.3 of \pycheops.}
\citep[][Maxted et al., submitted]{Benz2020ExA...tmp...53B},
a custom python package developed to manage and analyse CHEOPS datasets.

\paragraph{Single-visit analysis}
The fitting parameters of the single-visit transit model within \pycheops{} are:
the transit time ($T_0$),
the orbital period ($P$),
the transit depth ($D$)\footnote{The transit depth $D$ is defined as the square of the planet-star radius ratio ($k$):
$D = k^2 = \left( \frac{R_\mathrm{p}}{R_\star} \right)^2$.},  
the transit duration \citep[$W$, eq.~16 of][]{SeagerMallenOrnelas2003ApJ...585.1038S} in unit of $P$,
the impact parameter ($b$)\footnote{Impact parameter for the circular case $b = \frac{a}{R_\star}\cos i$},
the combination of eccentricity ($e$) and argument of pericenter ($\omega$) in the form
$\sqrt{e}\cos \omega$ and $\sqrt{e}\sin \omega$.
\pycheops{} implements the algorithm \texttt{qpower2} \citep{Maxted2019AA...622A..33M}
for the power-2 law for the limb-darkening (LD)
with parameters $h_1$ and $h_2$,
but constrained in the ($0, 1$) uniform space of 
the fitting parameters $q_1$ and $q_2$
\citep{Maxted2018AA...616A..39M,Short2019RNAAS...3..117S}.
The program takes into account trends and/or patterns using
detrending parameters, such as
first and second order derivative in time (linear $\dd f/\dd t$ and quadratic $\dd^2 f/\dd t^2$ term),
first and second order derivative of the centroid offset in $x$ and $y$ pixel coordinates
$(\dd f/\dd x,\ \dd^2 f/\dd x^2,\ \dd f/\dd y,\ \dd^2 f/\dd y^2)$,
background $(\dd f/\dd \mathrm{bg})$,
contamination $(\dd f/\dd \mathrm{contam})$,
and the first three harmonics of the roll angle (in $\cos \phi$ and $\sin \phi$).
It has an additional term called glint, that models the internal reflections 
as a smooth function of the roll angle; 
this parameter can be modelled measuring the roll angle relative to the apparent Moon distance
(that is the glint is caused by the moonlight). 
It also models the stellar activity, i.e., the stellar granulation, with
Gaussian process \citep[GP,][]{Rasmussen2006gpml.book.....R} 
with the \texttt{SHOTerm} kernel,
with a fixed quality factor $Q=1/\sqrt{2}$, 
implemented in \texttt{celerite} \citep{Harvey1985ESASP.235..199H,Kallinger2014AA...570A..41K,celerite,Barros2020AA...634A..75B}.
The \texttt{SHOTerm} kernel describes a stochastically-driven, damped harmonic oscillator,
characterised by a damping time scale equal to $\tau = 2 Q / \omega_0$ and 
a standard deviation of the process $\sigma_\mathrm{GP} = \sqrt{S_0 \omega_0 Q}$.
The fitting hyper-parameters used in the kernel are $\log S_0$ and $\log \omega_0$.
A jitter term has been always added in quadrature to the flux errors
and it was fitted as $\log \sigma_j$;
also a constant term ($c$) has been taken into account in the detrending model.
\par

During the single-visit analysis we did not fit all the parameters.
We fixed $P$, $\sqrt{e}\cos \omega$, and $\sqrt{e}\sin \omega$
to the values found in literature.
For each visit we compared
the Bayesian Information Criterion (BIC)
for two transit models,
i.e., fitting the parameters of the transit shape,
that is $D$, $W$, and $b$,
or fixing them.
The physical parameters of the planets taken from the literature
are used to compute the initial parameters and 
the Gaussian priors for the fitting parameters $D$, $W$, and $b$.
For all the detrending parameters we used Uniform priors between -1 and 1,
only the glint parameter was bounded between 0 and 2. 
From the determinant of the Jacobian matrix
we constrained the model to have uniform priors on
$\cos i$, $\log k$, and $\log a/R_\star$.
During the fit, \pycheops{} computes the $\log$ of the stellar density ($\log \rho_{\star}$)
from $k$, $b$, $W$, and $P$ and
it applies a prior determined from the stellar parameters, i.e. mass and radius.
Also the LD power-2 law coefficient values and priors are computed from the stellar parameters
in the form $h_1$ and $h_2$, defined in \citet{Maxted2018AA...616A..39M}.
\par

We did the analysis using as initial points the parameters in literature,
fitted with the Levenberg-Marquart \citep[based on MINPACK,][]{MINPACK} implemented in \texttt{lmfit}\footnote{\url{https://lmfit.github.io/lmfit-py/}},
and then we ran an MCMC analysis with
the affine-invariant sampler \citep{GoodmanWeare2010CAMCS...5...65G}
implemented in the \emcee{} package \citep{DFM2013ascl.soft03002F, DFM2019JOSS....4.1864F}.
First, we used only the detrending parameters without the GP,
then we fixed the transit shape (if fitted) and the $T_0$
training the GP on the residuals.
The posteriors of the hyperparameters obtained are then used to define the priors
of for the subsequent analyses as
twice the error computed from the posterior distribution.
We re-ran the full analysis (transit model, detrending parameters, and GP) 
with physical and hyperparameter priors.
So, for each visit we ran the analysis both fitting and fixing the transit shape, 
different combinations of detrending parameters of the same kind,
e.g. linear and quadratic trend in time, 
first and second order derivative of the $x$ and $y$ pixel offset, etc,
and an additional set of detrending parameters determined
with a least squares fit on the out-of-transit part,
and with and without the GP.
For each of these analysis we computed the BIC,
and we visually inspected each single fit to avoid overcompensation 
of the GP,
looking for transit-like feature (also upside-down).
In addition,
we computed the Pearson's correlation $r$\footnote{Implemented within \texttt{SciPy} at \url{https://docs.scipy.org/doc/scipy/reference/generated/scipy.stats.pearsonr.html}.}
between the flux and the best-fit transit model ($r_\mathrm{tra}$),
and the flux and the best-fit GP model ($r_\mathrm{GP}$) without the transit model.
We found that all the transit models are strongly and significantly
correlated with the flux
($r_\mathrm{tra} > 0.9,\,p\mathrm{-value} < 0.05$),
while $r_\mathrm{GP}$ did not show any correlation.
We also tried to evaluate the possible level of overcompensation by adding a scaled transit model
(from the best-fit without the GP) to the GP model and
computing the correlation coefficient $r_\mathrm{GP,scaled}$.
We tested a scale factor ranging from $2.5\%$ to $0.5\%$ in steps of $0.5\%$.
We found that all $r_\mathrm{GP}$ were lower than $r_\mathrm{GP,scaled}$ with the scale factor at $0.5\%$,
allowing us to conclude that all our GP models could contribute to the transit model for less than $0.5\%$.
Even if this analysis is not conclusive, it is a further indication we are not introducing a strong bias in our
transit model and parameter estimation.
This allowed us to determine the best-fit combination of transit, detrending, and GP parameters for each visit as the model with the lowest BIC.
In the \emcee{} analysis we used 128 walkers
and we fine-tuned the number of steps and burn-in for each visit,
that is repeating the analysis with an increased number of steps
if the chains did not converge 
(we checked it with visual inspections of all the chains).
\par

\paragraph{Multi-visit analysis}
For the targets already observed by CHEOPS multiple times,
with \pycheops{} we were able to combine
the best-fit results of the single-visit analysis.
This allowed us to analyse simultaneously the multiple visits.
We fitted the transit and LD common model,
as for the single visit 
($D$, $W$, $b$, $h_1$, and $h_2$).
We also used the detrending parameters of each single visit,
and a common \texttt{SHOTerm} GP kernel \citep{Foreman-Mackey2017AJ....154..220F}
with two common hyperparameters ($\log S_0$ and $\log \omega_0$).
The GP is able to take into account linear trends in time,
so if present, we used very tight priors on $\dd f/\dd t$.
The priors on the hyperparameters were determined as the average (with error propagation)
of the single-visit hyperparameters. 
We used the default values of the GP hyperparameters if not present in the single-visit.
In the multi-visit the roll angle model within \pycheops{}
is not part of the detrending model as in the single-visit analysis.
The detrending parameters of the roll angle (and its harmonics)
are treated as nuisance parameter following the recipe by \citet{Luger2017RNAAS...1....7L}
and they are marginalised away as a \texttt{celerite CosineTerm} kernel
(see Maxted et al., in prep. for further details)
added to the covariance matrix.
This method implicitly assumes that the roll angle is a linear function
of time for each visit,
that is the rate of change of the roll angle is constant.
\par
First, we fitted the common transit parameter,
the detrending and GP hyperparameters,
and a linear ephemeris with parameters 
the reference time ($T_{0,\mathrm{ref}}$) and the period ($P$).
Then, we took the best-fit parameters from the posterior distribution and
we repeated the analysis, but
fixing $T_{0,\mathrm{ref}}$ and $P$ and
fitting $\Delta T_{0,n}$\footnote{Also referred into \pycheops{} as $\mathrm{ttv}_n$, with $n$ the visit number.}
for each visit $n$,
that is the deviation from the calculated transit time from the linear ephemeris
$T_{0,n} = T_{0,\mathrm{ref}} + E \times P + \Delta T_{0,n}$,
with $E$ the epoch, an integer number that identifies the transit.
We found that using a number of walkers (or chains) between 64 and 128
(depending on the number of fitting parameters)
was enough to reach convergence
for the multi-visit analysis with \emcee{},
because we are starting from previous single-visit analysis.
\par

For targets with multiple visits we calculated the so-called Observed-Calculated ($O-C$) plot,
where $O$ is the observed $T_0$ and
the $C$ is the transit time computed from the linear ephemeris.
The $O-C$ diagram is a simple tool able to identify a possible TTV signal.
We computed two $O-C$ values,
one for the $T_{0,n}$ of single visits with the ephemeris obtained from multi-visit analysis,
and a second one as direct output of the multi-visit analysis, 
that is $(O-C)_{n} = \Delta T_{0,n}$.
In this way we were also able to assess visual improvement
on the transit timing measurement with simultaneous multi-visit analysis.
\par

For all the single and multiple visit analysis
we used as best-fit solution the maximum likelihood estimation (MLE), 
that is the set of parameters that maximises the likelihood of the posterior distribution.
We computed as error, $\sigma$, of the best-fit the semi-interval of
the high density interval (HDI\footnote{
Based on the implementation of \texttt{TraceAnalysis.hpd} within the \texttt{PyAstronomy} package,
available at \url{https://pyastronomy.readthedocs.io/en/latest/}.
})
at $68.27\%$ of the posterior,
which is equivalent to the semi-interval defined by the $16$-th and $84$-th percentiles
in case of Gaussian distribution\footnote{The error of the fitted parameters computed as the semi-difference between
the $84$-th and the $16$-th percentile is the default method within \pycheops.}.
\par

\subsection{HAT-P-17 b}
\label{sec:lc_hatp17}

HAT-P-17 is an early K dwarf (see Table~\ref{tab:hatp17} for stellar parameters)
that hosts two exoplanets,
it was the second multi-planet system detected by a ground-based facility \citep{Howard2012ApJ...749..134H}.
The outer planet, HAT-P-17 c, has a poorly constrained orbit with a period
that could be anywhere in range between 10 and 36 years \citep{Fulton2013ApJ...772...80F}.
It does not appear to transit.
HAT-P-17 b is a transiting exoplanet with mass and radius of about
$0.5\, M_\mathrm{Jup}$ and $1\, R_\mathrm{Jup}$, respectively,
and an orbital period of 10.3 days.
The planet has a high eccentricity $e = 0.342$ that
would suggest a perturbation process responsible for the formation of the system,
even if the spin-orbit misalignment ($\lambda = 19^{+14}_{-16}\degr$) was not significantly detected
by \citet{Fulton2013ApJ...772...80F}.
The same author, from adaptive optics analysis, ruled out the possible presence of
a distant ($> 50$~au) and massive object ($M \sim 80\ M_\mathrm{Jup}$).
This suggests that Kozai-Lidov process was not responsible for the formation of the system.
Detecting a TTV signal from a fourth lighter object on a mutually inclined orbit 
would be the evidence that the P-P scattering could be the main process
in the evolution of the HAT-P-17 system.
\par

CHEOPS observed HAT-P-17 from August 2020 to October 2020,
obtaining three transits of the planet b with \geff{} of $65.8\%$, $57.4\%$, and $48.5\%$, respectively.
The third visit covers almost 0\% of both ingress and egress,
lowering the precision on the transit time of this visit.
\par

Before observing we realised that there are two stars of magnitude
$G = 14.6$ and $15.7$ ($4-5$ magnitudes fainter than the target),
located close to the edge of the photometric aperture (aperture radius of $25\arcsec$).
These two stars are not physically bound to HAT-P-17, 
at a distance of about $26\arcsec$,
as we can infer from their parallax ($\pi$) and proper motions ($\mu_\alpha,\, \mu_\delta$)
from \gaia{} EDR3\footnote{
Only in this case we used EDR3 instead of DR2 because of the updated values of
parallaxes and proper motions of neighbour stars at time of writing.
Using EDR3 parallaxes in the stellar properties of all the targets
would have not affected the results of our work.
} 
\citep[][Gaia Collaboration et al., in prep.]{GaiaPLX_2020arXiv201201742L},
i.e.,
$\pi= 1.29\pm0.02$~mas, 
$\mu_\alpha= 5.21\pm0.02,\, \mu_\delta= -14.19\pm0.02$~mas/yr and
$\pi= 0.53\pm0.05$~mas, 
$\mu_\alpha= 6.31\pm0.04,\, \mu_\delta= -9.53\pm0.04$~mas/yr,
respectively, 
while HAT-P-17 has 
$\pi = 10.82 \pm 0.02$~mas and 
$\mu_\alpha= 80.28\pm0.02 ,\, \mu_\delta= -127.04\pm0.02$~mas/yr.
We estimated that the flux contribution from the contaminants is about $2.5\%$,
but we were able to model it with \pycheops.
\par

\begin{figure*}
\centering
	\includegraphics[width=0.99\columnwidth]{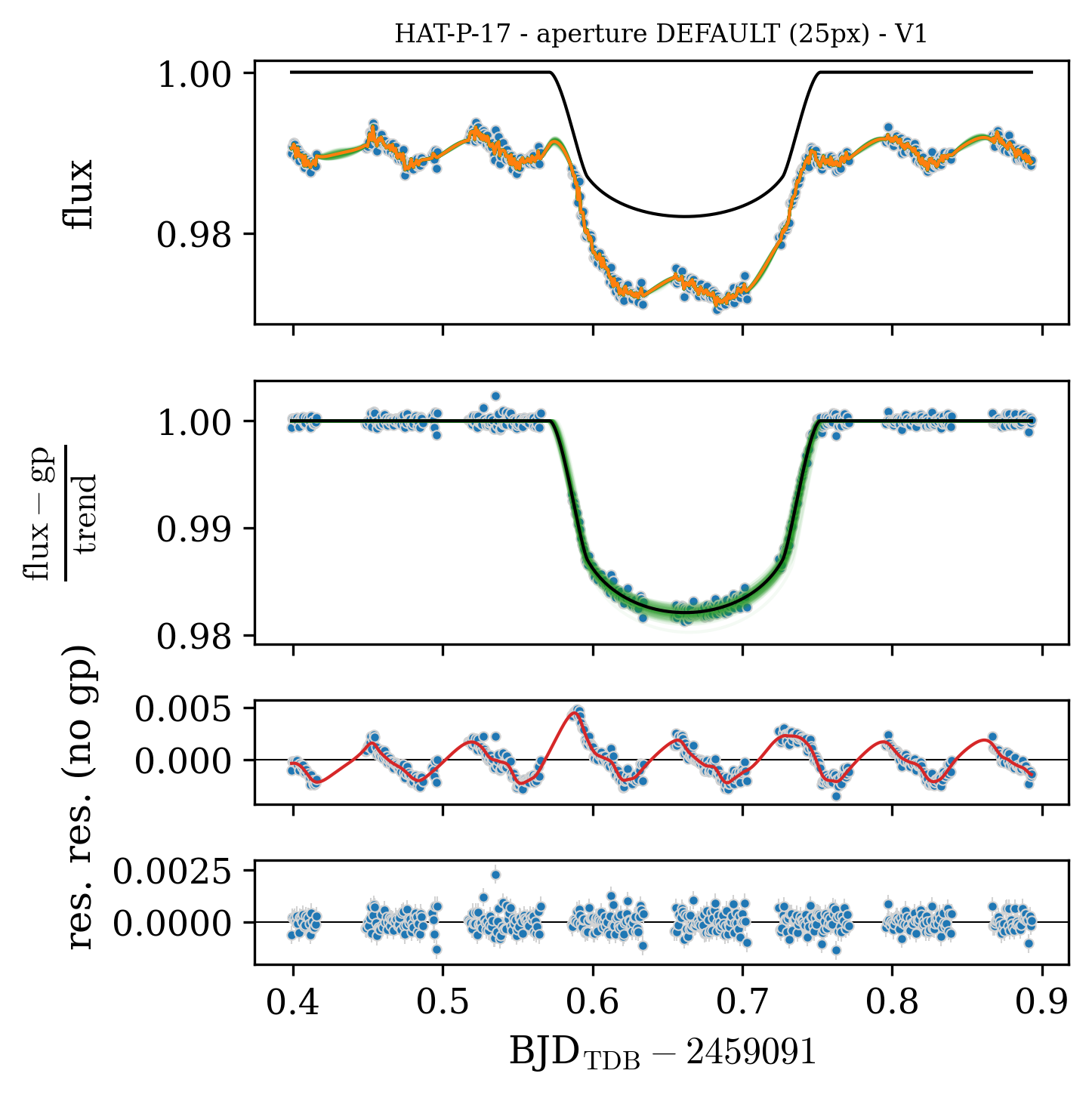}
	\includegraphics[width=0.99\columnwidth]{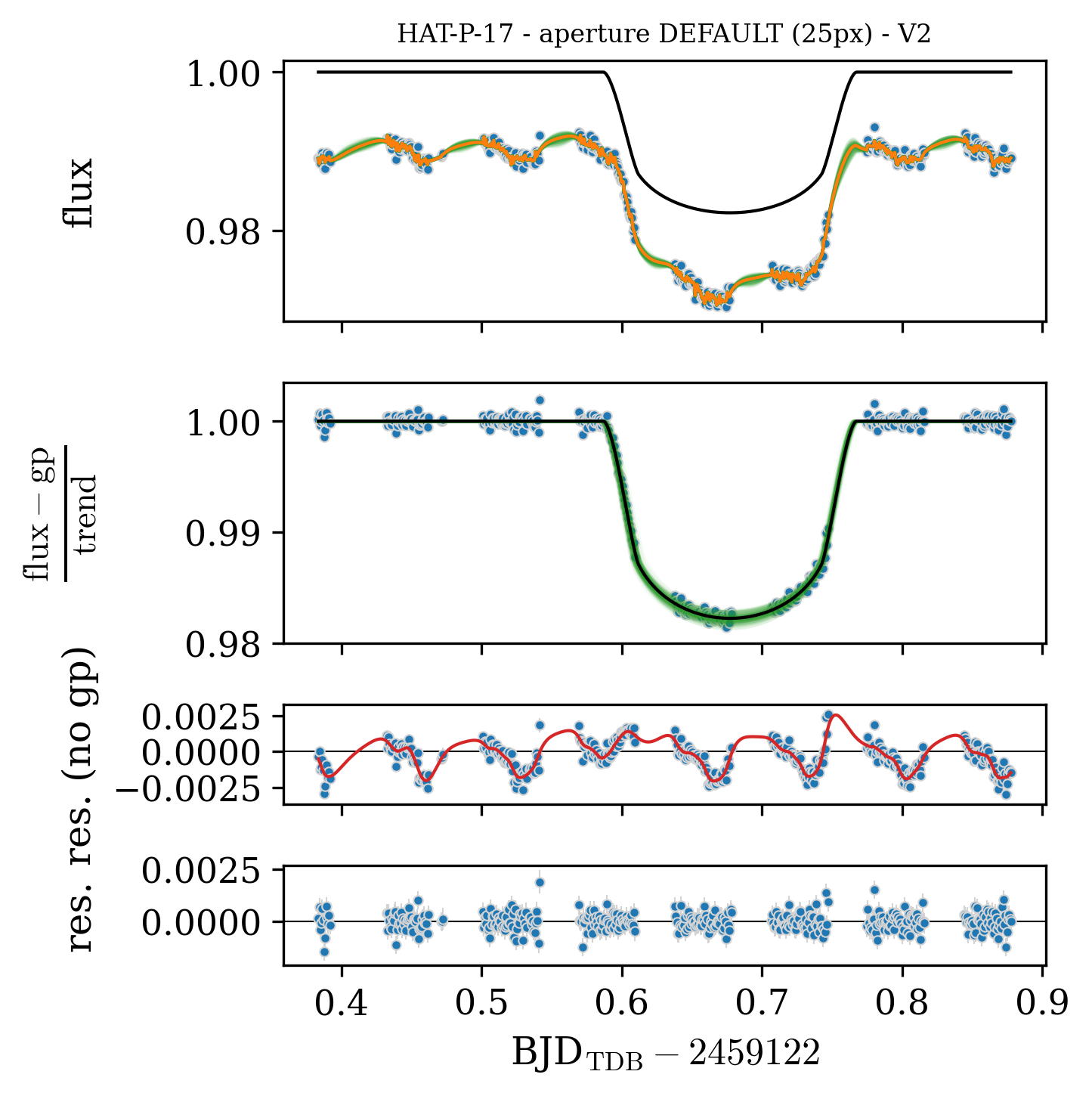}
	\includegraphics[width=0.99\columnwidth]{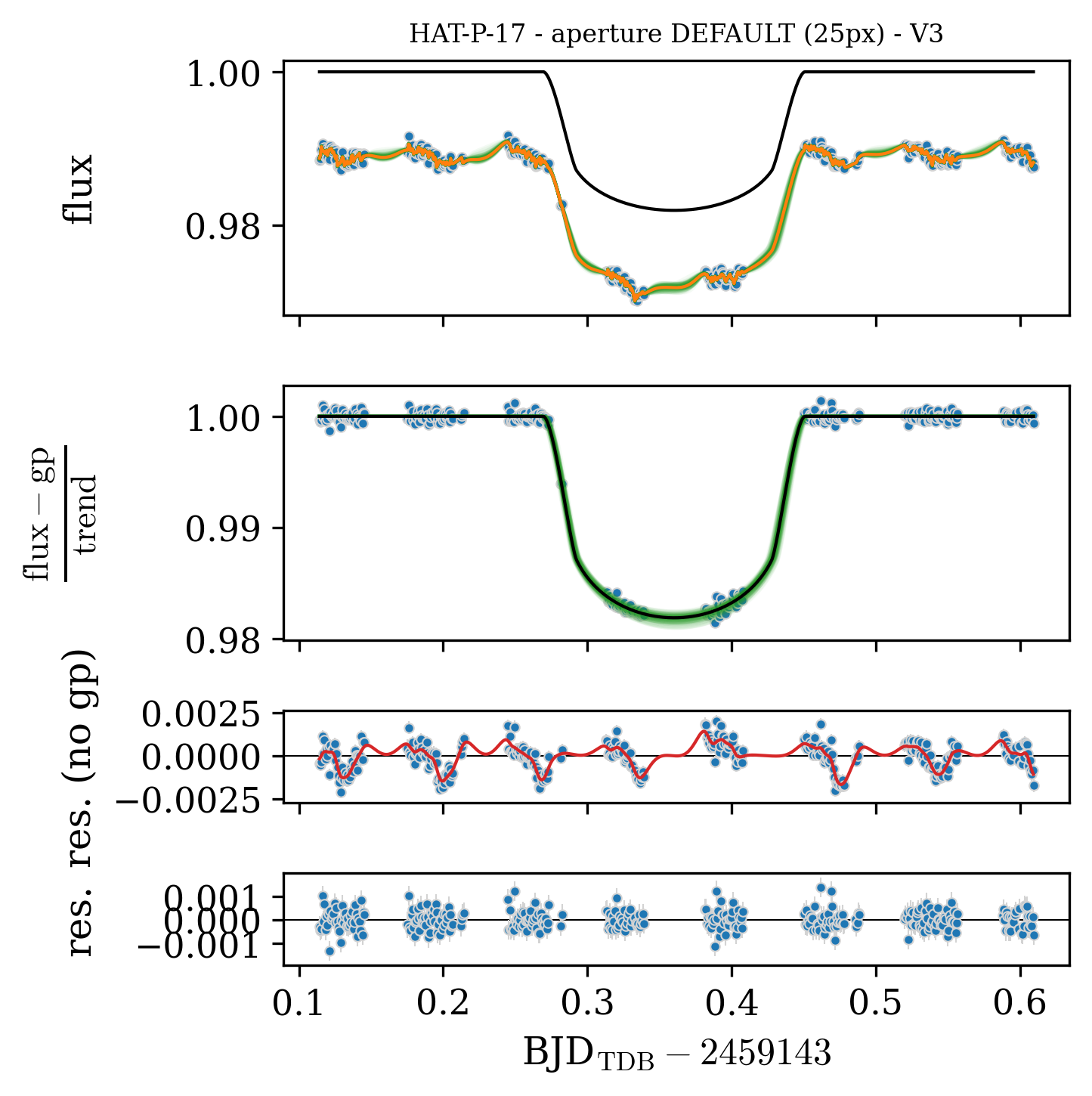}
    \caption{HAT-P-17 b single visit analysis.
    Maximum Likelihood Estimation (MLE, orange line) from the posterior distribution as the best-fit model 
    (lowest BIC)
    with 128 random samples as green lines
    (un-detrended and detrended in the first and second panel, respectively, of each figure);
    black line as the transit model (with out-of-transit set to 1 by default).
    If gaussian process (GP) has been used an additional panel shows the residuals with over-plotted the best-fit GP model (red line).
    The last panel shows the residuals with respect to the best-fit model
    with the photometric jitter term (fitted as $\log \sigma_j$) 
    added in quadrature to the photometric errors.
    \textit{Upper-left:} first visit, fitted transit shape and detrending against contaminants and GP;
    \textit{upper-right:} second visit, same fitting and detrending parameters of first visit;
    \textit{lower:} third visit, model parameters as first and second visit.
    }
    \label{fig:visits_hatp17}
\end{figure*}

\begin{figure*}
\centering
	\includegraphics[width=1.1\columnwidth]{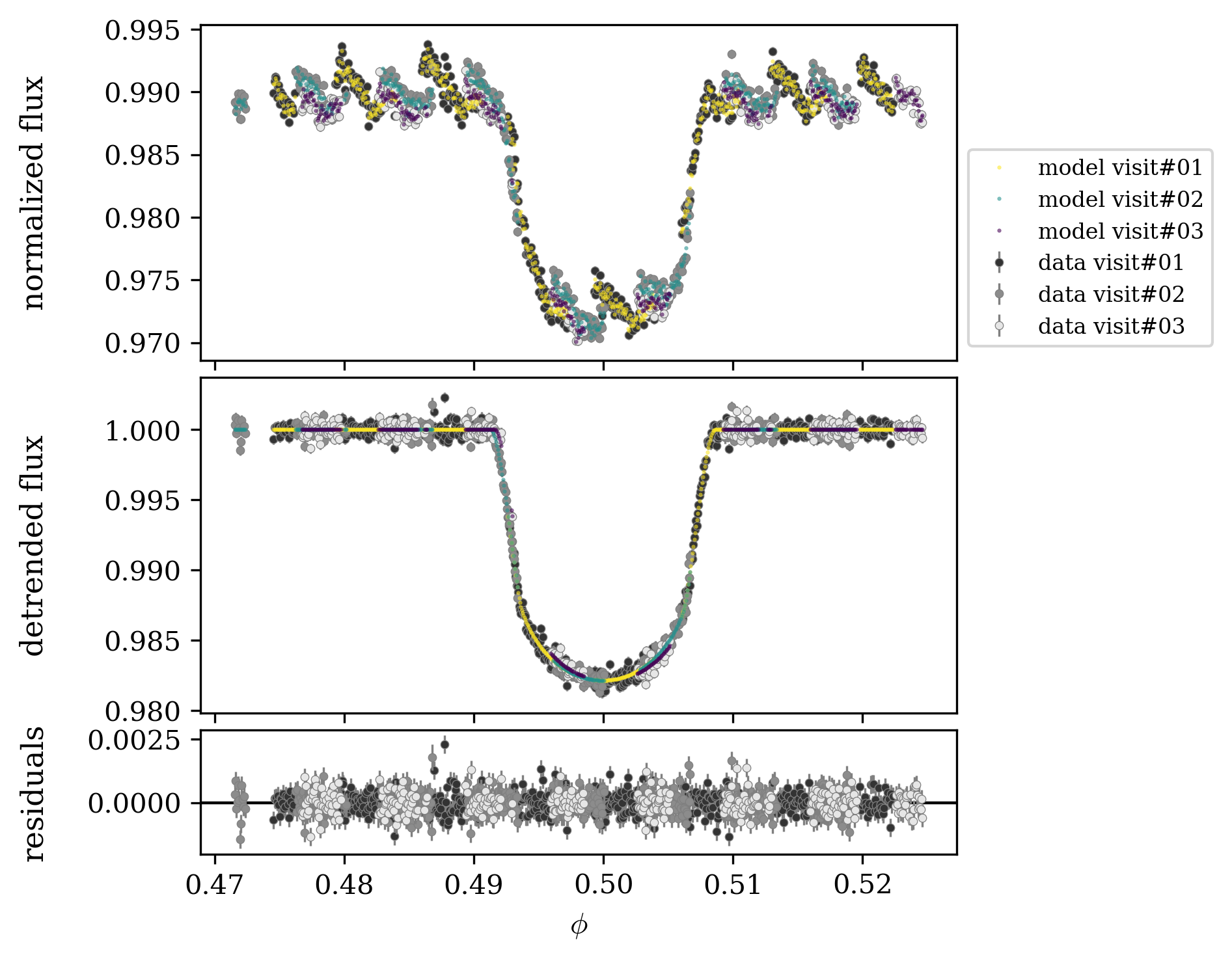}
	\includegraphics[width=0.9\columnwidth]{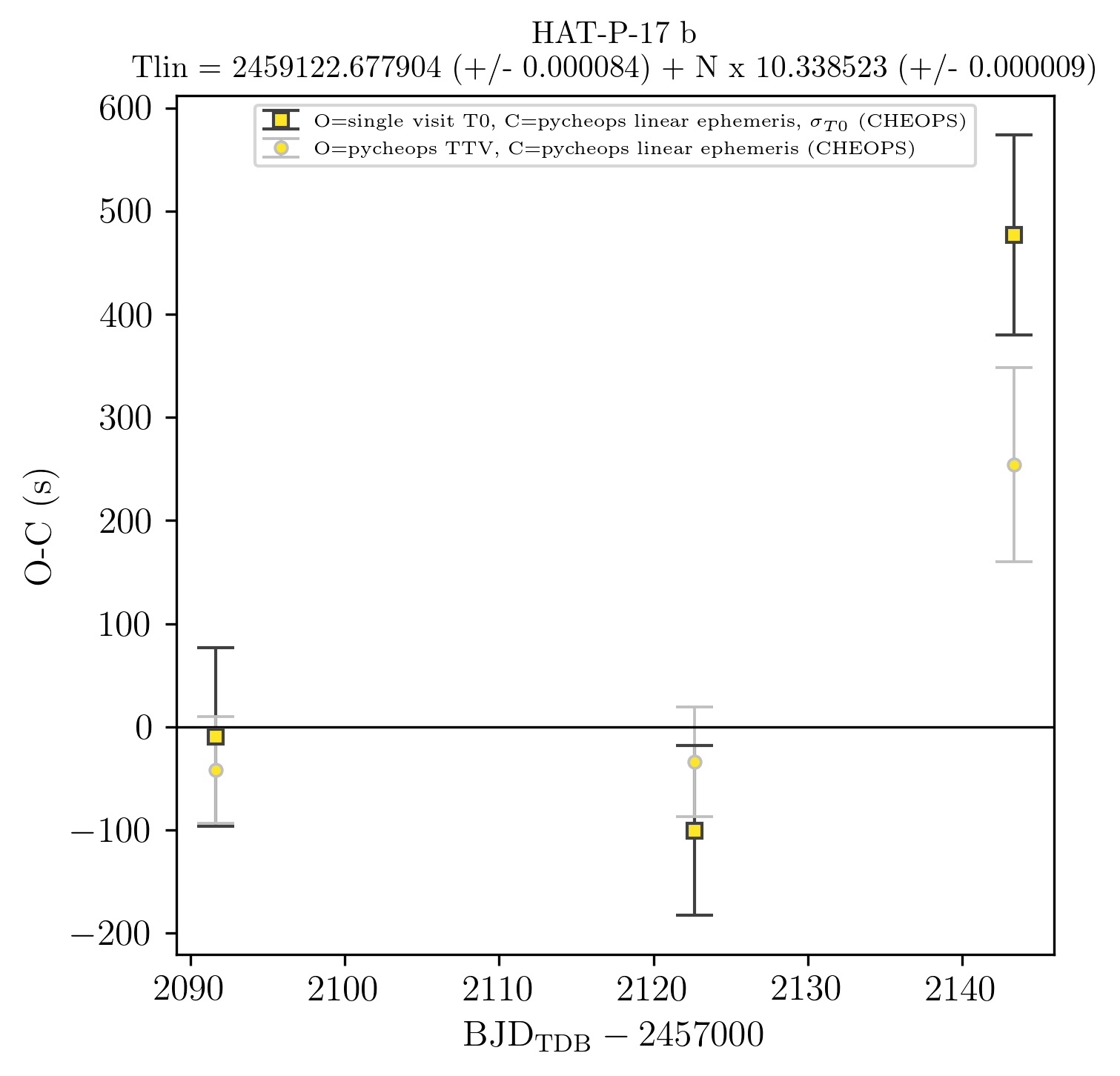}
    \caption{Multi-visit analysis of HAT-P-17 b.
    \textit{Left}: three CHEOPS visits in phase, $\phi$,
    with respect to the linear ephemeris and
    taking into account possible TTV signal 
    by fitting $\Delta T_{0,n}$; 
    data points plotted as white, gray and black circles for first, second, and third visit, respectively;
    coloured circles represent the model for different visit;
    from top to bottom panels: first panel shows the raw light curves,
    second panel shows the detrended light curves also corrected by gaussian process,
    third panel shows the residuals.
    \textit{Right}: $O-C$ diagram with values from
    the single-visit analysis (squares)
    and the multi-visit analysis (circles).
    We used a common linear ephemeris (on top of the figure)
    from the first iteration of the multi-visit analysis as calculated $C$
    and the $T_0$s of the sigle-visit analysis as observed $O$.
    The $O-C$ values for the multi-visit analysis correspond to the directly fitted $\Delta T_{0,n}$, with $k$ the visit number.
    }
    \label{fig:mv_hatp17}
\end{figure*}

For each visit we modelled the light curves fitting the shape of the transit and
the systematics with the contaminant parameter in the detrending model and 
the GP kernel. 
These were the models with the lowest BIC.
Table~\ref{tab:hatp17} lists literature values used as initial guess and priors.
See Fig.~\ref{fig:visits_hatp17} for the three single visits of HAT-P-17 b,
with best-fit model (transit, detrending, and GP).
We obtained from single-visit analysis an error on the transit time of 
$\sigma_{T_0} = 87$~s, 82~s, and 97~s, respectively.
We used the single-visit analysis as input for a simultaneous-combined multi-visit analysis (Fig.~\ref{fig:mv_hatp17}).
We reported in Table~\ref{tab:hatp17} the best-fit solution of the multi-visit analysis.
The $O-C$ plot of the three visits is shown in Fig.~\ref{fig:mv_hatp17} with
the linear ephemeris from the first iteration of the multi-visit analysis
(see $T_{0,\mathrm{ref}}$ and $P$ in Table~\ref{tab:hatp17}).
We found that the first two visits have an improvement on the $\sigma_{T_0}$ of $\sim30$~s 
($40\%$ for the first visit, $35\%$ for the second visit),
and they agree with the linear ephemeris at 1-$\sigma$.
On the other hand the third visit improved the $T_0$ only by 3~s and 
shows a deviation from the linear ephemeris.
Also the $T_0$ of the multi-visit analysis
agrees with the single-visit analysis only at 2-$\sigma$.
As shown by \citet{Barros2013MNRAS.430.3032B} the uncertainties from partial transits
are usually underestimated which explains the discrepancies found.
To confirm or rule out the TTV signal in the $O-C$ diagram of Fig.~\ref{fig:mv_hatp17},
we need to analyse the CHEOPS observations with literature and TESS data,
but this was not the purpose of this work.

\begin{table}
	\centering
	\caption{HAT-P-17 summary table of stellar and planetary (planet b) parameters.
	Input and priors planetary parameters from \citet{Howard2012ApJ...749..134H}
	and \citet{Fulton2013ApJ...772...80F}.
	Best-fit solution (MLE and semi-interval HDI at $68.27\%$) from the simultaneous three visits analysis.
	}
	\label{tab:hatp17}
	\resizebox{\fitbox}{!}{%
	\begin{tabular}{lcc}
		\hline
		 Parameters & Input/priors & Source\\
		\hline
		 HAT-P-17 & \multicolumn{2}{c}{Gaia DR2 1849786481031300608} \\
		 RA (J2000)  & 21:38:08.73 & Simbad\\
		 DEC (J2000) & +30:29:19.4 & Simbad\\
		 $\mu_\mathrm{\alpha}$~(mas/yr) & $-80.4 \pm	0.2$ & \gaia{} DR2\\
		 $\mu_\mathrm{\delta}$~(mas/yr) & $-127.0	\pm 0.2$ & \gaia{} DR2\\
		 age (Gyr) & $7 \pm 2$ & This work\\
		 parallax (mas) & $10.80 \pm 0.06$ & \gaia{} DR2\\
		 $V$~(mag) & 10.4 & Simbad\\
		 $G$~(mag) & 10.3 & \gaia{} DR2\\
		 $M_{\star}\, (M_{\sun})$ & $0.88 \pm 0.04$ & This work\\
		 $R_{\star}\, (R_{\sun})$ & $0.84 \pm 0.01$ & This work\\
		 $\rho_{\star}\, (\rho_{\sun})$ & $1.1 \pm 0.5$ & This work\\
		 $T_\mathrm{eff}$~(K) & $5332 \pm 55$ & SWEET-Cat\\
		 $\log g$ & $4.45 \pm 0.13$ & SWEET-Cat\\
		 \feh~(dex) & $+0.05 \pm 0.03$ & SWEET-Cat\\
		 \hline
		 HAT-P-17 b & & \\
		 Model & Input/priors & Multi-visit (MLE \& HDI) \\
		 $T_{0,\mathrm{ref}}^{(a)}$~(days) & $-2198.8306 \pm 0.0002$& $2122.67790 \pm  0.00008$ \\
		 $P$~(days) & $10.338523 \pm 0.000009$ & $10.338524 \pm 0.000009$ \\
		 $D = k^2$ &  & $ 0.0153 \pm 0.0002$\\
		 $W$~(unit of $P$) &  & $ 0.01609 \pm 0.00008$\\
		 $b$ & $0.31 \pm 0.07$ & $0.47 \pm 0.02$\\
		 $h_1$ & $0.71 \pm 0.01$ & $0.70 \pm 0.01$\\
		 $h_2$ & $0.44 \pm 0.05$ & $0.47 \pm 0.05$\\
		 $T_{0,1}^{(a)}$~(days) & $2091.66222 \pm 0.00100$ & $2091.66185 \pm 0.00060$ \\
		 $T_{0,2}^{(a)}$~(days) & $2122.67674 \pm 0.00095$ & $2122.67751 \pm 0.00062$ \\
		 $T_{0,3}^{(a)}$~(days) & $2143.36047 \pm 0.00112$ & $2143.35789 \pm 0.00109$ \\
 		 $\log \sigma_j$ & - & $-8.22 \pm 0.07$\\
		 Derived/physical &  &  \\
		 $k=R_\mathrm{b}/R_{\star}$ & $0.124 \pm 0.001$ & $0.1237 \pm 0.0007$\\
		 $R_\mathrm{b}\, (R_\mathrm{Jup})$ & - & $1.04 \pm 0.02$ \\
		 $a/R_{\star}$ & $22.6 \pm 0.5$ & $20.2 \pm 0.3$\\
		 $i\, (\degr)$ & $89.2 \pm 0.2$ & $88.7 \pm 0.1$\\
		 $T_{14}^{(b)}$~(days) & $0.1690 \pm 0.0009$ & $0.1664 \pm 0.0008$\\
		 $e$ & $0.342 \pm 0.005$ & fixed \\
		 $\omega\, (\degr)$ & $201.5 \pm 1.6$ & fixed \\
		 $K_\mathrm{RV}$~(ms$^{-1}$)) & $58.6 \pm 0.7$ & - \\
		 $M_\mathrm{b}\, (M_\mathrm{Jup})$ & - & $0.54 \pm 0.02$\\
		 $\rho_\mathrm{b}$~(gcm$^{-3}$) & - & $0.44 \pm 0.02$\\
		 $\lambda^{(c)}\, (\degr)$ & $19^{+14}_{-16}$& \\
		 GP hyperparameters & & \\
		 $\log S_0$ & - & $-18.9 \pm 0.2$ \\
		 $\log \omega_0$ & - & $4.78 \pm 0.07$\\
		\hline
	\end{tabular}
	}
	\\
	\textbf{Notes:}
	$^{(a)}$: Transit times in BJD$_\mathrm{TDB}-2457000$. 
	$T_{0,n}$ single visit output in the input/priors column, 
	while they are the linear ephemeris plus $\Delta T_{0,n}$ from multi-visit analysis.
	$^{(b)}$: Total duration. The eq. used depends on the literature.
	The multi-visit duration is equal to $T_{14} = W \times P$.
	$^{(c)}$: spin-orbit angle measured from the Rossiter-McLaughlin effect.
\end{table}

\subsection{KELT-6 b}
\label{sec:lc_kelt6}

KELT-6 is a late F-type star that hosts two exoplanets,
one transiting, KELT-6 b \citep{Collins2014AJ....147...39C}, with a period of 7.85~d
and an outer more massive non-transiting planet, KELT-6 c \citep{Damasso2015AA...581L...6D}.
\citet{Damasso2015AA...581L...6D} proposed that
the main formation process of the system can be the result of
a P-P scattering of more than two planets and 
a successive coplanar high-eccentricity migration \citep[CHE][]{Petrovich2015ApJ...805...75P}.
Detecting a TTV signal induced by a lighter planet on an outer coplanar orbit and in MMR (or close to)
with planet b would imply a disk-driven migration,
instead of a P-P scattering, that would result in a perturber on a mutually inclined orbit,
and outside a MMR.
\par

\begin{figure}
\centering
	\includegraphics[width=0.99\columnwidth]{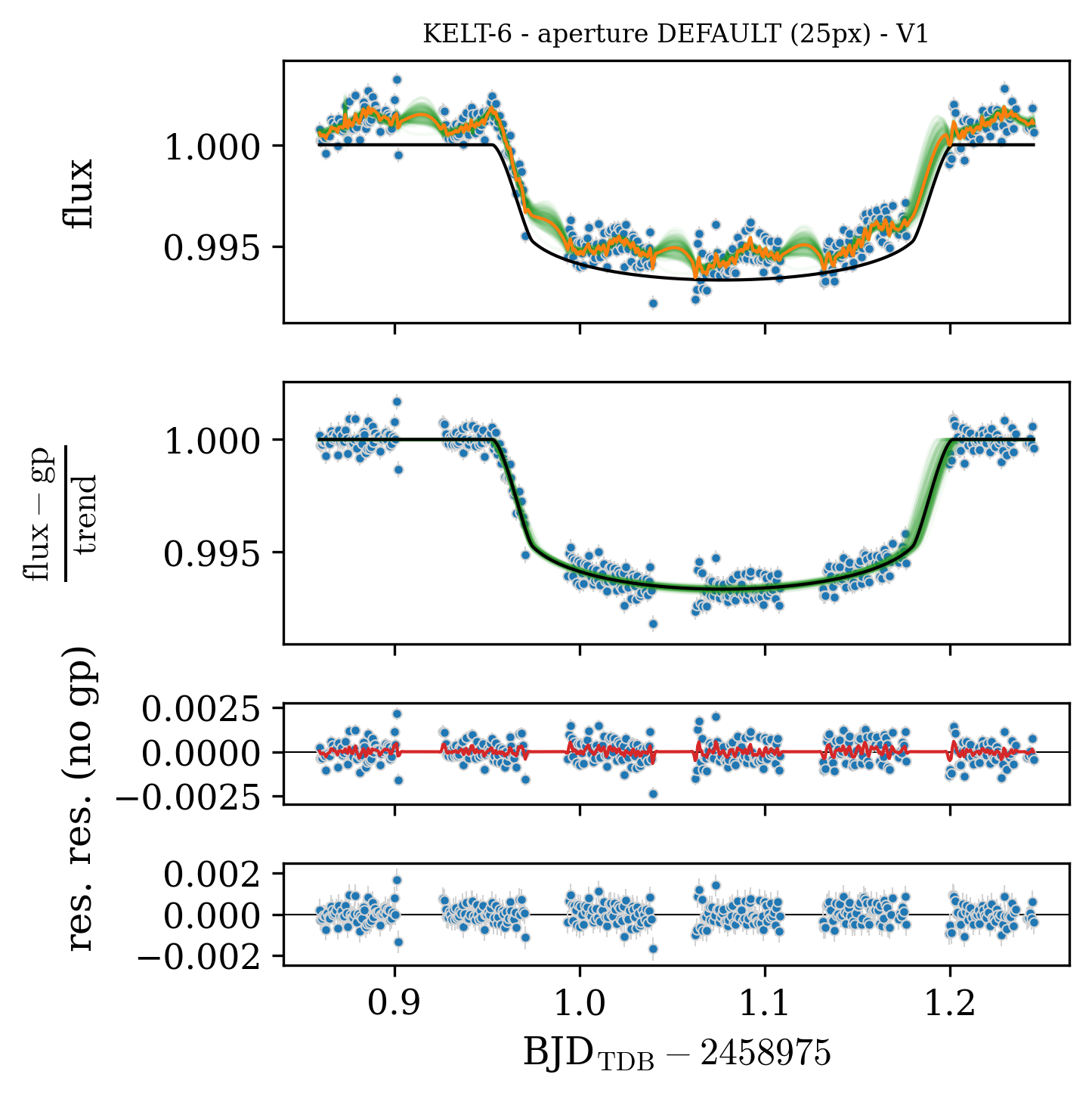}
    \caption{KELT-6 b single visit analysis (see Fig.~\ref{fig:visits_hatp17} for description).
    The model, with the lowest BIC,
    contains the fitted transit shape and detrending against first three harmonics of the satellite roll angle.
    }
    \label{fig:visits_kelt6}
\end{figure}

We collected only one CHEOPS visit of KELT-6 b on May 6, 2020,
with a \geff{} of about $69\%$.
The visit duration was too short to sample correctly the post-egress part,
and also the egress phase was completely missed (Fig.~\ref{fig:visits_kelt6}).
In this case we run only the single-visit analysis, 
and the BIC analysis favoured the model
fitting for the transit shape and 
detrending for the first three harmonics of the roll angle
(as $\dd f/\dd \cos (n \times \phi)$,  $\dd f/\dd \sin(n \times \phi)$ with
$n = 1, 2, 3$ that identifies the harmonic)
and with the GP.
We obtained an error on the $T_0$ of about 114~s,
totally dominated by the lack of points in the egress.
However, with only one CHEOPS visit
we were able to improve the parameters of KELT-6 b (see Table~\ref{tab:kelt6}).
More transits are needed to run a combined analysis covering all the phases of the transit 
to improve the precision on the transit time.
\par

\begin{table}
	\centering
	\caption{KELT-6 summary table of stellar and planetary (planet b) parameters.
	Input and priors planetary parameters from \citet{Collins2014AJ....147...39C}
	and \citet{Damasso2015AA...581L...6D}.
	Best-fit solution (MLE and semi-interval HDI at $68.27\%$) from the one single-visit analysis.
	}
	\label{tab:kelt6}
	\resizebox{\fitbox}{!}{%
	\begin{tabular}{lcc}
		 \hline
		 Parameters & Input/priors & Source\\
		 \hline
		 KELT-6 & \multicolumn{2}{c}{Gaia DR2 1464700950221781504} \\
		 RA (J2000)  & 13:03:55.65 & Simbad \\
		 DEC (J2000) & +30:38:24.28 & Simbad \\
		 $\mu_\mathrm{\alpha}$~(mas/yr) & $-5.11 \pm 0.05$ & \gaia{} DR2\\
		 $\mu_\mathrm{\delta}$~(mas/yr) & $15.64 \pm 0.05$ & \gaia{} DR2\\
         age (Gyr) & $5 \pm 1$ & This work\\
		 parallax (mas) & $4.13 \pm 0.03$ & \gaia{} DR2\\
		 $V$~(mag) & 10.3 & Simbad \\
		 $G$~(mag) & 10.2 & \gaia{} DR2\\
		 $M_{\star}\, (M_{\sun})$ & $1.11 \pm 0.06$ & This work\\
		 $R_{\star}\, (R_{\sun})$ & $1.34 \pm	0.06$ & This work\\
		 $\rho_{\star}\, (\rho_{\sun})$ & $0.4 \pm 0.2$ & This work\\
		 $T_\mathrm{eff}$~(K) & $6246 \pm  88$ & SWEET-Cat\\
		 $\log g$ & $4.22 \pm 0.09$ & SWEET-Cat\\
		 \feh~(dex) & $-0.22 \pm 0.06$ & SWEET-Cat\\
		 \hline
		 KELT-6 b & & \\
		 Model & Input/priors & Single-visit (MLE \& HDI) \\
		 $P$~(days) & $7.845582 \pm 0.000007$ & fixed \\
		 $D=k^2$ & $ 0.0060 \pm 0.0002$ & $0.0058 \pm 0.0001$\\
		 $W$~(unit of $P$) & $0.0311 \pm 0.003$ & $0.0310 \pm 0.0004$ \\
		 $b$ & $0.22 \pm 0.17$ & $0.43 \pm 0.07$ \\
		 $h_1$ & $0.76 \pm 0.01$ & $0.76 \pm 0.01$\\
		 $h_2$ & $0.46 \pm 0.05$ & $0.46 \pm 0.05$\\
		 $T_{0,1}^{(a)}$~(days) & - & $1976.0773 \pm 0.0013$\\
     	 $\log \sigma_j$ & - & $-7.78 \pm 0.07$\\
		 Derived/physical &  &  \\
		 $k=R_\mathrm{b}/R_{\star}$ & $0.077 \pm 0.001$ & $0.0764 \pm 0.0008$\\
		 $R_\mathrm{b}\, (R_\mathrm{Jup})$ & - & $1.02 \pm 0.05$\\
		 $a/R_{\star}$ & $10.8 \pm 0.9$ & $10.1 \pm 0.4$\\
		 $i\, (\degr)$ & $88.8 \pm 0.9$ & $87.6 \pm 0.5$\\
		 $T_{14}^{(b)}$~(days) & - & $0.243 \pm 0.003$\\
		 $e$ & $0.029 \pm 0.016$ & fixed \\
		 $\omega\, (\degr)$ & $308 \pm 272$ & fixed \\
		 $K_\mathrm{RV}$~(ms$^{-1}$)) & $41.8 \pm 1.1$ & - \\
		 $M_\mathrm{b}\, (M_\mathrm{Jup})$ & - & $0.44 \pm 0.02$\\
		 $\rho_\mathrm{b}$~(gcm$^{-3}$) & - & $0.27 \pm 0.04$\\
		 $\lambda^{(c)}\, (\degr)$ & $-36 \pm 11$ & \\
		 GP hyperparameters & & \\
		 $\log S_0$ & - & $-23 \pm 2$\\
		 $\log \omega_0$ & - & $8 \pm 2$\\
		\hline
	\end{tabular}
	}
	\\
	\textbf{Notes:}
	$^{(a)}$: Transit times in BJD$_\mathrm{TDB}-2457000$. 
	$T_{0,n}$ single visit output.
	$^{(b)}$: Total duration equal to $T_{14} = W \times P$.
	$^{(c)}$: spin-orbit angle measured from the Rossiter-McLaughlin effect.
\end{table}

\subsection{WASP-8 b}
\label{sec:lc_wasp8}

WASP-8 b is an exoplanet with a radius similar to Jupiter and a mass of about 2~$M_\mathrm{Jup}$.
It has an eccentric, retrograde orbit ($\lambda = - 143\degr$),
with a period of about 8.16~d
\citep{Queloz2010AA...517L...1Q,Knutson2014ApJ...785..126K,Bourrier2017AA...599A..33B}.
\par

The host star, WASP-8 A, has a physical stellar companion, B,
at about $4\farcs5$ \citep{GaiaDR2_2018yCat.1345....0G}.
WASP-8 B lies within the CHEOPS point spread function of WASP-8 A,
but is four magnitudes fainter than A (in $G$-band),
and its contribution to the flux (less than $2\%$) in the aperture is almost negligible.
The presence of the stellar companion impacts the depth of transit,
but without changing the symmetry with respect to the $T_0$ and its measurement.
For these reasons, we did not take into account a dilution factor in the transit analysis,
but it will be done in future works.
\par

\citet{Knutson2014ApJ...785..126K} found that WASP-8 B,
having a mass of about $0.5\ M_{\sun}$
and a sky-projected separation greater than 390~au,
was not sufficient to explain the RV trend and modulation.
The authors suggested that two massive planets on outer orbits are needed.
So this system cannot be the result of disk-driven migration.
Instead a Kozai or a P-P scattering mechanism was invoked \citep{Knutson2014ApJ...785..126K},
making WASP-8 b a good candidate for our purpose.
\par

\begin{figure*}
\centering
	\includegraphics[width=0.99\columnwidth]{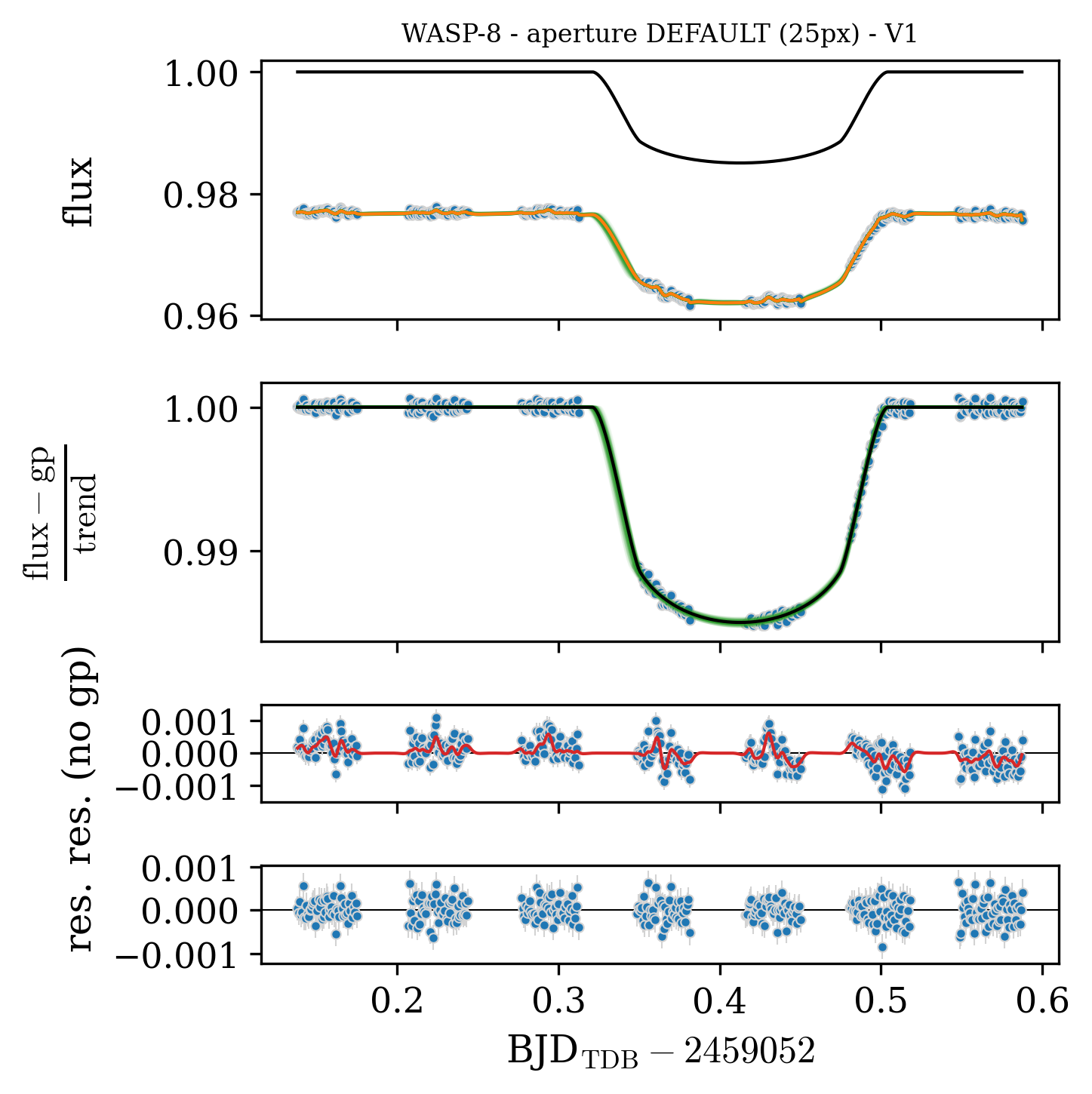}
	\includegraphics[width=0.99\columnwidth]{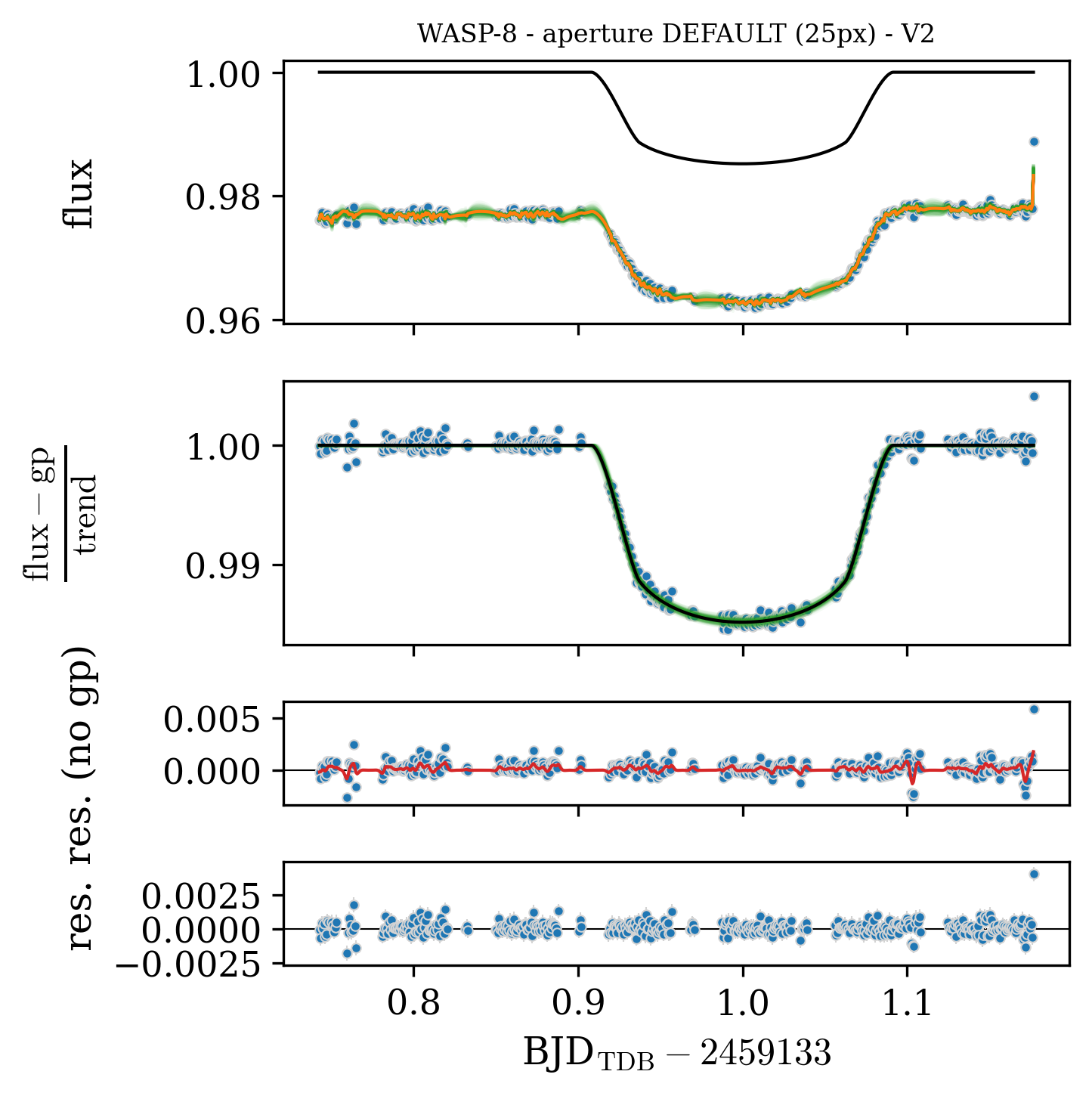}
    \caption{WASP-8 b single visits analysis (see Fig.~\ref{fig:visits_hatp17} for description).
    \textit{Left:} first visit, fitted transit shape and detrending against background and GP;
    \textit{right:} second visit, fitted transit shape and detrending against all parameters (but glint effect) and GP.
    }
    \label{fig:visits_wasp8}
\end{figure*}

\begin{figure*}
\centering
	\includegraphics[width=1.1\columnwidth]{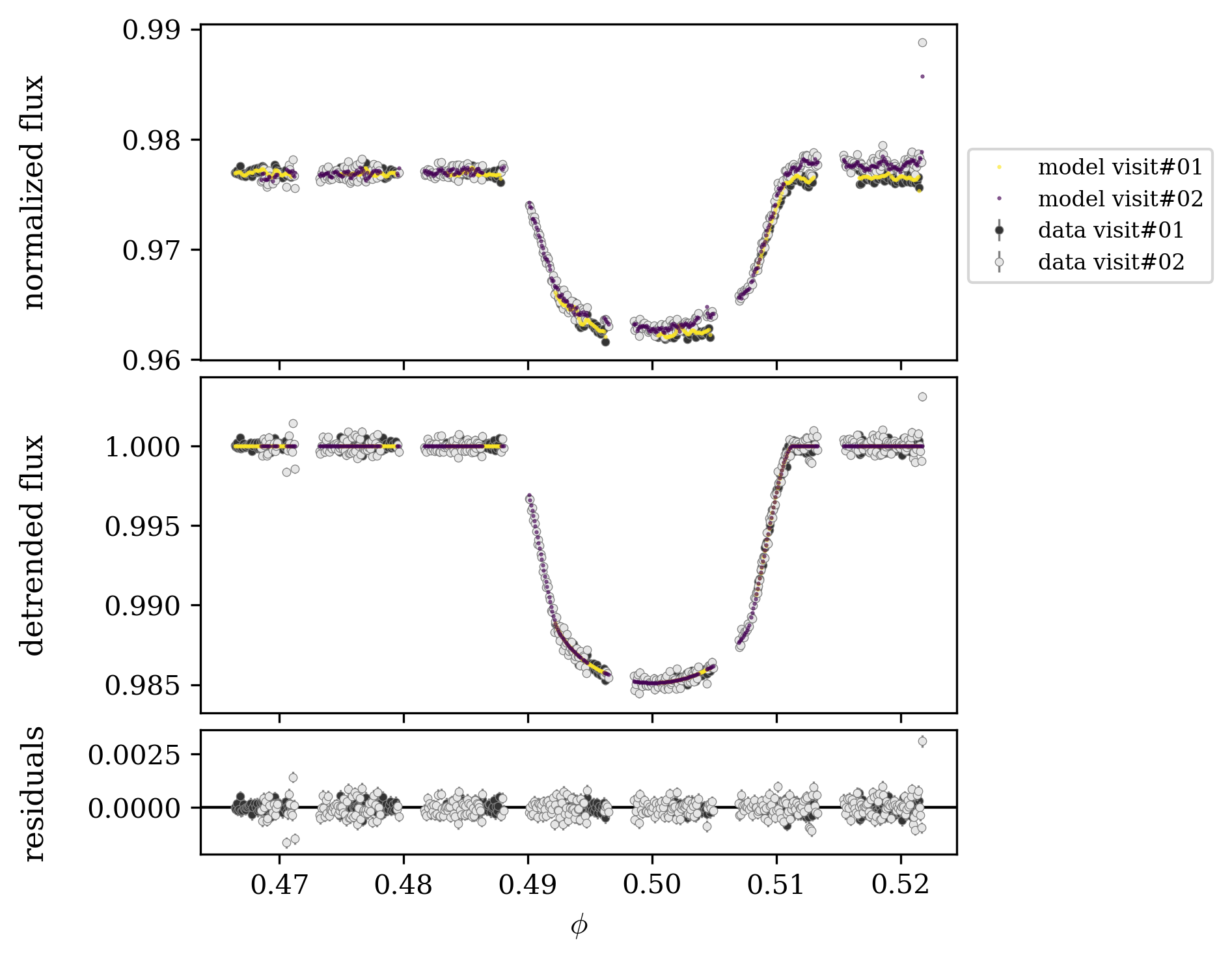}
	\includegraphics[width=0.9\columnwidth]{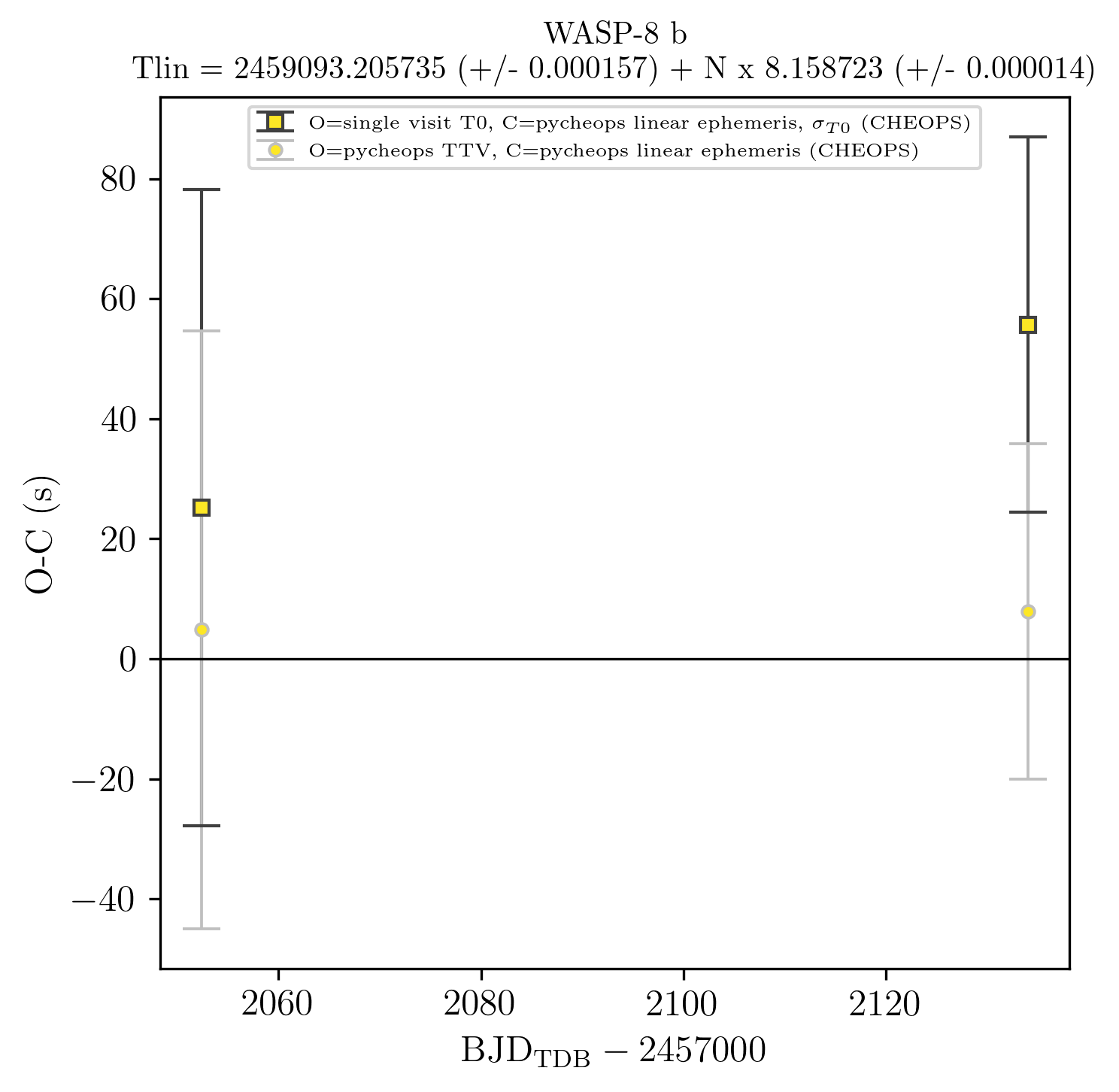}
    \caption{As in Fig~\ref{fig:mv_hatp17}, but for WASP-8 b.
    \textit{Left:} multi-visit phase plot of two CHEOPS visits;
    \textit{right:} O-C diagram.
    }
    \label{fig:mv_wasp8}
\end{figure*}

The transit of WASP-8 b was observed twice by CHEOPS, 
with one visit in July and one in October 2020,
with a \geff{} of $57.8\%$
and $67.8\%$, respectively.
The first visit shows a low coverage (almost null) of the ingress phase and good egress,
while the second visit has a \cpreff~$> 50\%$ of both ingress and egress.
We run the analysis and found that the BIC favoured models
fitting the shape of both transits and
detrending for the background and GP the first visit and 
for all the parameters (but glint) and GP the second visit.
The single-visit analysis provides a $\sigma_{T_0}$ of 53~s and 31~s
for the first and second visit, respectively (Fig.~\ref{fig:visits_wasp8}).
From the multi-visit analysis we had an improvement of about 4~s for both visits
(see Table~\ref{tab:wasp8} for the summary of the results),
taking into account that the gaps of the two visits are in phase,
lowering the effective sample timing of the transit,
i.e., the start of the ingress phase is missing.
With further visits with different \geff, \cpreff, and gap phases,
we will be able to reach a higher precision on the $T_0$,
improving also the preliminary result of the $O-C$ (see Fig.~\ref{fig:mv_wasp8}).
\par

\begin{table}
	\centering
	\caption{WASP-8 summary table of stellar and planetary (planet b) parameters.
	Input and priors planetary parameters from \citet{Queloz2010AA...517L...1Q}, \citet{Knutson2014ApJ...785..126K} and \citet{Bourrier2017AA...599A..33B}.
	Best-fit solution (MLE and semi-interval HDI at $68.27\%$) from the two multi-visit analysis.
	}
	\label{tab:wasp8}
	\resizebox{\fitbox}{!}{%
	\begin{tabular}{lcc}
		\hline
		 Parameters & Input/priors & Source\\
		\hline
		 WASP-8 & \multicolumn{2}{c}{Gaia DR2 2312679845530628096} \\
		 RA (J2000)  & 23:59:36.07 & Simbad\\
		 DEC (J2000) & -35:01:52.92 & Simbad\\
		 $\mu_\mathrm{\alpha}$~(mas/yr) & $109.75 \pm 0.06$ & \gaia{} DR2\\
		 $\mu_\mathrm{\delta}$~(mas/yr) & $7.61 \pm 0.06$ & \gaia{} DR2\\
         age (Gyr) & $3 \pm 1$ & This work\\
		 parallax (mas) & $11.09 \pm 0.05$ & \gaia{} DR2\\
		 $V$~(mag) & 9.9 & Simbad\\
		 $G$~(mag) & 9.6 & \gaia{} DR2\\
		 $M_{\star}\, (M_{\sun})$ & $1.07 \pm 0.04$ & This work\\
		 $R_{\star}\, (R_{\sun})$ & $0.96  \pm 0.03$ & This work\\
		 $\rho_{\star}\, (\rho_{\sun})$ & $0.9 \pm 0.6$ & This work\\
		 $T_\mathrm{eff}$~(K) & $5690 \pm 36$ & SWEET-Cat\\
		 $\log g$ & $4.42 \pm 0.15$ & SWEET-Cat\\
		 \feh~(dex) & $0.29 \pm 0.03$ & SWEET-Cat\\
		 \hline
		 WASP-8 b & & \\
		 Model & Input/priors & Multi-visit (MLE \& HDI) \\
		 $T_{0,\mathrm{ref}}^{(a)}$~(days) & $-2320.6661 \pm 0.0005$ & $2093.20574 \pm 0.00016$ \\
		 $P$~(days) & $8.15872 \pm 0.00002$ & $8.15872 \pm 0.00001$ \\
		 $D=k^2$ & $0.0127 \pm 0.0003$ & $0.0136 \pm 0.0002$\\
		 $W$~(unit of $P$) & $0.018 \pm 0.001$ & $0.0179 \pm 0.0003$\\
		 $b$ & $0.46 \pm 0.06$ & $0.47 \pm 0.02$\\
		 $h_1$ & $0.72 \pm 0.01$ & $0.72 \pm 0.01$\\
		 $h_2$ & $0.44 \pm 0.05$ & $0.48 \pm 0.05$\\
		 $T_{0,1}^{(a)}$~(days) & $2052.41241 \pm +0.00061$ & $2052.41218 \pm +0.00058$ \\
		 $T_{0,2}^{(a)}$~(days) & $2134.00000 \pm +0.00036$ & $2133.99944 \pm +0.00032$ \\
		 $\log \sigma_j$ & - & $-8.1 \pm 0.1$ \\		 
		 Derived/physical &  &  \\
		 $k=R_\mathrm{b}/R_{\star}$ & $0.1130 \pm 0.0015$ & $0.11685 \pm 0.0009$\\
		 $R_\mathrm{b}\, (R_\mathrm{Jup})$ & - & $1.12 \pm 0.04$\\
		 $a/R_{\star}$ & $18.2 \pm 0.8$ & $18.3 \pm 0.5$\\
		 $i\, (\degr)$ & $88.6 \pm 0.2$ & $88.5 \pm 0.1$\\
		 $T_{14}^{(b)}$~(days) & $0.144 \pm 0.008$ & $0.146 \pm 0.003$\\
		 $e$ & $0.304 \pm 0.004$ & fixed \\
		 $\omega\, (\degr)$ & $274.2 \pm 0.1$ & fixed \\
		 $K_\mathrm{RV}$~(ms$^{-1}$)) & $221.1 \pm 1.2$ & - \\
		 $M_\mathrm{b}\, (M_\mathrm{Jup})$ & - & $2.19 \pm 0.06$\\
		 $\rho_\mathrm{b}$~(gcm$^{-3}$) & - & $2.1 \pm 0.2$\\
		 $\lambda^{(c)}\, (\degr)$ & $-143.0^{+1.6}_{-1:5}$ & \\
		 GP hyperparameters & & \\
		 $\log S_0$ & - & $-21.8 \pm 0.1$\\
		 $\log \omega_0$ & - & $6.7 \pm 0.1$\\
		\hline
	\end{tabular}
	}
	\\
	\textbf{Notes:}
	$^{(a)}$: Transit times in BJD$_\mathrm{TDB}-2457000$. 
	$T_{0,n}$ single visit output in the input/priors column, 
	while they are the linear ephemeris plus $\Delta T_{0,n}$ from multi-visit analysis.
	$^{(b)}$: Total duration is equal to $T_{14} = W \times P$.
	$^{(c)}$: spin-orbit angle measured from the Rossiter-McLaughlin effect.
\end{table}

\subsection{WASP-38 b}
\label{sec:lc_wasp38}

WASP-38 is the brightest star of our current sample, with $G=9.2$ and $V=9.4$.
It hosts a quite massive ($2.7\ M_\mathrm{Jup}$) warm-Jupiter, WASP-38 b,
on a slightly eccentric orbit ($e = 0.028\pm0.003$) with a period of about 6.9 days
\citep{Barros2011AA...525A..54B, Simpson2011MNRAS.414.3023S, Brown2012ApJ...760..139B, Bonomo2017AA...602A.107B}.
WASP-38 b orbit is aligned (within $2\sigma$)
with the stellar spin \citep{Brown2012ApJ...760..139B},
even if it was expected to be misaligned
due to its eccentricity and mass \citep{Simpson2011MNRAS.414.3023S}.
Table~\ref{tab:wasp38} summarises the parameters from literature
that we used in our analysis.
The lack of an RV trend due to an external massive planet or stellar companion would rule out
the Kozai and P-P scattering mechanisms in the formation process,
making this system the result of a disk-driven migration or 
of a more complex scenario.
\par

\begin{figure*}
\centering
	\includegraphics[width=0.99\columnwidth]{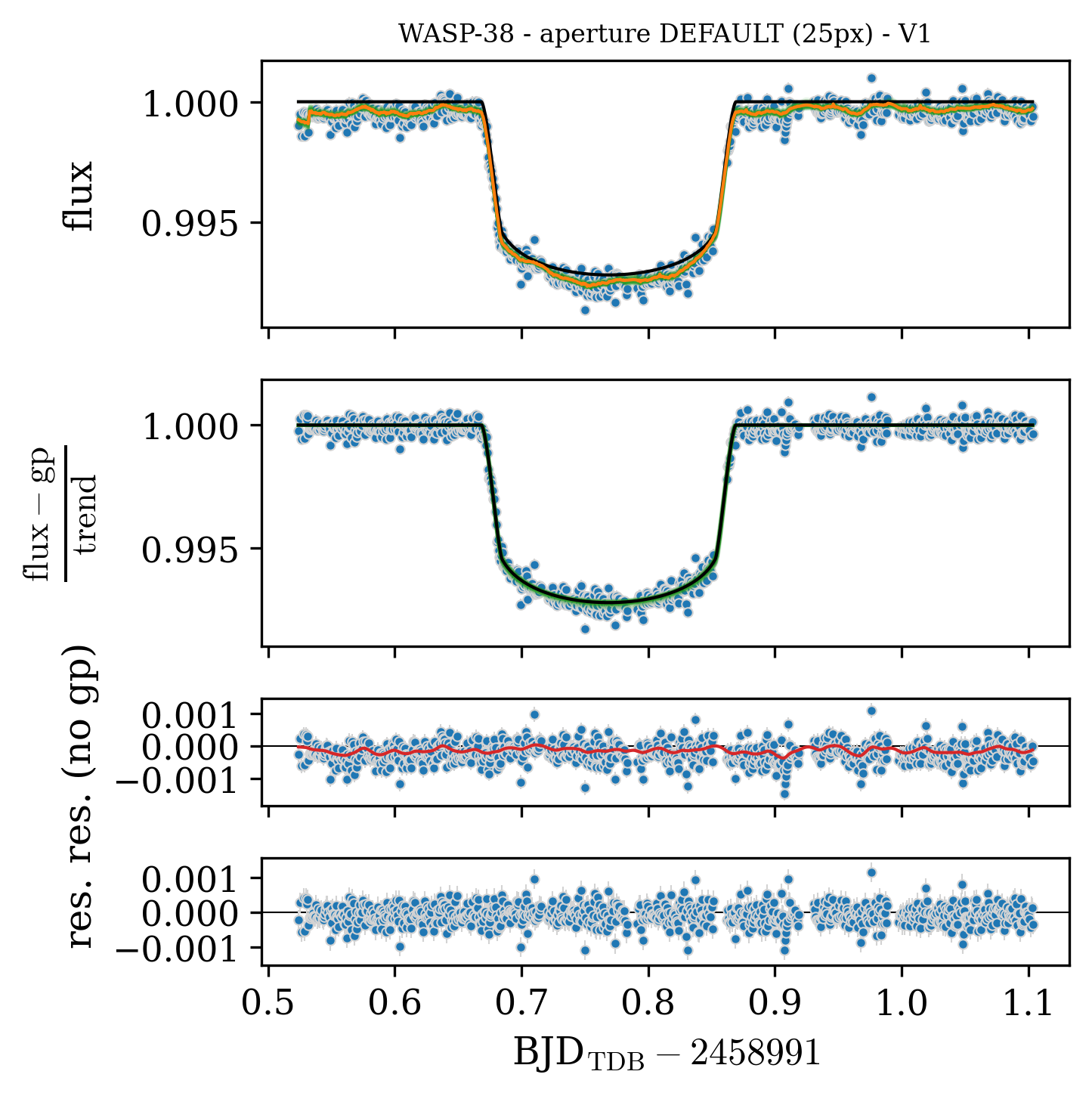}
	\includegraphics[width=0.99\columnwidth]{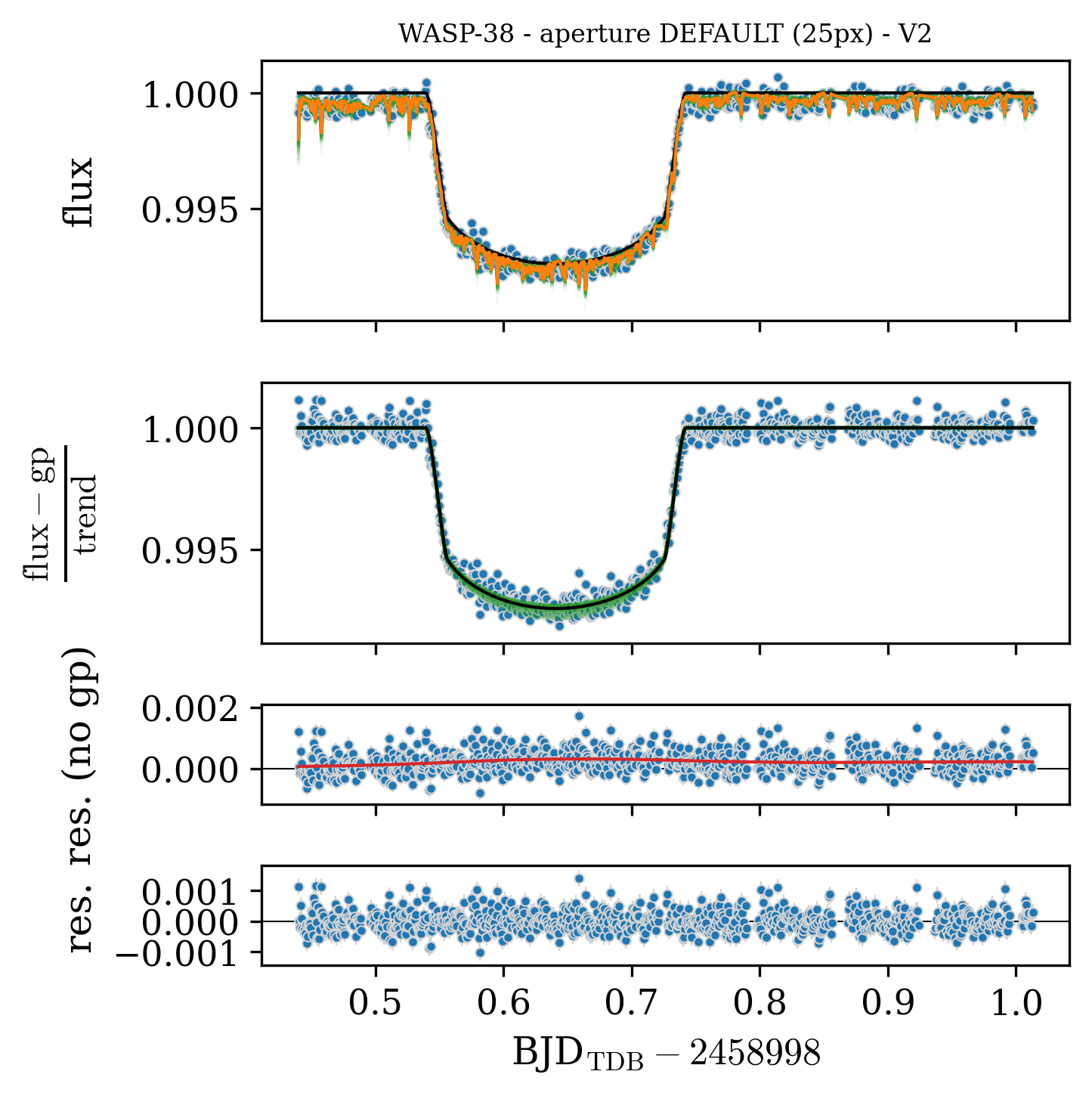}\\
	\includegraphics[width=0.99\columnwidth]{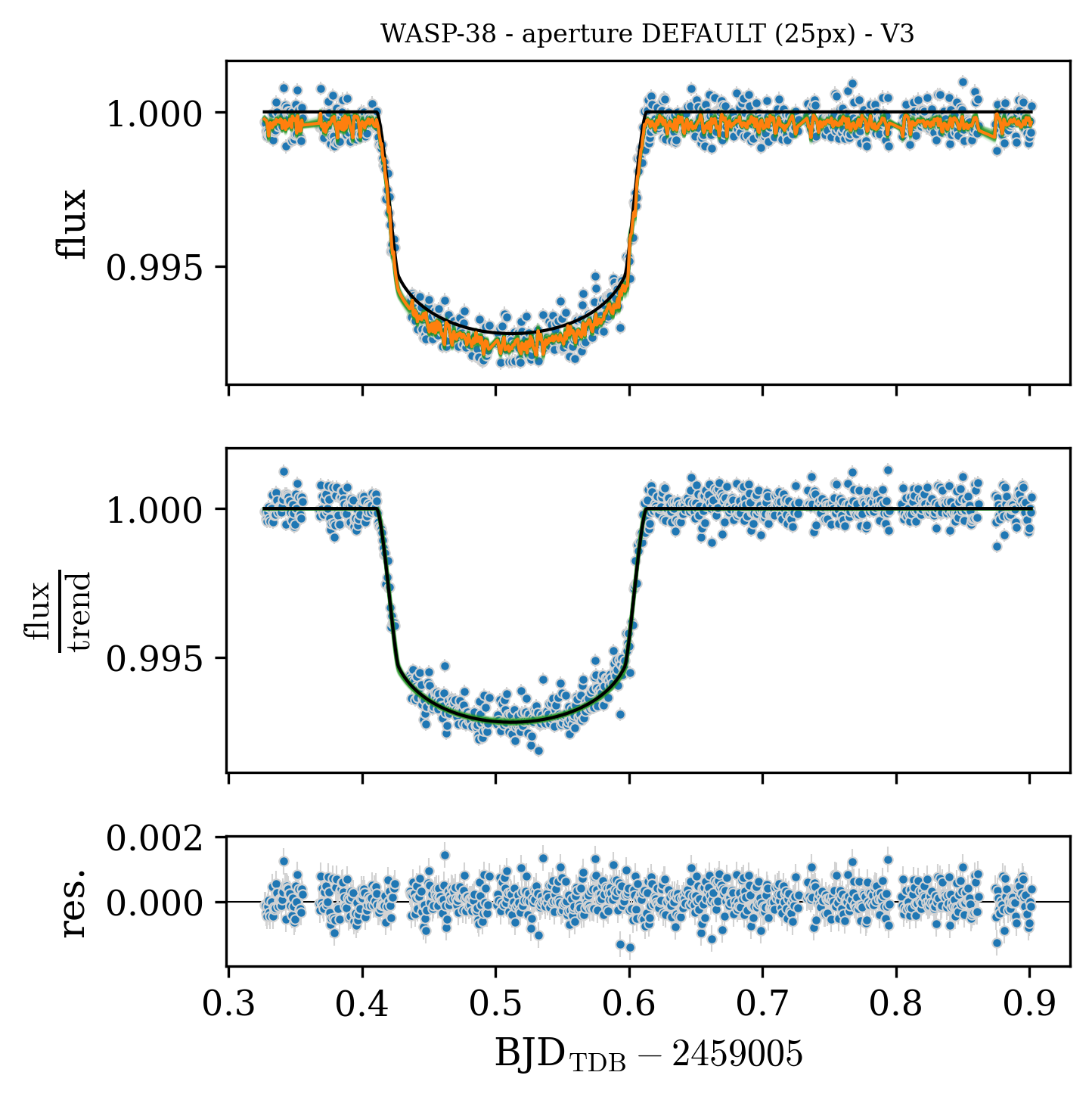}
	\includegraphics[width=0.99\columnwidth]{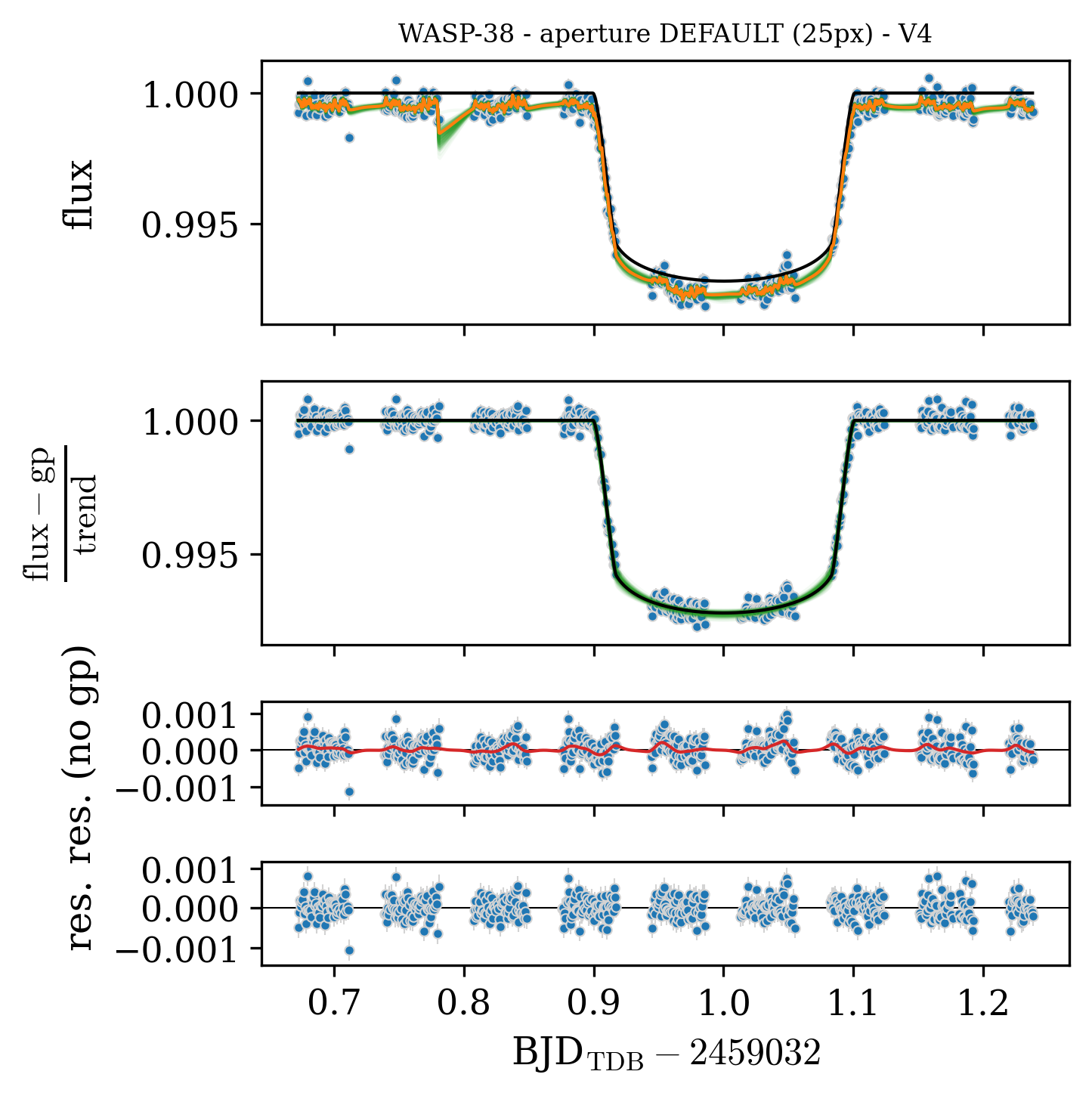}
    \caption{WASP-38 b single visits analysis (see Fig.~\ref{fig:visits_hatp17} for description).
    \textit{Upper-left:} first visit, fitted transit shape and detrending against
    linear trend, $x$ and $y$ pixel offset, first harmonic of the roll angle, and GP;
    \textit{upper-right:} second visit, fitted transit shape and detrending against
    the background, contaminants, quadratic term, second order of $x$ and $y$ pixel offset,
    two harmonics of the roll angle, and GP;
    \textit{lower-left:} third visit, fitted transit shape and detrending against
    the linear trend and $x$ and $y$ pixel offset without the GP;
    \textit{lower-right:} fourth visit, fitted transit shape and detrending against
    the background, contaminants, $y$ pixel offset, first harmonics of roll angle, and GP.
    }
    \label{fig:visits_wasp38}
\end{figure*}

\begin{figure*}
\centering
	\includegraphics[width=1.1\columnwidth]{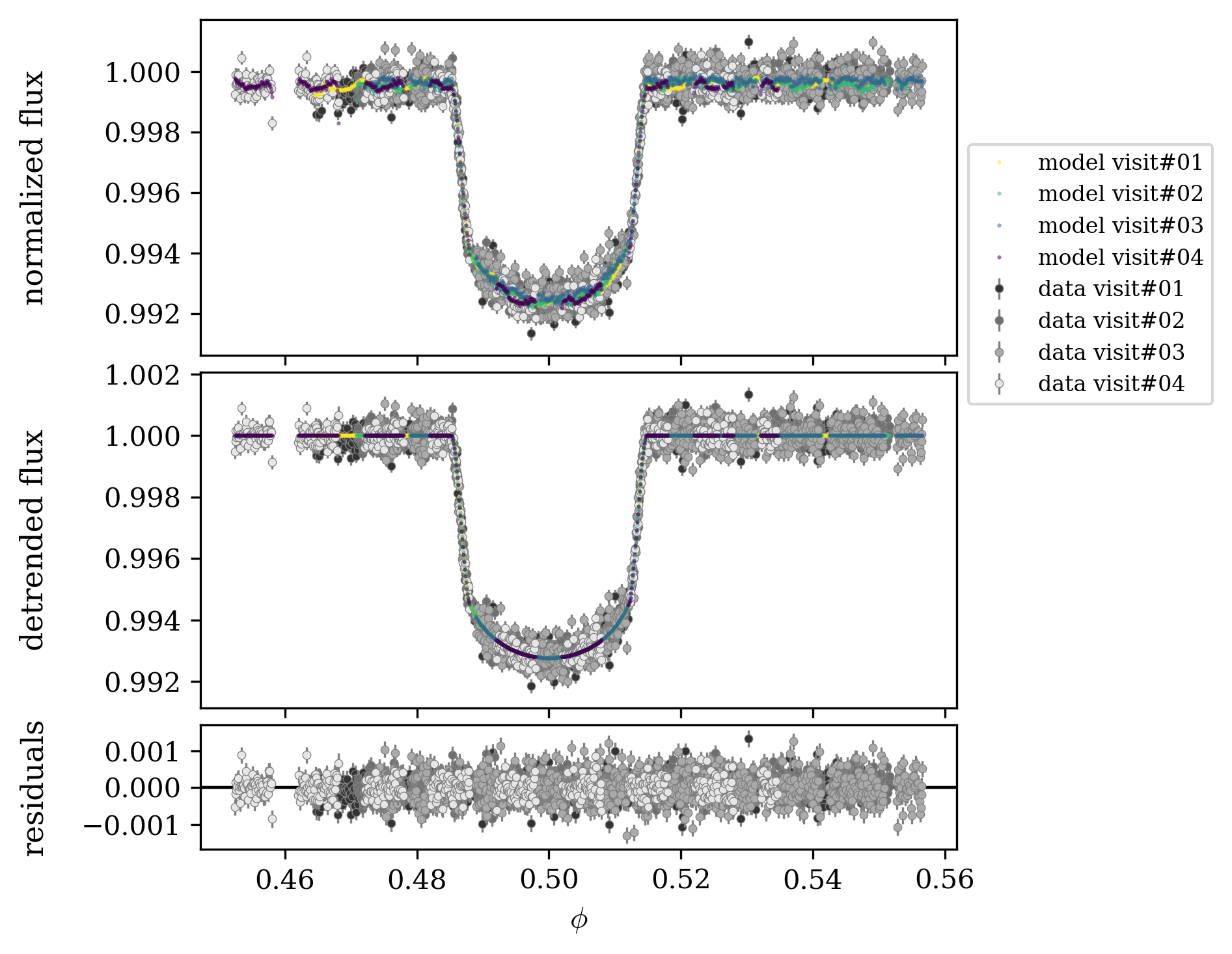}
	\includegraphics[width=0.9\columnwidth]{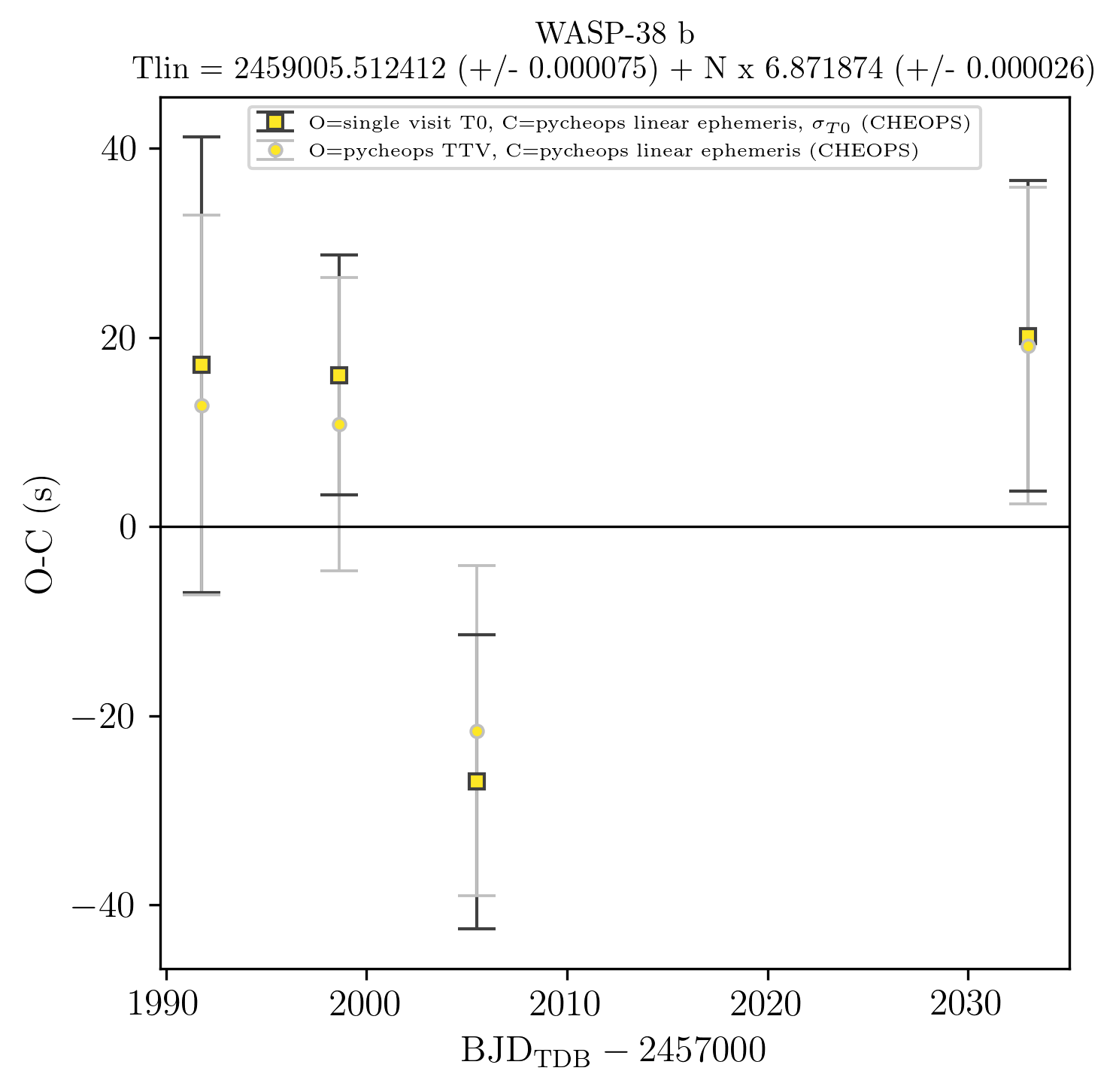}
    \caption{As in Fig~\ref{fig:mv_hatp17}, but for WASP-38 b.
    \textit{Left:} multi-visit phase plot of four CHEOPS visits;
    \textit{right:} O-C diagram.
    }
    \label{fig:mv_wasp38}
\end{figure*}

We collected four visits with CHEOPS, 
spanning an observing period of only two months from May to July 2020.
The first three visits have very high \geff{} ($>91\%$) and
high temporal sampling of both ingress and egress.
Only the egress of the first visit has a low coverage ($\sim 30\%$).
The fourth visit has a \geff{} of $62.2\%$,
but both ingress and egress were sampled with a high \cpreff.
The BIC favoured the analysis fitting the shape of all the four transits and
detrending  for the linear trend,
$x$ and $y$ pixel offset, first harmonic of the roll angle, and GP the first ($\sigma_{T_0} = 24$~s) and
for the background, contaminants, quadratic term, second order of $x$ and $y$ pixel offset,
two harmonics of the roll angle, and GP the second visit ($\sigma_{T_0} = 13$~s), 
for the linear trend and $x$ and $y$ pixel offset without the GP the third ($\sigma_{T_0} = 16$~s) and
for the background, contaminants, $y$ pixel offset, first harmonics of roll angle, and GP the fourth visit ($\sigma_{T_0} = 16$~s).
See Fig.~\ref{fig:visits_wasp38} for the single-visit plots and fits.
This analysis allowed us to determine the $T_0$ of the transits with the highest precision
of our current whole data sample.
From the multi-visit analysis
(see Fig.~\ref{fig:mv_wasp38} and Table~\ref{tab:wasp38} for the summary of the results),
we obtained $\sigma_{T_0} = 20$, $16$, $17$, and $17$~s for the four visits, respectively.
Only for the first visit we had a slightly improved $\sigma_{T_0}$ ($\sim 17\%$)
due to the partial egress, whose phase is covered in the joint analysis.
We had a worsening on $\sigma_{T_0}$ of the latest three visits
($-22\%$, $-12\%$, and $-2\%$, respectively).
The third visit was not detrended with the GP 
(see lower-left plot in Fig.~\ref{fig:visits_wasp38}), 
so, we suspect that
the common GP kernel in multi-visit could have introduced more noise
due to an overfitting.
However, this aspect will be analysed in detail in a future work.
The second single-visit analysis used the GP,
but it appears (see upper-right plot in Fig.~\ref{fig:visits_wasp38}) to
be a modulation more than a short timescale variation 
(i.e., the stellar granulation),
and, as for the third visit, the common GP kernel could have introduced some noise,
increasing the uncertainty in the transit time determination.
\par

Unfortunately, the first three visits have been scheduled as consecutive,
reducing the time-span needed to identify TTV signal.
The third visit shows a slight departure from the linear ephemeris 
(see $O-C$ plot in Fig.~\ref{fig:mv_wasp38}),
but it is still within $2\sigma$.
We cannot draw any conclusion on the existence of a TTV signal
based on the current dataset,
and we need to extend the temporal baseline of the observations.
\par

\begin{table}
	\centering
	\caption{WASP-38 summary table of stellar and planetary (planet b) parameters.
	Input and priors planetary parameters from \citet{Brown2012ApJ...760..139B}
	and \citet{Bonomo2017AA...602A.107B}.
	Best-fit solution (MLE and semi-interval HDI at $68.27\%$) from the four multi-visit analysis.
	}
	\label{tab:wasp38}
	\resizebox{\fitbox}{!}{%
	\begin{tabular}{lcc}
		\hline
		 Parameters & Input/priors & Source\\
		\hline
		 WASP-38 & \multicolumn{2}{c}{Gaia DR2 4453211899986180352} \\
		 RA (J2000)  & 16:15:50.37 & Simbad\\
		 DEC (J2000) & +10:01:57.28 & Simbad\\
		 $\mu_\mathrm{\alpha}$~(mas/yr) & $-31.07 \pm 0.05$ & \gaia{} DR2\\
		 $\mu_\mathrm{\delta}$~(mas/yr) & $-39.17 \pm 0.04$ & \gaia{} DR2\\
         age (Gyr) & $2.8 \pm 0.6$ & This work\\
		 parallax (mas) & $7.31 \pm 0.04$ & \gaia{} DR2\\
		 $V$~(mag) & 9.4 & Simbad\\
		 $G$~(mag) & 9.2 & \gaia{} DR2\\
		 $M_{\star}\, (M_{\sun})$ & $1.28 \pm 0.05$ & This work\\
		 $R_{\star}\, (R_{\sun})$ & $1.35 \pm 0.03$ & This work\\
		 $\rho_{\star}\, (\rho_{\sun})$ & $0.52 \pm 0.04$ & This work\\
		 $T_\mathrm{eff}$~(K) & $6436 \pm 60$ & SWEET-Cat\\
		 $\log g$ & $ 4.8 \pm 0.07$ & SWEET-Cat\\
		 \feh~(dex) & $0.06 \pm 0.04$ & SWEET-Cat\\
		 \hline
		 WASP-38 b & & \\
		 Model & Input/priors & Multi-visit (MLE \& HDI) \\
		 $T_{0,\mathrm{ref}}^{(a)}$~(days) & $-1664.0795 \pm 0.0007$ & $2005.51241 \pm 0.00008$\\
		 $P$~(days) & $6.87182 \pm 0.00005$ & $6.87187 \pm 0.00003$\\
		 $D=k^2$ & $0.0069 \pm 0.0001$ & $0.00633 \pm 0.00003$\\
		 $W$~(unit of $P$) & $0.02865 \pm 0.00015$ & $0.02915 \pm 0.00004$ \\
		 $b$ & $0.12 \pm 0.08$ & $0.35 \pm 0.02$\\
		 $h_1$ & $0.8 \pm 0.1$ & $0.78 \pm 0.01$\\
		 $h_2$ & $0.5 \pm 0.1$ & $0.29 \pm 0.05$\\
		 $T_{0,1}^{(a)}$~(days) & $1991.76886 \pm +0.00028$ & $1991.76881 \pm +0.00023$ \\
		 $T_{0,2}^{(a)}$~(days) & $1998.64072 \pm +0.00015$ & $1998.64066 \pm +0.00018$ \\
		 $T_{0,3}^{(a)}$~(days) & $2005.51210 \pm +0.00018$ & $2005.51216 \pm +0.00020$ \\
		 $T_{0,4}^{(a)}$~(days) & $2033.00014 \pm +0.00019$ & $2033.00013 \pm +0.00019$ \\
		 $\log \sigma_j$ & - & $-8.47 \pm 0.03$ \\
		 Derived/physical &  &  \\
		 $k=R_\mathrm{b}/R_{\star}$ & $0.0831 \pm 0.0006$ & $0.0796 \pm 0.0002$ \\
		 $R_\mathrm{b}\, (R_\mathrm{Jup})$ & - & $1.07 \pm 0.02$ \\
		 $a/R_{\star}$ & $12.1 \pm 0.1$ & $11.17 \pm 0.07$ \\
		 $i\, (\degr)$ & $89.4 \pm 0.4$ & $88.2 \pm 0.1$ \\
		 $T_{14}^{(b)}$~(days) & $0.197 \pm 0.001$ & $0.2003 \pm 0.0003$ \\
		 $e$ & $0.028 \pm 0.003$ & fixed \\
		 $\omega\, (\degr)$ & $338 \pm 9$ & fixed \\
		 $K_\mathrm{RV}$~(ms$^{-1}$)) & $246.6 \pm 1.2$ & - \\
		 $M_\mathrm{b}\, (M_\mathrm{Jup})$ & - & $2.7 \pm 0.1$ \\
		 $\rho_\mathrm{b}$~(gcm$^{-3}$) & - & $2.3 \pm 0.1$ \\
		 $\lambda^{(c)}\, (\degr)$ & $7.5^{+4.7}_{-6.1}$ & \\
		 GP hyperparameters & & \\
		 $\log S_0$ & - & $-24.0 \pm 0.3$ \\
		 $\log \omega_0$ & - & $5.0 \pm 0.2$ \\
		\hline
	\end{tabular}
	}
	\\
	\textbf{Notes:}
	$^{(a)}$: Transit times in BJD$_\mathrm{TDB}-2457000$. 
	$T_{0,n}$ single visit output in the input/priors column, 
	while they are the linear ephemeris plus $\Delta T_{0,n}$ from multi-visit analysis.
	$^{(b)}$: Total duration. The eq. used depends on the literature.
	The multi-visit duration is equal to $T_{14} = W \times P$.
	$^{(c)}$: spin-orbit angle measured from the Rossiter-McLaughlin effect.
\end{table}

\subsection{WASP-106 b}
\label{sec:lc_wasp106}

WASP-106 is the faintest target in the G band ($G = 11.4$ and $V=11.2$) of our sample,
and it hosts a warm-Jupiter planet (b) with a mass about double that of Jupiter,
and a radius slightly larger than Jupiter.
WASP-106 b has been discovered by \citet{Smith2014AA...570A..64S}
and it has a circular orbit with a period of about 9.3 days.
The same authors found that the planetary orbit cannot be circularised by tidal forces,
so the orbit remained almost circular for the system lifetime.
This could be a hint of a disk-driven migration as the main process of
the evolution of the system \citep{Smith2014AA...570A..64S}.
\par

\begin{figure}
\centering
	\includegraphics[width=0.99\columnwidth]{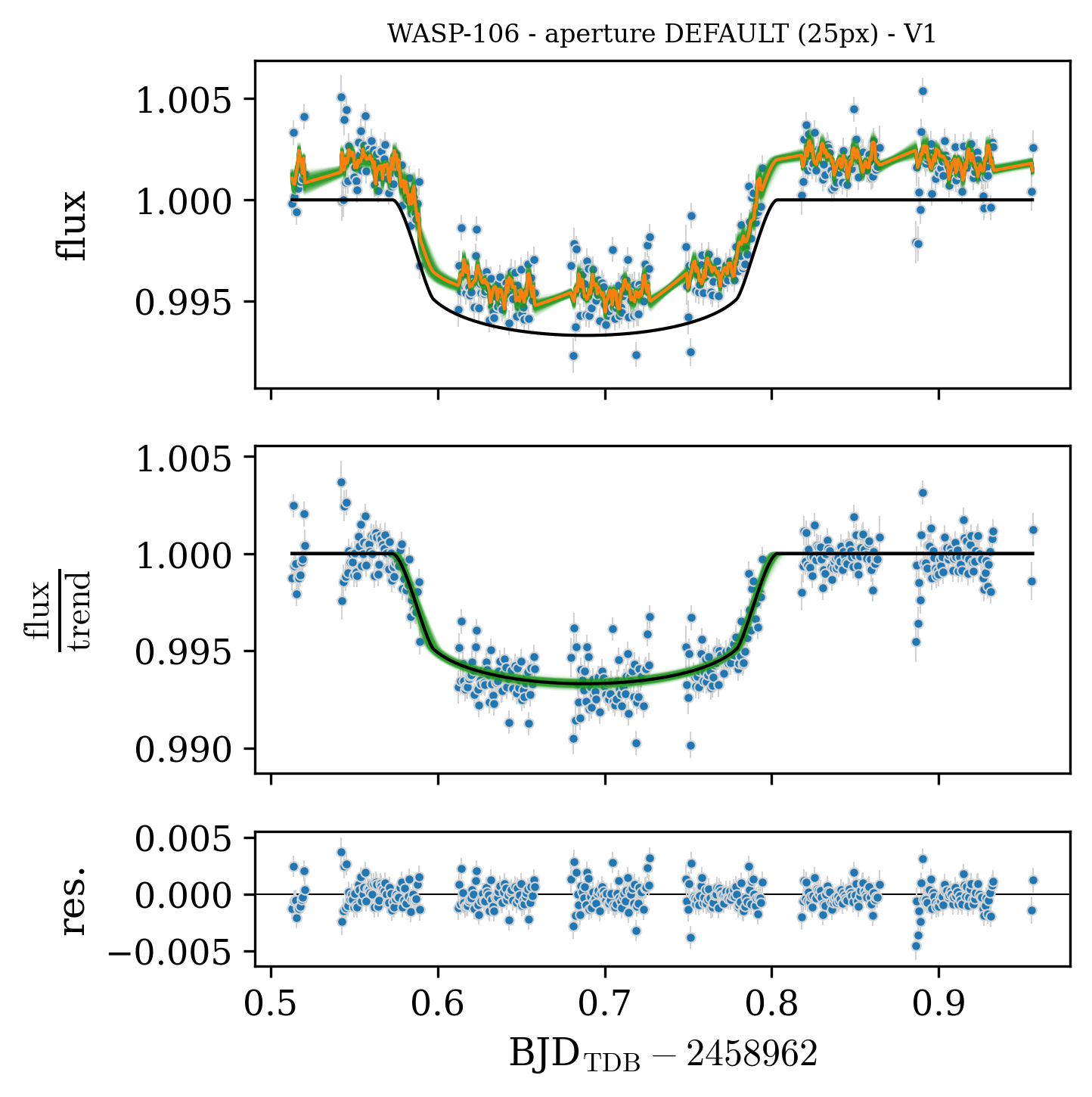}
    \caption{WASP-106 b single visit analysis (see Fig.~\ref{fig:visits_hatp17} for description).
    The model contains the fitted transit shape and detrending against $x$ and $y$ pixel offset.
    }
    \label{fig:visits_wasp106}
\end{figure}

We observed the transit of WASP-106 b only once with CHEOPS in April 2020.
We obtained a light curve with a \geff{} of about $66.3\%$ and
with ingress and egress sampled with an efficiency of about $56\%$ and $60\%$, respectively.
We modelled the light curve, based on BIC statistics,
fitting the shape of the transit and
detrending for the $x$ and $y$ pixel offset, without GP
(see Fig.~\ref{fig:visits_wasp106} and Table~\ref{tab:wasp106}).
We obtained $\sigma_{T_0} = 60$~s, probably due to the noisy data
and due to the short visit and bad sampling of the pre-ingress phase,
making it difficult to properly constrain the detrending parameters
during the model fit.
We need more visits to better understand the possibility to detect a TTV for this target.
\par

\begin{table}
	\centering
	\caption{WASP-106 summary table of stellar and planetary (planet b) parameters.
	Input and priors planetary parameters from \citet{Smith2014AA...570A..64S}.
	Best-fit solution (MLE and semi-interval HDI at $68.27\%$) from the one single-visit analysis.
	}
	\label{tab:wasp106}
	\resizebox{\fitbox}{!}{%
	\begin{tabular}{lcc}
		\hline
		 Parameters & Input/priors & Source\\
		\hline
		 WASP-106 & \multicolumn{2}{c}{Gaia DR2 3788394461991295488} \\
		 RA (J2000)  & 11:05:43.14 & Simbad\\
		 DEC (J2000) & -05:04:45.94 & Simbad\\
		 $\mu_\mathrm{\alpha}$~(mas/yr) & $-24.818 \pm 0.077$ & \gaia{} DR2\\
		 $\mu_\mathrm{\delta}$~(mas/yr) & $-13.294 \pm 0.060$ & \gaia{} DR2\\
         age (Gyr) & $2.5 \pm 0.6$ & This work\\
		 parallax (mas) & $2.81 \pm 0.05$ & \gaia{} DR2\\
		 $V$~(mag) & 11.2 & Simbad\\
		 $G$~(mag) & 11.4 & \gaia{} DR2\\
		 $M_{\star}\, (M_{\sun})$ & $1.26 \pm 0.05$ & This work\\
		 $R_{\star}\, (R_{\sun})$ & $1.42 \pm 0.02$ & This work\\
		 $\rho_{\star}\, (\rho_{\sun})$ & $0.81 \pm 0.15$ & This work\\
		 $T_\mathrm{eff}$~(K) & $6265 \pm 36$ & SWEET-Cat\\
		 $\log g$ & $4.38 \pm 0.04$ & SWEET-Cat\\
		 \feh~(dex) & $+0.15 \pm 0.03$ & SWEET-Cat\\
		 \hline
		 WASP-106 b & & \\
		 Model & Input/priors & Single-visit (MLE \& HDI) \\
		 $P$~(days) & $9.28972 \pm 0.00001$ & fixed\\
		 $D=k^2$ & $0.00642 \pm 0.00018$ & $0.00607 \pm 0.00016$\\
		 $W$~(unit of $P$) & $0.0240 \pm 0.0008$ & $0.0247 \pm 0.0003$\\
		 $b$ & $0.13 \pm 0.16$ & $0.57 \pm 0.07$\\
		 $h_1$ & $0.75 \pm 0.01$ & $0.75 \pm 0.01$\\
		 $h_2$ & $0.46 \pm 0.05$ & $0.46 \pm 0.05$\\
		 $T_{0,1}^{(a)}$~(days) & - & $1962.68825 \pm 0.00069$ \\
 		 $\log \sigma_j$ & - & $-7.27 \pm 0.06$\\
		 Derived/physical &  &  \\
		 $k=R_\mathrm{b}/R_{\star}$ & $0.080 \pm 0.001$ & $0.078 \pm 0.001$\\
		 $R_\mathrm{b}\, (R_\mathrm{Jup})$ & - & $1.10 \pm 0.02$\\
		 $a/R_{\star}$ & $14.2 \pm 0.4$ & $11.8 \pm 0.7$\\
		 $i\, (\degr)$ & $89.5 \pm 0.6$ & $87.2 \pm 0.5$\\
		 $T_{14}^{(b)}$~(days) & $0.223 \pm 0.008$ & $0.229 \pm 0.003$\\
		 $e$ & 0 & fixed \\
		 $\omega\, (\degr)$ & $90$ & fixed \\
		 $K_\mathrm{RV}$~(ms$^{-1}$)) & $165.3 \pm 4.3$ & - \\
		 $M_\mathrm{b}\, (M_\mathrm{Jup})$ & - & $2.00 \pm 0.08$\\
		 $\rho_\mathrm{b}$~(gcm$^{-3}$) & - & $1.14 \pm 0.22$\\
		 $\lambda^{(c)}\, (\degr)$ & - & \\
		\hline
	\end{tabular}
	}
	\\
	\textbf{Notes:}
	$^{(a)}$: Transit times in BJD$_\mathrm{TDB}-2457000$. 
	$T_{0,n}$ single visit output.
	$^{(b)}$: Total duration equal to $T_{14} = W \times P$.
	$^{(c)}$: spin-orbit angle measured from the Rossiter-McLaughlin effect.
\end{table}

\subsection{WASP-130 b}
\label{sec:lc_wasp130}

WASP-130 was classified
as a metal-rich G6 star, with magnitude V=11.1,
by \citet{Hellier2017MNRAS.465.3693H}.
The same authors discovered WASP-130 b,
a warm-Jupiter with period of about 11.6~d and a circular orbit.
There is no evidence of a RV trend due to a planetary or stellar companion.
So, also this target will be part of the sample
for testing the disk-driven migration process.
\par

\begin{figure*}
\centering
	\includegraphics[width=0.99\columnwidth]{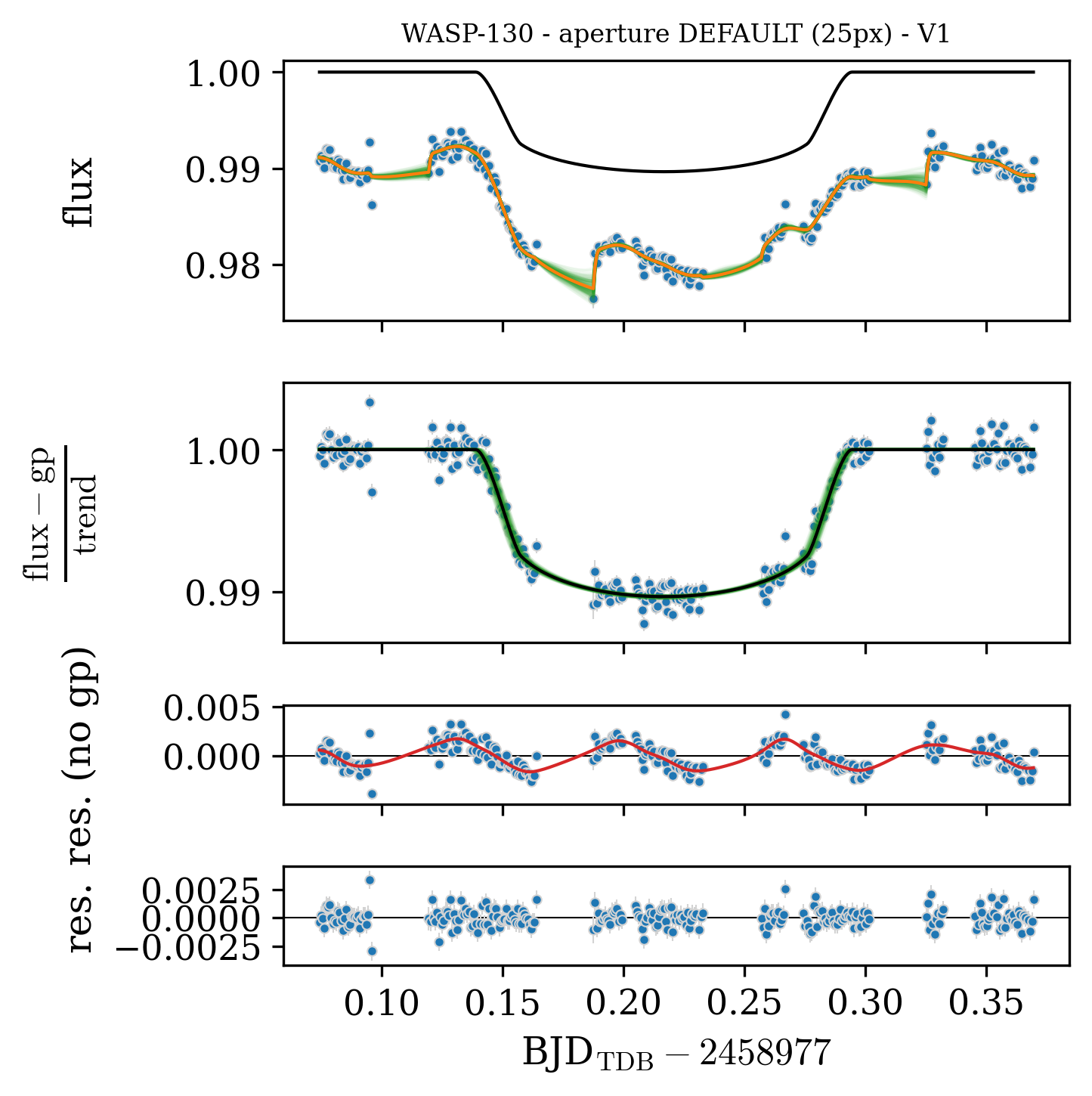}
	\includegraphics[width=0.99\columnwidth]{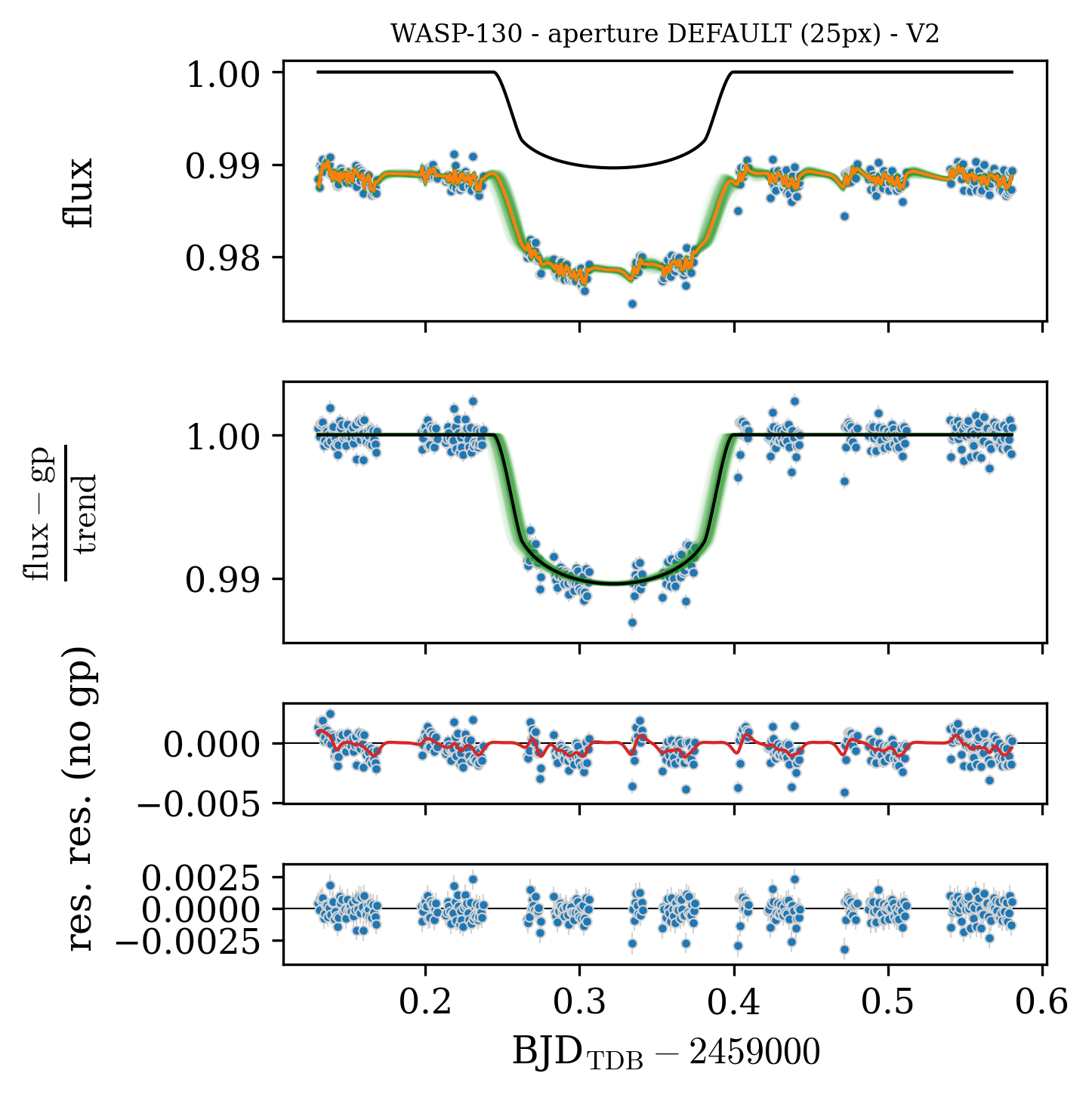}
	\includegraphics[width=0.99\columnwidth]{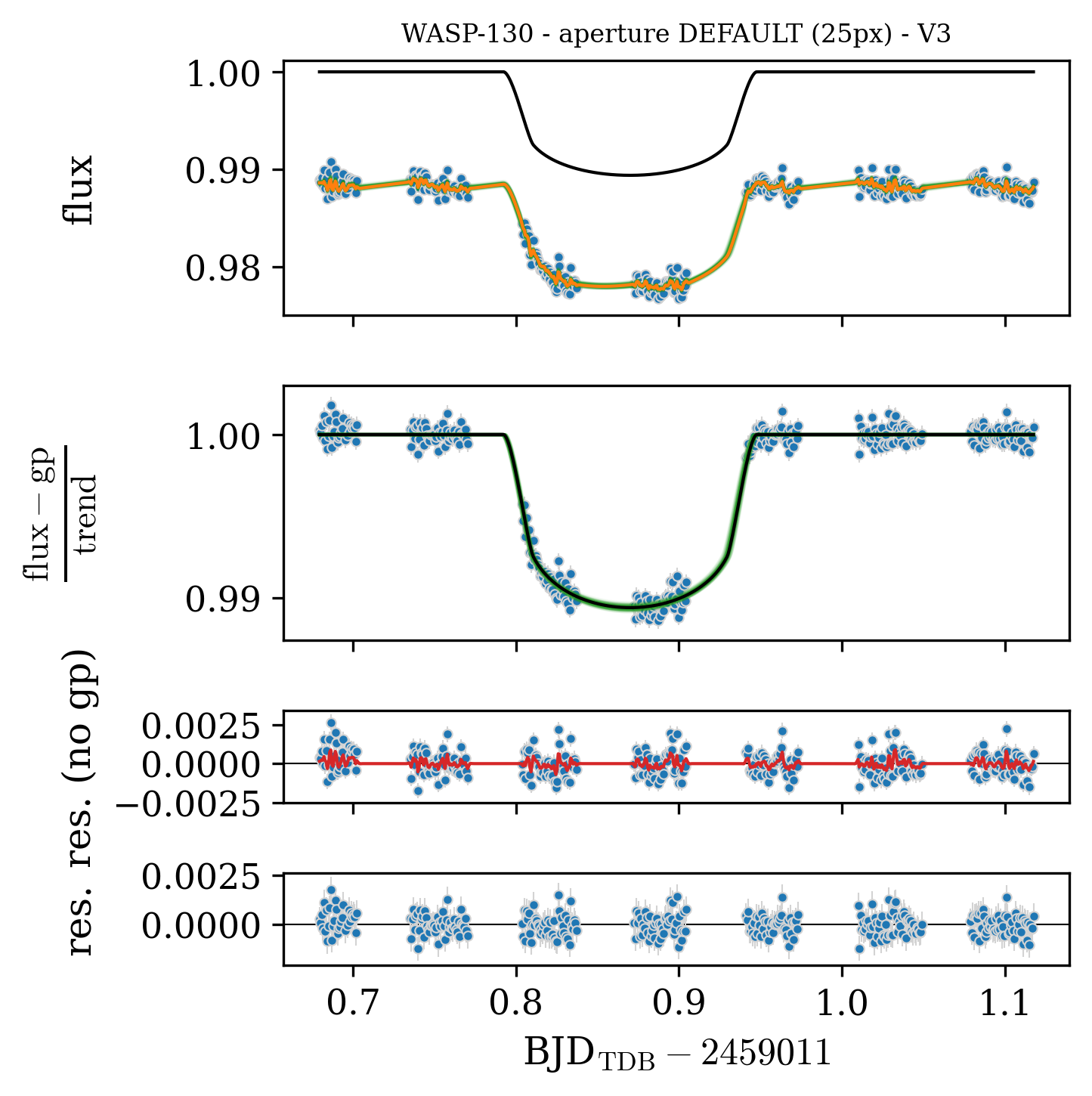}
    \caption{WASP-130 b single visits analysis (see Fig.~\ref{fig:visits_hatp17} for description).
    \textit{Upper-left:} first visit, fixed transit shape and detrending against background and GP;
    \textit{upper-right:} second visit, fixed transit shape and detrending against $x$ and $y$ pixel offset and GP; 
    \textit{lower:} third visit, fitted transit shape and detrending against the first harmonics of the satellite roll angle and GP.
    }
    \label{fig:visits_wasp130}
\end{figure*}

\begin{figure*}
\centering
	\includegraphics[width=1.1\columnwidth]{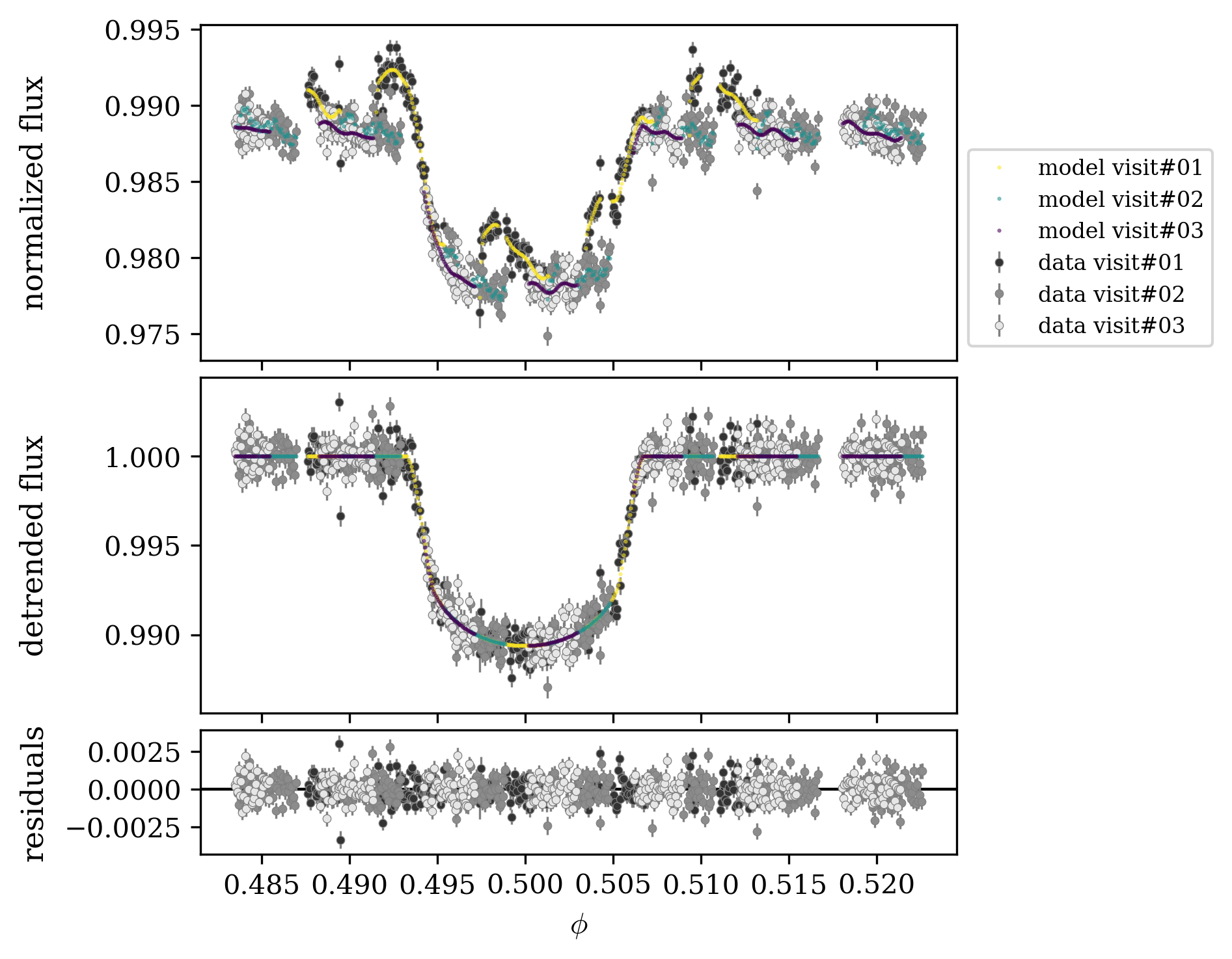}
	\includegraphics[width=0.9\columnwidth]{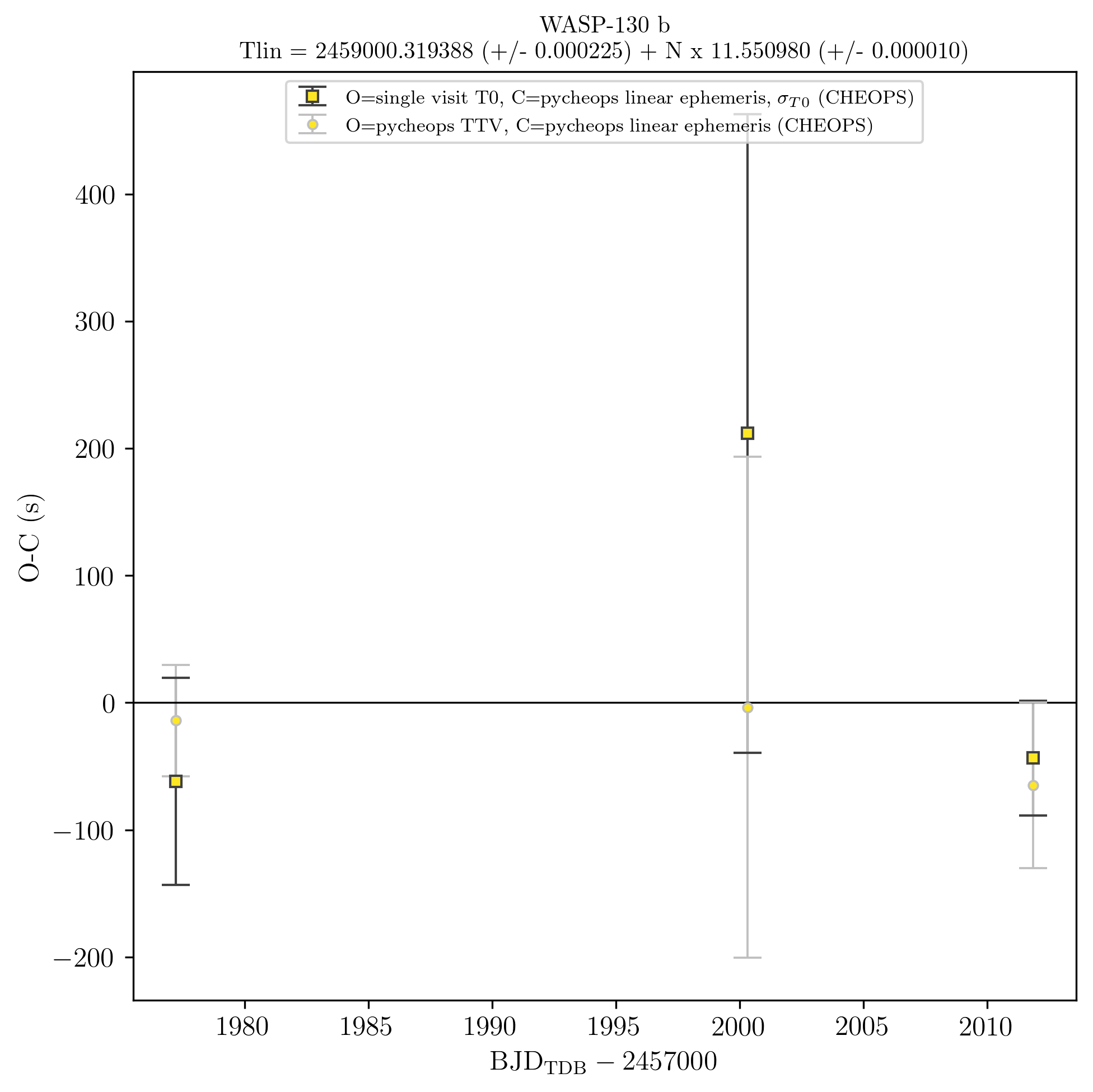}
    \caption{As in Fig~\ref{fig:mv_hatp17}, but for WASP-130 b.
    \textit{Left:} multi-visit phase plot of three CHEOPS visits;
    \textit{right:} O-C diagram.
    }
    \label{fig:mv_wasp130}
\end{figure*}

We obtained three visits with CHEOPS in May and June 2020.
The first visit has a \geff{} of $61.8\%$
and good sampling of ingress and egress,
but it is too short and strongly affected by systematic effects.
The \geff{} of the second and third visit is of $54.3\%$ for both.
Furthermore, the second visit is characterised by an empty sampling of ingress and egress,
and the third visit covered only about $50\%$ of the ingress (see Fig.~\ref{fig:visits_wasp130}).
Due to these reasons, for the first
and the second visit in the single-visit analysis
we obtained the best-fit transit model with fixed shape parameters.
For these two visits we used as detrending parameters the background with GP (first visit) and
the $x$ and $y$ pixel offset with GP (second visit).
We fitted the shape of the transit of the third visit, 
detrending for the first harmonic of the roll angle with the GP.
See Fig.~\ref{fig:visits_wasp130} for the single-visit light curves with models.
From the single-visit analysis we obtained $\sigma_{T_0} = 82$, 251, and 45 seconds,
for the first, the second, and third visits, respectively.
In the multi-visit analysis we fit the shape of the transit 
(as already mentioned in Sec.~\ref{sec:data_analysis}), 
and used the detrending parameters and GP information from the single-visit analysis
(see the phase folded light curve of the multi-visit analysis in Fig.~\ref{fig:mv_wasp130} and
the summary of the results in Table~\ref{tab:wasp130}).
We obtained an improvement on the $\sigma_{T_0}$ of
the first ($\sigma_{T_0}=44$~s) and
second visit ($\sigma_{T_0}=198$~s),
and a worsening by 20~s of the third visit.
This is due to the fact that in the detrending model of the multi-visit 
the roll angle harmonic of the third visit is not used, 
because the multi-visit GP kernel should already incorporate it,
but not so efficiently in this case.
A more careful and detailed analysis is mandatory.
This effect of the large $\sigma_{T_0}$ is clearly visible in
the $O-C$ diagram in Fig.~\ref{fig:mv_wasp130}, 
that does not show any hint of TTV with the current dataset.
\par

\begin{table}
	\centering
	\caption{WASP-130 summary table of stellar and planetary (planet b) parameters.
	Input and priors planetary parameters from \citet{Hellier2017MNRAS.465.3693H}.
	Best-fit solution (MLE and semi-interval HDI at $68.27\%$) from the three multi-visit analysis.
	}
	\label{tab:wasp130}
	\resizebox{\fitbox}{!}{%
	\begin{tabular}{lcc}
		\hline
		 Parameters & Input/priors & Source\\
		\hline
		 WASP-130 & \multicolumn{2}{c}{Gaia DR2 6112606840179716096} \\
		 RA (J2000)  & 13:32:25.44 & Simbad\\
		 DEC (J2000) & -42:28:30.97 & Simbad\\
		 $\mu_\mathrm{\alpha}$~(mas/yr) & $6.11 \pm 0.08$ & \gaia{} DR2\\
		 $\mu_\mathrm{\delta}$~(mas/yr) & $-1.24 \pm 0.08$ & \gaia{} DR2\\
         age (Gyr) & $3.2 \pm 0.7$ & This work\\
		 parallax (mas) & $5.78 \pm 0.05$ & \gaia{} DR2\\
		 $V$~(mag) & 11.1 & Simbad\\
		 $G$~(mag) & 11.0 & \gaia{} DR2\\
		 $M_{\star}\, (M_{\sun})$ & $1.06 \pm 0.04$ & This work\\
		 $R_{\star}\, (R_{\sun})$ & $1.02 \pm 0.01$ & This work\\
		 $\rho_{\star}\, (\rho_{\sun})$ & $1.0 \pm 0.2$ & This work\\
		 $T_\mathrm{eff}$~(K) & $5667 \pm 34$ & SWEET-Cat\\
		 $\log g$ & $4.43 \pm 0.05$ & SWEET-Cat\\
		 \feh~(dex) & $0.31 \pm 0.03$ & SWEET-Cat\\
		 \hline
		 WASP-130 b & & \\
		 Model & Input/priors & Multi-visit (MLE \& HDI) \\
		 $T_{0,\mathrm{ref}}^{(a)}$~(days) & $-78.85693 \pm 0.00025$ & $2000.31939 \pm 0.00023$ \\
		 $P$~(days) & $11.55098 \pm 0.00001$ & $11.55098 \pm 0.00001$\\
		 $D=k^2$ & $0.00916 \pm 0.00014$ & $0.0092 \pm 0.0001$\\
		 $W$~(unit of $P$) & $0.01342 \pm 0.00009$ & $0.01347 \pm 0.00007$\\
		 $b$ & $0.53 \pm 0.03$ & $0.49 \pm 0.02$\\
		 $h_1$ & $0.72 \pm 0.01$ & $0.72 \pm 0.01$\\
		 $h_2$ & $0.44 \pm 0.05$ & $0.46 \pm 0.05$\\
		 $T_{0,1}^{(a)}$~(days) & $1977.21671 \pm +0.00094$ & $1977.21727 \pm +0.00051$ \\
		 $T_{0,2}^{(a)}$~(days) & $2000.32184 \pm +0.00291$ & $2000.31935 \pm +0.00228$ \\
		 $T_{0,3}^{(a)}$~(days) & $2011.86986 \pm +0.00052$ & $2011.86962 \pm +0.00075$ \\
		 $\log \sigma_j$ & - & $-7.40 \pm 0.04$\\
		 Derived/physical &  &  \\
		 $k=R_\mathrm{b}/R_{\star}$ & $0.0957 \pm 0.0007$ & $0.0961 \pm 0.0005$\\
		 $R_\mathrm{b}\, (R_\mathrm{Jup})$ & - & $0.98 \pm 0.01$\\
		 $a/R_{\star}$ & $22.7 \pm 2.4$ & $23.2 \pm 0.3$\\
		 $i\, (\degr)$ & $88.66 \pm 0.12$ & $88.79 \pm 0.07$\\
		 $T_{14}^{(b)}$~(days) & $0.155 \pm 0.001$ & $0.1556 \pm 0.0008$ \\
		 $e$ & $0$ & fixed \\
		 $\omega\, (\degr)$ & $90$ & fixed \\
		 $K_\mathrm{RV}$~(ms$^{-1}$)) & $108 \pm 2$ & - \\
		 $M_\mathrm{b}\, (M_\mathrm{Jup})$ & - & $1.25 \pm 0.04$\\
		 $\rho_\mathrm{b}$~(gcm$^{-3}$) & - & $2.2 \pm 0.1$\\
		 $\lambda^{(c)}\, (\degr)$ & - & \\
		 GP hyperparameters & & \\
		 $\log S_0$ & - & $-21.1 \pm 0.2$ \\
		 $\log \omega_0$ & - & $5.5 \pm 0.1$ \\
		\hline
	\end{tabular}
	}
	\\
	\textbf{Notes:}
	$^{(a)}$: Transit times in BJD$_\mathrm{TDB}-2457000$. 
	$T_{0,n}$ single visit output in the input/priors column, 
	while they are the linear ephemeris plus $\Delta T_{0,n}$ from multi-visit analysis.
	$^{(b)}$: Total duration. The eq. used depends on the literature.
	The multi-visit duration is equal to $T_{14} = W \times P$.
	$^{(c)}$: spin-orbit angle measured from the Rossiter-McLaughlin effect.
\end{table}

\subsection{K2-287 b}
\label{sec:lc_k2-287}

K2-287 is a V=11.3 star (the faintest in the V band, $G=11.1$)
observed by \kepler/\textit{K2} \citep{Howell2014PASP..126..398H}
during campaign 15.
This star hosts K2-287 b, 
a warm-Saturn ($M_\mathrm{b}=0.3\ M_\mathrm{Jup}$, $R_\mathrm{b}=0.8\ R_\mathrm{Jup}$)
recently discovered by \citet{Jordan2019AJ....157..100J}.
Even if this planet has been classified as warm-Saturn,
we included it in our sample because
it lies on an eccentric ($e=0.478$) orbit with a period of about 15 days.
The authors suggested that this planet needs more follow-up observations
to better understand the evolution process responsible for its orbital configuration.
In particular, they suggested long-term RV monitoring,
RM analysis, and search for TTV signal due to close companions that
migrated with K2-287 b.
The long period and transit duration of K2-287 b makes it difficult
to schedule and observe with ground-based facilities.
\par

We obtained three visits spanning two months of CHEOPS observations,
with \geff{} of $88\%$, $71.9\%$, and $57.3\%$ for the first, second, and third visit, respectively.
We observed many strong dips in the first visit with amplitude greater than the transit depth,
and we found that they were caused by the background.
We decided to remove these points with $5\sigma$-clipping above the median of the background flux,
reducing the effective \geff{} of about $30\%$.
These dips did impact also the \cpreff{} of egress, lowering it to less than $30\%$.
Furthermore, the pre-ingress part is very short in the first visit.
We did not find the background features in the second and third visit.
The \cpreff{} of both ingress and egress of the second visit is below $30\%$,
as also the \cpreff{} of the ingress of the third visit.
\par

\begin{figure*}
\centering
	\includegraphics[width=0.99\columnwidth]{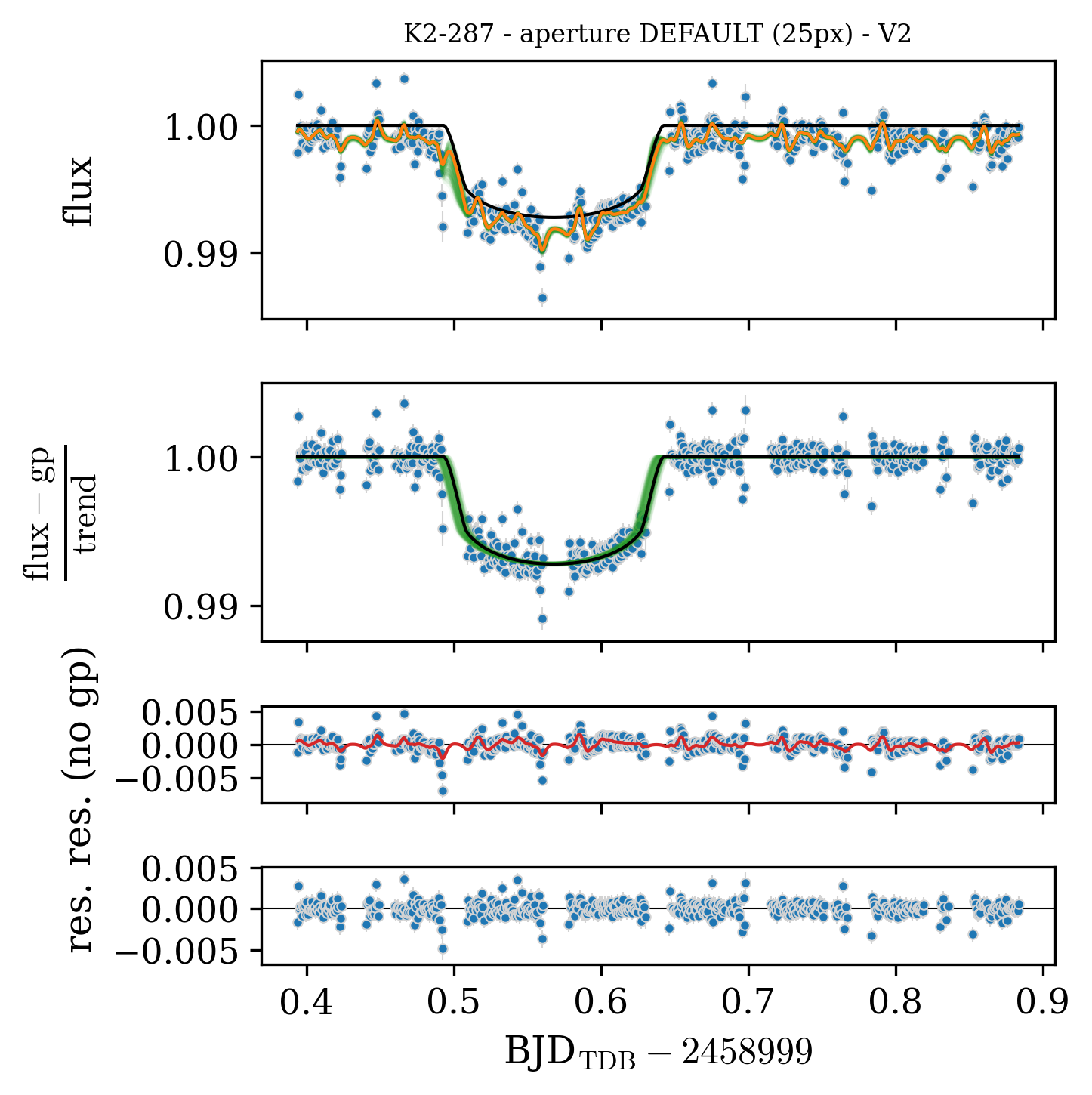}
	\includegraphics[width=0.99\columnwidth]{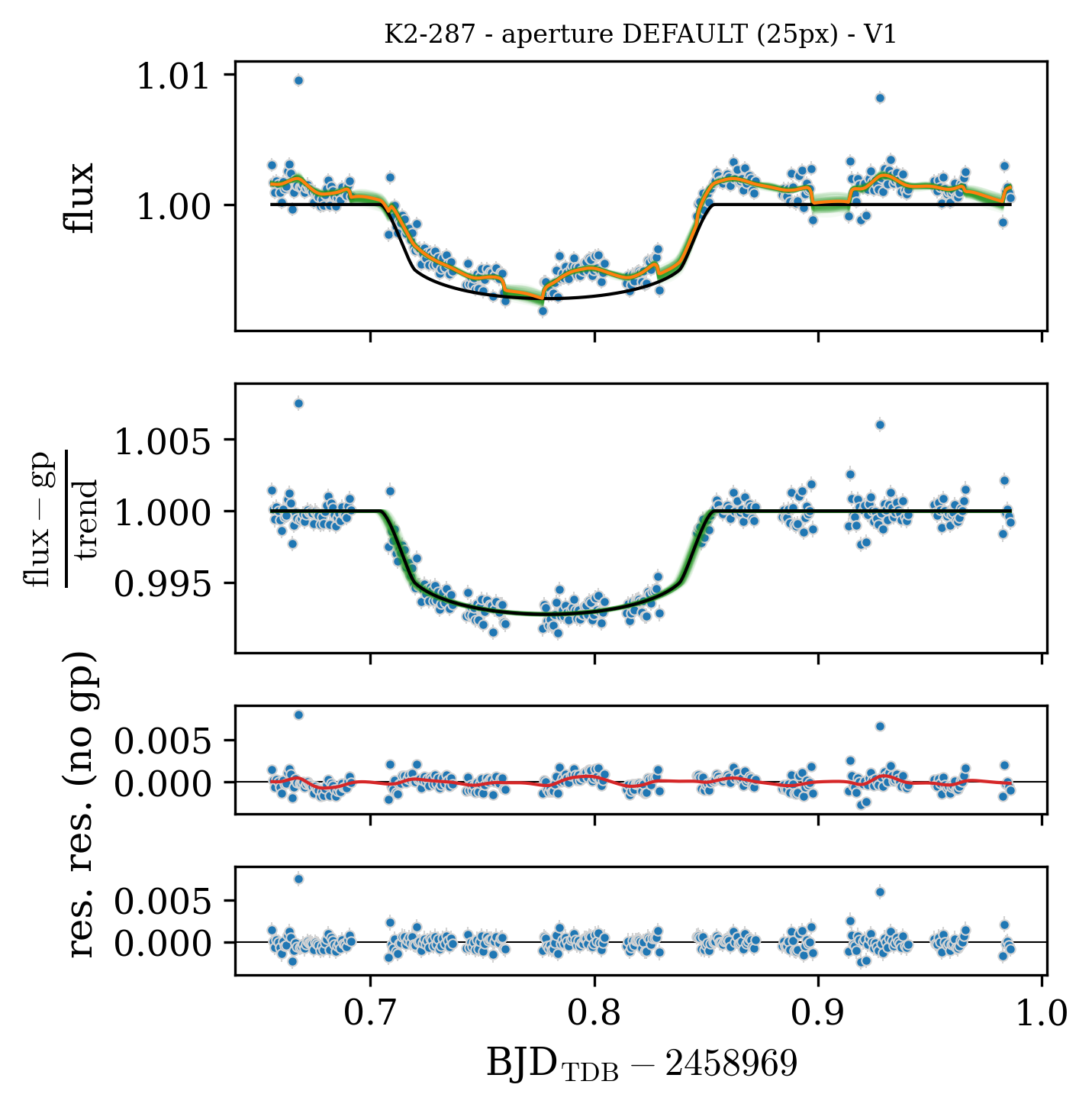}
	\includegraphics[width=0.99\columnwidth]{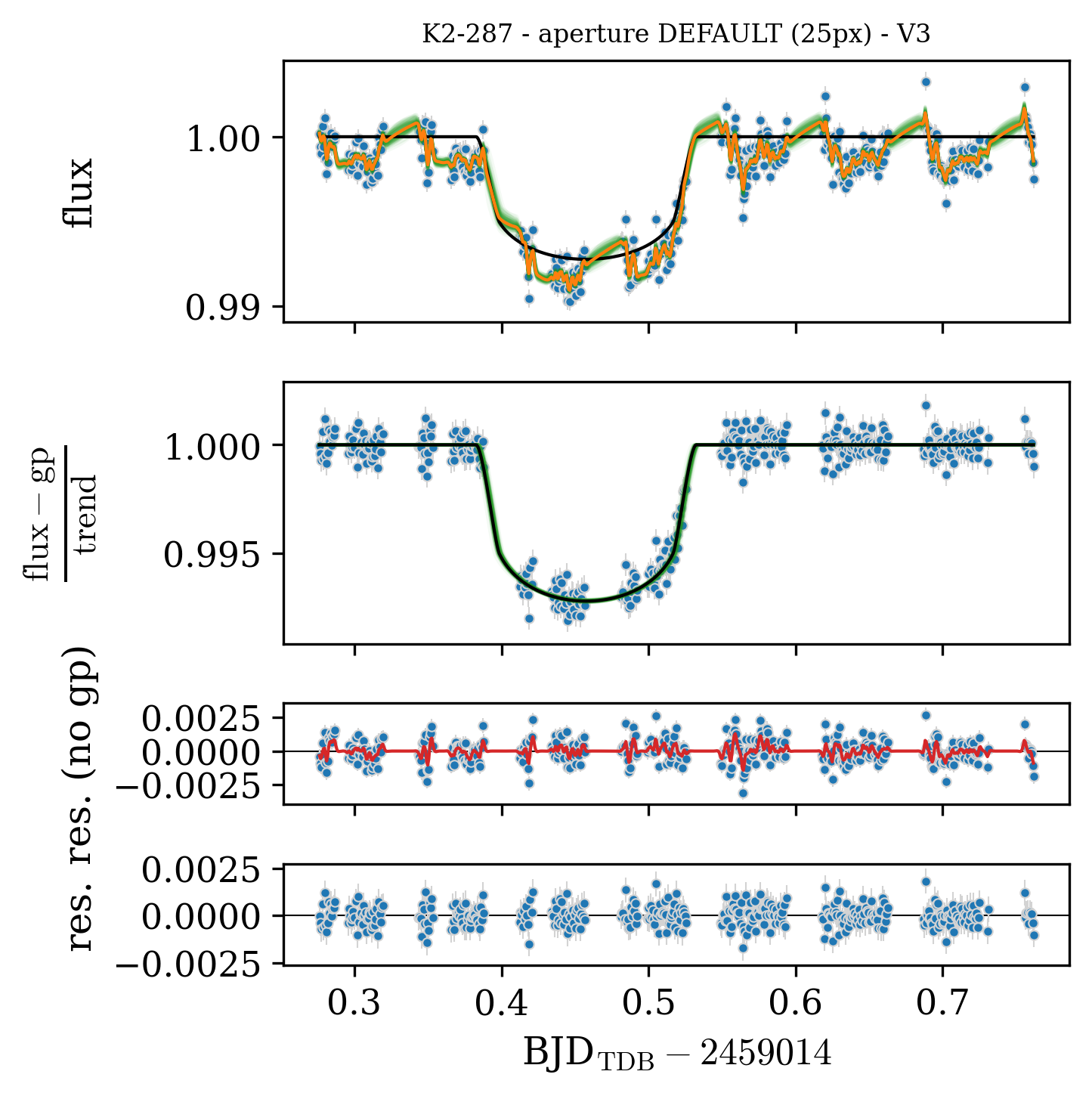}
    \caption{K2-287 b single visits analysis (see Fig.~\ref{fig:visits_hatp17} for description).
    \textit{Upper-left:} first visit, fixed transit shape and detrending against background and GP;
    \textit{upper-right:} second visit, fixed transit shape and detrending with only GP;
    \textit{lower:} third visit, fixed transit shape and detrending against the first two harmonics of the satellite roll angle and GP.
    }
    \label{fig:visits_k2-287}
\end{figure*}

\begin{figure*}
\centering
	\includegraphics[width=1.1\columnwidth]{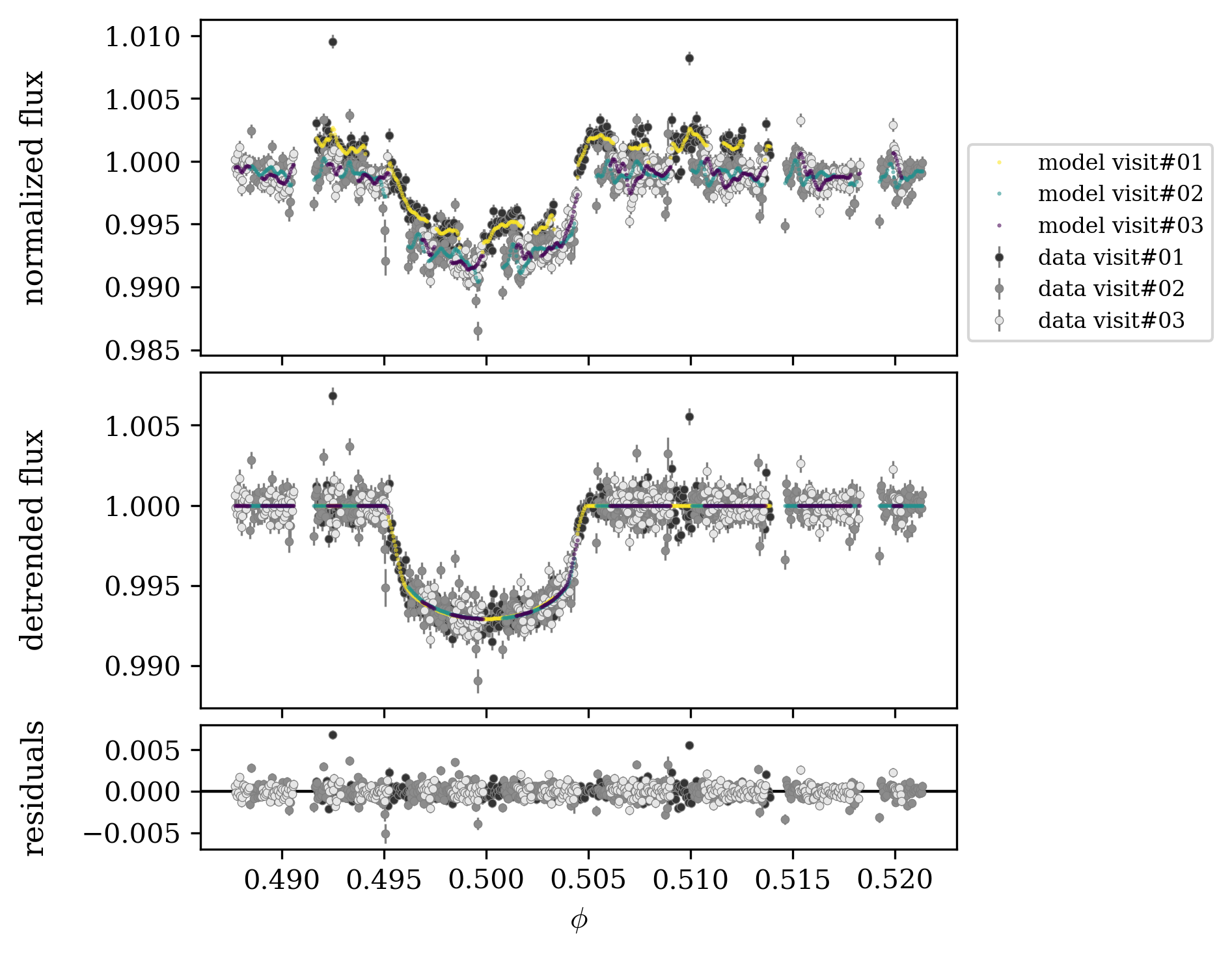}
	\includegraphics[width=0.9\columnwidth]{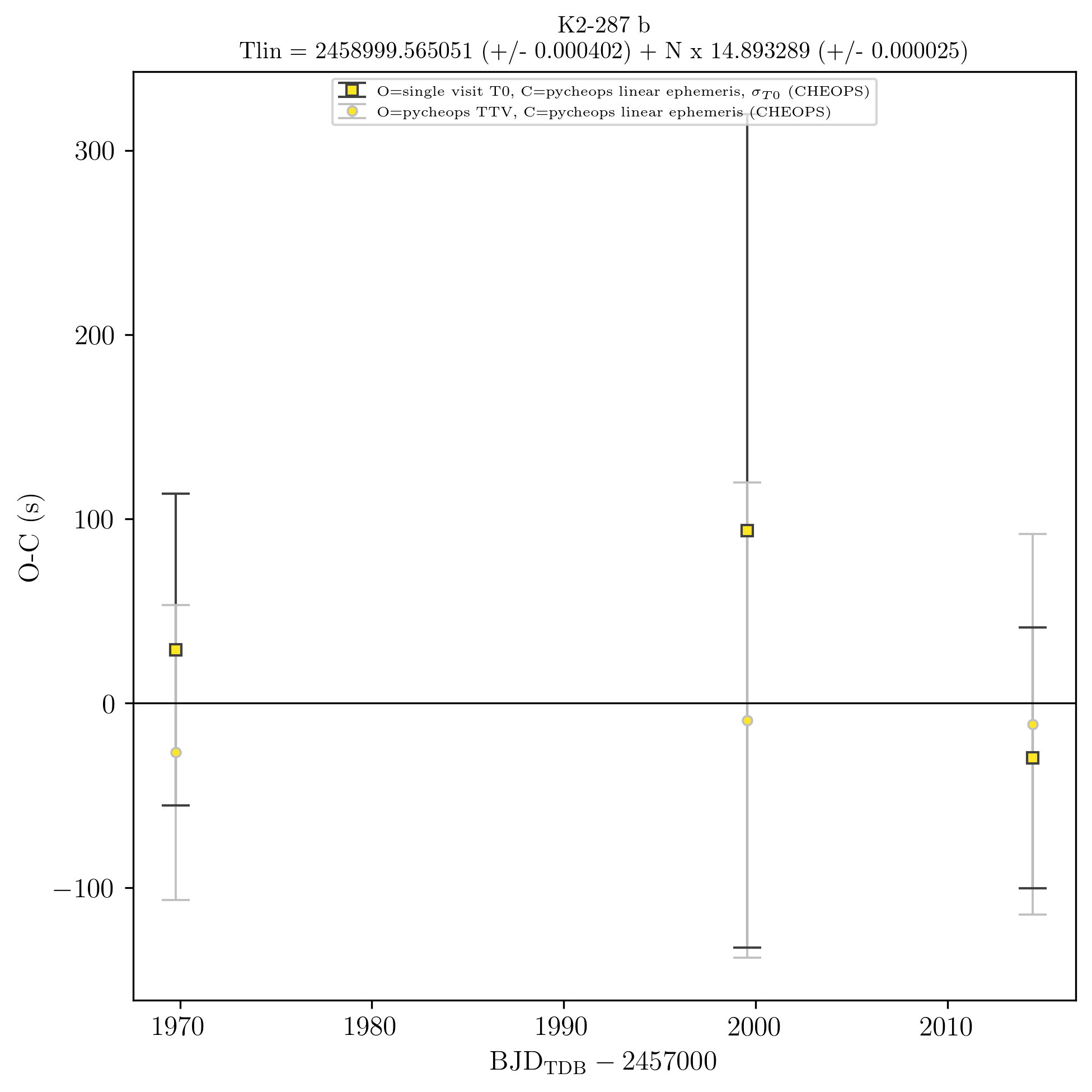}
    \caption{As in Fig~\ref{fig:mv_hatp17}, but for K2-287 b.
    \textit{Left:} multi-visit phase plot of three CHEOPS visits;
    \textit{right:} O-C diagram.
    }
    \label{fig:mv_k2-287}
\end{figure*}

In the best-fit model of the single-visit analysis
we fixed the transit shape for all three visits.
We used as detrending the background with GP in the first visit, 
only the GP in the second visit,
and the first two harmonics of the roll angle with GP for the third visit.
See the best-fit modelling in Fig.~\ref{fig:visits_k2-287}.
We obtained a precision $\sigma_{T_0}=85$~s, 226~s, and 71~s, for the first,
second, and third visit, respectively.
The lack of both ingress and egress and the low \geff{} of the second visits
have a huge impact on the determination of the transit time.
In the multi-visit analysis we fitted the transit shape,
the background of the first visit, 
and GP incorporates the roll angle harmonics of the third visit
(see best-fit model in Fig.~\ref{fig:mv_k2-287} and final parameters in Table~\ref{tab:k2-287}).
We obtained $\sigma_{T_0} = 80$~s, $129$, and $103$~s, 
with a slight improvement of about 5~s ($\sim6\%$) on the $\sigma_{T_0}$ of the first transit,
a huge improvement of 97~s ($\sim43\%$) for the second visit,
and a worsening by 32~s ($\sim46\%$) for the third transit.
As seen for WASP-130 b, the implementation of the GP kernel in the multi-visit analysis
cannot properly model the roll angle of the third visit,
reducing the precision on the $T_0$.
However, the $T_0$ values of the single-visit and of the multi-visit analysis
are all consistent within $1\sigma$, as shown in the $O-C$ plot in Fig.~\ref{fig:mv_k2-287}.
There is not evidence of a TTV, 
because of short baseline and consecutive visits (second and third).
So, it is still too early to draw any conclusion.
\par

\begin{table}
	\centering
	\caption{K2-287 summary table of stellar and planetary (planet b) parameters.
	Input and priors planetary parameters from \citet{Jordan2019AJ....157..100J}.
	Best-fit solution (MLE and semi-interval HDI at $68.27\%$) from the three multi-visit analysis.
	}
	\label{tab:k2-287}
	\resizebox{\fitbox}{!}{%
	\begin{tabular}{lcc}
		\hline
		 Parameters & Input/priors & Source\\
		\hline
		 K2-287 & \multicolumn{2}{c}{Gaia DR2 6239702034929248512} \\
		 RA (J2000)  & 15:32:17.85 & Simbad\\
		 DEC (J2000) & -22:21:29.76 & Simbad\\
		 $\mu_\mathrm{\alpha}$~(mas/yr) & $-4.59 \pm 0.11$ & \gaia{} DR2\\
		 $\mu_\mathrm{\delta}$~(mas/yr) & $-17.90 \pm 0.07$ & \gaia{} DR2\\
         age (Gyr) & $6.6 \pm 1.5$ & This work\\
		 parallax (mas) & $6.29 \pm 0.05$ & \gaia{} DR2\\
		 $V$~(mag) & 11.3 & Simbad\\
		 $G$~(mag) & 11.1 & \gaia{} DR2\\
		 $M_{\star}\, (M_{\sun})$ & $1.03 \pm 0.04$ & This work\\
		 $R_{\star}\, (R_{\sun})$ & $1.10 \pm 0.01$ & This work\\
		 $\rho_{\star}\, (\rho_{\sun})$ & $0.7 \pm 0.3$ & This work\\
		 $T_\mathrm{eff}$~(K) & $5625 \pm 64$ & SWEET-Cat\\
		 $\log g$ & $4.32 \pm 0.11$ & SWEET-Cat\\
		 \feh~(dex) & $+0.27 \pm 0.04$ & SWEET-Cat\\
		 \hline
		 K2-287 b & & \\
		 Model & Input/priors & Multi-visit (MLE \& HDI) \\
		 $T_{0,\mathrm{ref}}^{(a)}$~(days) & $1001.72138 \pm 0.00015$ & $1999.5651 \pm 0.0004$ \\
		 $P$~(days) & $14.893291 \pm 0.000025$ & $14.893289 \pm 0.000025$\\
		 $D=k^2$ & $0.00642 \pm 0.00016$ & $0.0064 \pm 0.0001$\\
		 $W$~(unit of $P$) & $0.0100 \pm 0.0006$ & $0.0098 \pm 0.0003$\\
		 $b$ & $0.78 \pm 0.04$ & $0.80 \pm 0.03$\\
		 $h_1$ & $0.72 \pm 0.01$ & $0.72 \pm 0.01$\\
		 $h_2$ & $0.45 \pm 0.05$ & $0.42 \pm 0.05$\\
		 $T_{0,1}^{(a)}$~(days) & $1969.77881 \pm +0.00098$ & $1969.77816 \pm +0.00093$ \\
		 $T_{0,2}^{(a)}$~(days) & $1999.56614 \pm +0.00262$ & $1999.56494 \pm +0.00149$ \\
		 $T_{0,3}^{(a)}$~(days) & $2014.45800 \pm +0.00082$ & $2014.45821 \pm +0.00119$ \\
		 $\log \sigma_j$ & - & $-7.31 \pm 0.04$\\
		 Derived/physical &  &  \\
		 $k=R_\mathrm{b}/R_{\star}$ & $0.08014 \pm 0.00098$ & $0.0799 \pm 0.0006$\\
		 $R_\mathrm{b}\, (R_\mathrm{Jup})$ & - & $0.88 \pm 0.01$\\
		 $a/R_{\star}$ & $23.87 \pm 0.31$ & $23.6 \pm 0.6$\\
		 $i\, (\degr)$ & $88.1 \pm 0.1$ & $88.1 \pm 0.1$\\
		 $T_{14}^{(b)}$~(days) & $0.15 \pm 0.01$ & $0.146 \pm 0.005$\\
		 $e$ & $0.478 \pm 0.026$ & fixed \\
		 $\omega\, (\degr)$ & $10.1 \pm 4.6$ & fixed \\
		 $K_\mathrm{RV}$~(ms$^{-1}$)) & $28.8 \pm 2.3$ & - \\
		 $M_\mathrm{b}\, (M_\mathrm{Jup})$ & - & $0.31 \pm 0.03$\\
		 $\rho_\mathrm{b}$~(gcm$^{-3}$) & - & $0.63 \pm 0.07$\\
		 $\lambda^{(c)}\, (\degr)$ & - & \\
		 GP hyperparameters & & \\
		 $\log S_0$ & - & $-21.5 \pm 0.1$\\
		 $\log \omega_0$ & - & $6.5 \pm 0.1$\\
		\hline
	\end{tabular}
	}
	\\
	\textbf{Notes:}
	$^{(a)}$: Transit times in BJD$_\mathrm{TDB}-2457000$. 
	$T_{0,n}$ single visit output in the input/priors column, 
	while they are the linear ephemeris plus $\Delta T_{0,n}$ from multi-visit analysis.
	$^{(b)}$: Total duration is equal to $T_{14} = W \times P$.
	$^{(c)}$: spin-orbit angle measured from the Rossiter-McLaughlin effect.
\end{table}

\section{Results and discussion}
\label{sec:results}

From the current dataset of 17 transits of seven warm-Jupiters
we obtained a wide range of timing precision $\sigma_{T_0}$, 
summarised in Table~\ref{tab:results}.
The best timing is of about 13~s, for the brightest target WASP-38,
and the worst case is of about 250~s (for the single-visit analysis),
for WASP-130.
Beyond the stellar brightness and the global efficiency,
another major contributor, or limiting factor,
to the precision of the transit time is the efficiency of the critical phase ranges (\cpreff),
that is the coverage of the transit ingress and egress.
In case of small temporal sampling of the ingress/egress phases (\cpreff~$< 30\%-50\%$)
we improved the timing precision combining multiple visits.
In the case of WASP-8 b, we had almost no improvement from the multi-visit analysis,
because the combined transits (see Fig.~\ref{fig:mv_wasp8}) did not fully cover
both ingress and egress phases.
\par

We have to take into account that
the higher the requested \cpreff{} in both ingress and egress,
the lower the probability to schedule with CHEOPS that particular visit,
simply because there are less visits available for scheduling
that actually satisfy the stringent constraints on the critical phase ranges.
To ensure an appropriate time sampling of the TTV signal
we have to request for visits
with high efficiencies in the critical phase ranges.
We compared the expected \geff{} and \cpreff{} from the FC 
with the observed \geff{} and \cpreff{} in actual CHEOPS visits.
We remind the reader that the FC was meant as a statistically indicative tool
and not as a planning tool for the mission.
A few early visits have been scheduled without checking the \cpreff,
but we computed the critical phase ranges (cpr) for all the targets
with the parameters from the literature propagating the errors
\footnote{We used \textit{Uncertainties:
a Python package for calculations with uncertainties},
Eric O. LEBIGOT,
\url{http://pythonhosted.org/uncertainties/.}}
and we ran the FC to obtain the expected \cpreff.
We computed the observed \geff{} of a visit as the ratio of the number of data-points,
that is the number of real exposures,
over the maximum possible number of exposures due to the visit duration.
We computed the observed \cpreff{} in the same way as the \geff,
but taking into account only the length and the data within the phase ranges of the ingress and egress.
Then, we computed the maximum value of 
the absolute difference
between the expected and the observed efficiency.
For cpr timescales of the order of $\sim30$~min,
which is the typical duration of the ingress and/or egress
of the transit of a warm-Jupiter,
we found that the predicted \cpreff{}s agree with the observed ones
within $\sim10\%$.
Also the difference of the \geff{} is of the same order.
We expect that these differences will increase with time,
because the orbit file in the FC will not be updated.
\par

We found that for some targets 
the cpr of our visits do not cover the observed ingress and egress phases.
We re-computed the cpr with the updated linear ephemeris and parameters from this work.
We found that all cpr match exactly the ingress and egress of all the visits.
The mismatch on the positions of the cpr does not seem
to depend on the difference between the FC's orbit and the actual orbit,
but rather on the accuracy and precision of the ephemeris and transit parameters,
which are fundamental to prepare CHEOPS observations.
\par

In our cases, the best timing would allow us
to detect all the expected range of the TTV signals ($\geq 1$~min)
probing all the possible, and realistic, regions of the parameter space of a perturber.
Our worse cases, WASP-106 and K2-287,
have an average $\sigma_{{T_0},\mathrm{multi}}$ of less than 2~min with only three transits.
This would limit the range of the detectable TTV signals,
but the possible orbital configurations
of the system with a further planet (see Fig.~\ref{fig:grid_wasp106} and \ref{fig:grid_k2-287})
are so numerous and extended that the current study is still feasible.
We can affirm that, in general, CHEOPS will be able to detect TTV signals
with amplitude less than 1~min for target brighter than $G = 11-12$,
if the multiple visits could cover with high efficiency the ingress and egress phases.
\par

It is worth noting that one of the few hot Jupiters hosts known to have planetary companions,
WASP-47 \citep{Becker2015ApJ...812L..18B}, also falls within this magnitude range and
is well observable by CHEOPS.
It is actually included in another GTO subprogram (Nascimbeni et al., in prep.).
By applying the same techniques described in this work, its 40-s TTV \citep{Becker2015ApJ...812L..18B}
is expected to be detectable.
\par

\begin{table}
	\centering
	\caption{Summary of the $\sigma_{T_0}$ in seconds of all targets and visits
	(columns V1, V2, V3, and V4).
	In case of multi-visit analysis: 
	$\sigma_{{T_0},\mathrm{multi}}\ (\sigma_{{T_0},\mathrm{single}})$;
	if only single-visit analysis: $\sigma_{{T_0},\mathrm{single}}$.
	}
	\label{tab:results}
	\begin{tabular}{lcccc}
		\hline
		           & \multicolumn{4}{c}{$\sigma_{T_0}$~(seconds)}\\
		 target    & V1 & V2 & V3 & V4 \\
		\hline
		HAT-P-17 b & $52\ (87)$ & $53\ (82)$ & $94\ (97)$ &  \\
		KELT-6 b   & $114$ &  &  &  \\
		WASP-8 b   & $50\ (53)$ & $28\ (31)$ &  &  \\
		WASP-38 b  & $20\ (24)$ & $16\ (13)$ & $17\ (16)$ & $17\ (16)$ \\
		WASP-106 b & $60$ &  &  &  \\
		WASP-130 b & $44\ (81)$ & $197\ (251)$ & $65\ (45)$ &  \\
		K2-287 b   & $80\ (85)$ & $128\ (226)$ & $103\ (71)$ &  \\
		\hline
	\end{tabular}
\end{table}

\section{Conclusions}
\label{sec:conclusions}

The main purpose of this work was to demonstrate CHEOPS capability to schedule multiple observations
and obtain transit times with sufficient accuracy to allow detection of TTV signals.
In this context, we present one of the CHEOPS GTO programs
aimed at the detection of possible TTV signals
with amplitude of the order of a few minutes of warm-Jupiter exoplanets
due to gravitational interaction with a planetary companion
on outer orbit.
\par

We collected 17 light curves of transits of seven out of eight targets of our sample,
and presented the observing strategy and the data analysis.
We demonstrated the impact and the importance of a good sampling of the ingress and egress phases of a transit
on the precision of the transit time,
but also of the pre- and post-transit portions to properly detrend the light curve
for the systematic effects.
We showed improvement on timing precision $\sigma_{T_0}$ combining the multiple visits of five targets:
HAT-P-17 b, WASP-8 b, WASP-38 b, WASP-130 b, and K2-287 b.
The precision $\sigma_{T_0}$ ranges from about ten seconds (i.e., WASP-38 b)
to a couple of minutes (i.e., WASP-130 b and K2-287 b)
for visits with high and low temporal sampling
of both ingress and egress phases,
respectively.
\par

These observations were very helpful to understand how to properly prepare next observations,
how to precisely set the visit duration and the required efficiency of each transit phase.
A simulation of the feasible visits with updated linear ephemeris and 
stellar and planetary parameters is mandatory to increase the efficiency
of the CHEOPS observations
\par

With the current dataset, 
we cannot draw any conclusions about the existence of a TTV signal
in our target sample due to the short temporal span of our observations,
but this was not the purpose of this work,
focused on the demonstration of the timing capabilities of the CHEOPS mission.
We aim to collect further visits for each target
to reach at least five visits covering about a year of CHEOPS mission,
with the goal of 15 visits in the total nominal mission duration of 3.5~yr.
For each target we will analyse CHEOPS data simultaneously 
with literature photometric and spectroscopy data to detect a TTV signal
on a long temporal baseline.
This will help us to improve the planetary parameters and
to reduce the error on the ephemeris,
necessary to increase the efficiency of further follow-up
with current and future ground- and space-based
facilities,
i.e., 
HARPS \citep{Mayor2003Msngr.114...20M},
HARPS-N \citep{Cosentino2012SPIE.8446E..1VC},
ESPRESSO \citep{Pepe2010SPIE.7735E..0FP,Pepe2020arXiv201000316P},
the European Extremely Large Telescope (ELT),
JWST \citep{Gardner2006SSRv..123..485G},
and
ARIEL \citep{Pascale2018SPIE10698E..0HP,Pilbratt2019ESS.....450304P,Puig2018ExA....46..211P,Tinetti2018ExA....46..135T}.
\par

\section*{Acknowledgements}

CHEOPS is an ESA mission in partnership with Switzerland 
with important contributions to the payload and the ground segment from
Austria, Belgium, France, Germany, Hungary, Italy, Portugal, Spain, Sweden, and the United Kingdom.
The CHEOPS Consortium would like to gratefully acknowledge the support received
by all the agencies, offices, universities, and industries involved. 
Their flexibility and willingness to explore new approaches
were essential to the success of this mission.
KGI is the ESA CHEOPS Project Scientist and
is responsible for the ESA CHEOPS Guest Observers Programme.
She does not participate in, or contribute to,
the definition of the Guaranteed Time Programme of the CHEOPS mission
through which observations described in this paper have been taken,
nor to any aspect of target selection for the programme.
The Swiss participation to CHEOPS has been supported by
the Swiss Space Office (SSO) in the framework of the Prodex Programme and
the Activite Nationales Complementaires (ANC), the Universities of Bern and Geneva
as well as of the NCCR PlanetS and the Swiss National Science Foundation.
The early support for CHEOPS by Daniel Neuenschwander is gratefully acknowledged.
GPi, VN, GSs, IPa, LBo, GLa, and RRa acknowledge the funding support from Italian Space Agency (ASI)
regulated by ``Accordo ASI-INAF n. 2013-016-R.0 del 9 luglio 2013 e integrazione del 9 luglio 2015 CHEOPS Fasi A/B/C''.
GLa acknowledges support by CARIPARO Foundation, according to the agreement
CARIPARO-Universit{\`a} degli Studi di Padova (Pratica n. 2018/0098), 
and scholarship support by the ``Soroptimist International d'Italia'' association (Cortina d'Ampezzo Club).
VVG is an FRS-FNRS Research Associate.
VVG, LD and MG thank the Belgian Federal Science Policy Office (BELSPO) 
for the provision of financial support in the framework of the PRODEX Programme
of the European Space Agency (ESA) under contract number PEA 4000131343.
DG, MF, SC, XB, and JL acknowledge their roles as ESA-appointed CHEOPS science team members.
ZG was supported by the Hungarian NKFI grant No. K-119517 and the GINOP
grant No. 2.3.2-15-2016-00003 of the Hungarian National Research
Development and Innovation Office, by the City of Szombathely under
agreement No. 67.177-21/2016, and by the VEGA grant of the Slovak
Academy of Sciences No. 2/0031/18.
This work was supported by FCT - Funda\c{c}\~ao para a Ci\^encia e a Tecnologia
through national funds and by FEDER through COMPETE2020 - 
Programa Operacional Competitividade e Internacionaliza\c{c}\~ao by these grants:
UID/FIS/04434/2019; UIDB/04434/2020; UIDP/04434/2020; PTDC/FIS-AST/32113/2017 \&
POCI-01-0145-FEDER-032113; PTDC/FIS-AST/28953/2017 \&
POCI-01-0145-FEDER-028953; PTDC/FIS-AST/28987/2017 \&
POCI-01-0145-FEDER-028987. 
A.C.C. and T.G.W. acknowledge support from STFC consolidated grant number ST/M001296/1.
SH acknowledges CNES funding through the grant 837319.
O.D.S.D. is supported in the form of work contract (DL 57/2016/CP1364/CT0004)
funded by national funds through Funda\c{c}\~{a}o para a Ci\^{e}ncia e Tecnologia (FCT).
This project has received funding from the European Research Council (ERC)
under the European Union’s Horizon 2020
research and innovation programme (project {\sc Four Aces}; grant agreement No 724427)

\section*{Data Availability}
Data will be available at CDS.
Data type: default aperture data and best-fit model in ascii file.



\bibliographystyle{mnras}
\bibliography{biblio} 



\section*{Affiliations}
$^{1}$INAF, Osservatorio Astronomico di Padova, Vicolo dell'Osservatorio 5, 35122 Padova, Italy\\
$^{2}$Dipartimento di Fisica e Astronomia "Galileo Galilei", Universià degli Studi di Padova, Vicolo dell'Osservatorio 3, 35122 Padova, Italy\\
$^{3}$INAF, Osservatorio Astrofisico di Torino, via Osservatorio 20, 10025 Pino Torinese, Italy\\
$^{4}$Observatoire de Gen\`eve, Universit\'e de Gen\`eve, Chemin Pegasi, 51 1290 Versoix, Switzerland\\
$^{5}$Astrophysics Group, Keele University, Staffordshire, ST5 5BG, United Kingdom\\
$^{6}$Instituto de Astrof\'isica e Ci\^encias do Espa\c{c}o, Universidade do Porto, CAUP, Rua das Estrelas, 4150-762 Porto, Portugal\\
$^{7}$School of Physics and Astronomy, Physical Science Building, North Haugh, St Andrews, United Kingdom\\
$^{8}$Space Research Institute, Austrian Academy of Sciences, Schmiedlstrasse 6, A-8042 Graz, Austria\\
$^{9}$ELTE E\"otv\"os Lor\'and University, Gothard Astrophysical Observatory, 9700 Szombathely, Szent Imre h. u. 112, Hungary\\
$^{10}$MTA-ELTE Exoplanet Research Group, 9700 Szombathely, Szent Imre h. u. 112, Hungary\\
$^{11}$Astronomical Institute, Slovak Academy of Sciences, 05960 Tatransk\'a Lomnica, Slovakia\\
$^{12}$Physikalisches Institut, University of Bern, Gesellsschaftstrasse 6, 3012 Bern, Switzerland\\
$^{13}$Instituto de Astrof\'\i{}sica de Canarias (IAC), 38200 La Laguna, Tenerife, Spain\\
$^{14}$Departamento de Astrof\'\i{}sica, Universidad de La Laguna (ULL), E-38206 La Laguna, Tenerife, Spain\\
$^{15}$ESTEC, European Space Agency, Keplerlaan 1, 2201 AZ Noordwijk, The Netherlands\\
$^{16}$Admatis, Miskolc, Hungary\\
$^{17}$Center for Space and Habitability, Gesellsschaftstrasse 6, 3012 Bern, Switzerland\\
$^{18}$Depto. de Astrof\'\i{}sica, Centro de Astrobiologia (CSIC-INTA), ESAC campus, 28692 Villanueva de la Cãda (Madrid), Spain\\
$^{19}$Departamento de F\'isica e Astronomia, Faculdade de Ci\^encias, Universidade do Porto, Rua do Campo Alegre, 4169-007 Porto, Portugal\\
$^{20}$Universit\'e Grenoble Alpes, CNRS, IPAG, 38000 Grenoble, France\\
$^{21}$Department of Astronomy, Stockholm University, AlbaNova University Center, 10691 Stockholm, Sweden\\
$^{22}$Institute of Planetary Research, German Aerospace Center (DLR), Rutherfordstrasse 2, 12489 Berlin, Germany\\
$^{23}$Institut de Physique du Globe de Paris (IPGP), 1 rue Jussieu, 75005 Paris, France\\
$^{24}$Lund Observatory, Dept. of Astronomy and Theoreical Physics, Lund University, Box 43, 22100 Lund, Sweden\\
$^{25}$Aix Marseille Univ, CNRS, CNES, LAM, Marseille, France\\
$^{26}$Space sciences, Technologies and Astrophysics Research (STAR) Institute, Universit\'e de Li\`ege, All\'ee du six Ao\^ut 19C, 4000 Li\`ege, Belgium\\
$^{27}$Astrobiology Research Unit, Universit\'e de Li\`ege, All\'ee du six Ao\^ut 19C, 4000 Li\`ege, Belgium\\
$^{28}$Institut d’astrophysique de Paris, UMR7095 CNRS, Universit\'e Pierre \& Marie Curie, 98bis blvd. Arago, 75014 Paris, France\\
$^{29}$Institut de Ci\`encies de l'Espai (ICE, CSIC), Campus UAB, C/CanMagrans s/n, 08193 Bellaterra, Spain\\
$^{30}$Institut d’Estudis Espacials de Catalunya (IEEC), Barcelona, Spain\\
$^{31}$Leiden Observatory, University of Leiden, PO Box 9513, 2300 RA Leiden, The Netherlands\\
$^{32}$Department of Space, Earth and Environment, Chalmers University of Technology, Onsala Space Observatory, 43992 Onsala, Sweden\\
$^{33}$University of Vienna, Department of Astrophysics, T\"urkenschanzstrasse 17, 1180 Vienna, Austria\\
$^{34}$Department of Physics, University of Warwick, Gibbet Hill Road, Coventry CV4 7AL, United Kingdom\\
$^{35}$Konkoly Observatory, Research Centre for Astronomy and Earth Sciences, 1121 Budapest, Konkoly Thege Mikl\'os út 15-17, Hungary\\
$^{36}$IMCEE, UMR8028 CNRS, Observatoire de Paris, PSL Univ., Sorbonne Univ., 77 av. Denfert-Rochereau, 75014 Paris, France\\
$^{37}$INAF, Osservatorio Astrofisico di Catania, Via S. Sofia 78, 95123 Catania, Italy\\
$^{38}$Astrophysics Group, Cavendish Laboratory, J.J. Thomson Avenue, Cambridge CB3 0He, United Kingdom\\
$^{39}$Center for Astronomy and Astrophysics, Technical University Berlin, Hardenberstrasse 36, 10623 Berlin, Germany\\
$^{40}$Institut f\"ur Geologische Wissenschaften, Freie Universit\"at Berlin, 12249 Berlin, Germany\\


\appendix



\section{Maps of the expected TTV signals}
\label{apdx:ttv}

Maps of the expected TTV signals for each target.
Each map has been created as described in Sec~\ref{sec:warmJup}.

\begin{figure*}
    \centering
    \includegraphics[width=0.33\textwidth]{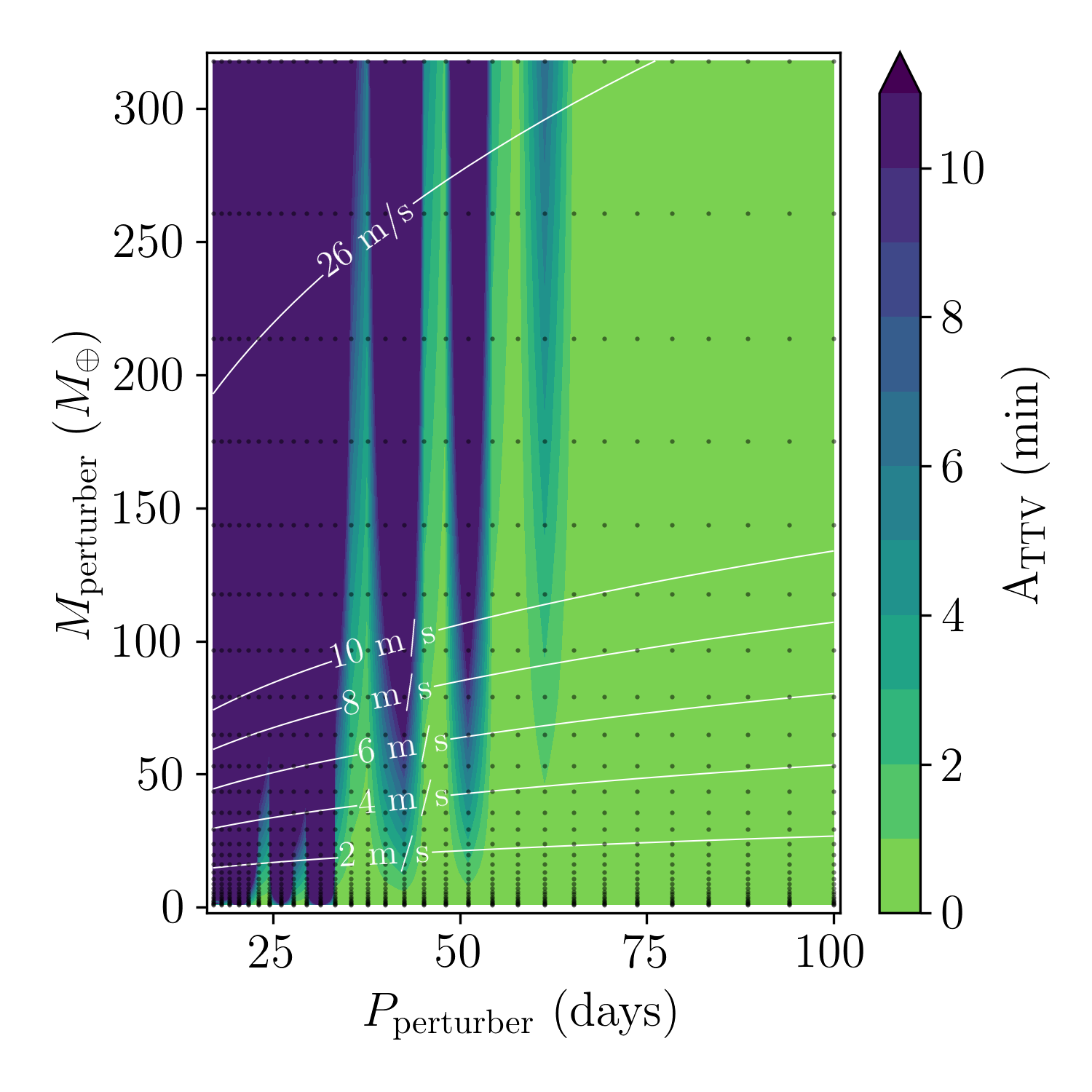}
    \includegraphics[width=0.33\textwidth]{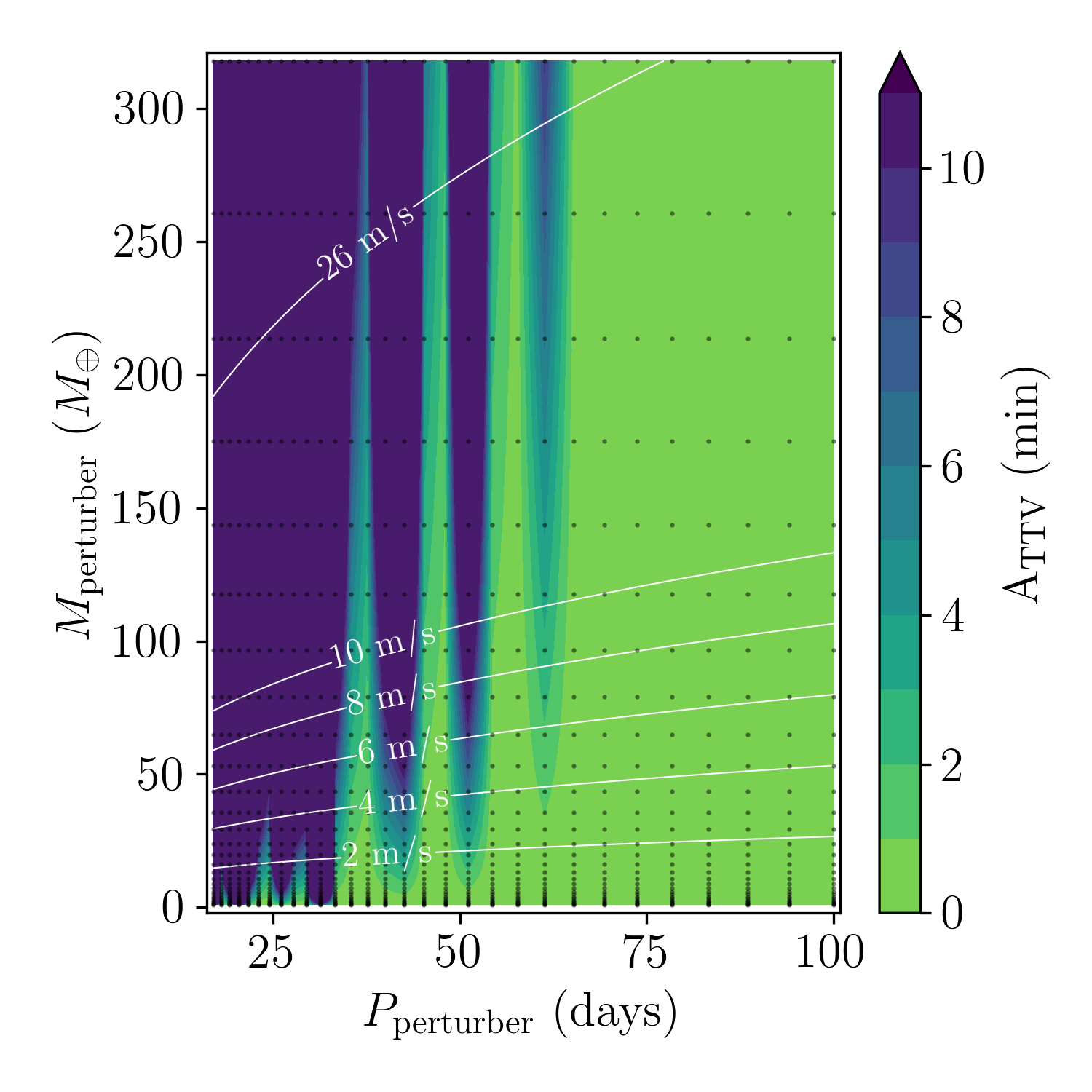}
    \includegraphics[width=0.33\textwidth]{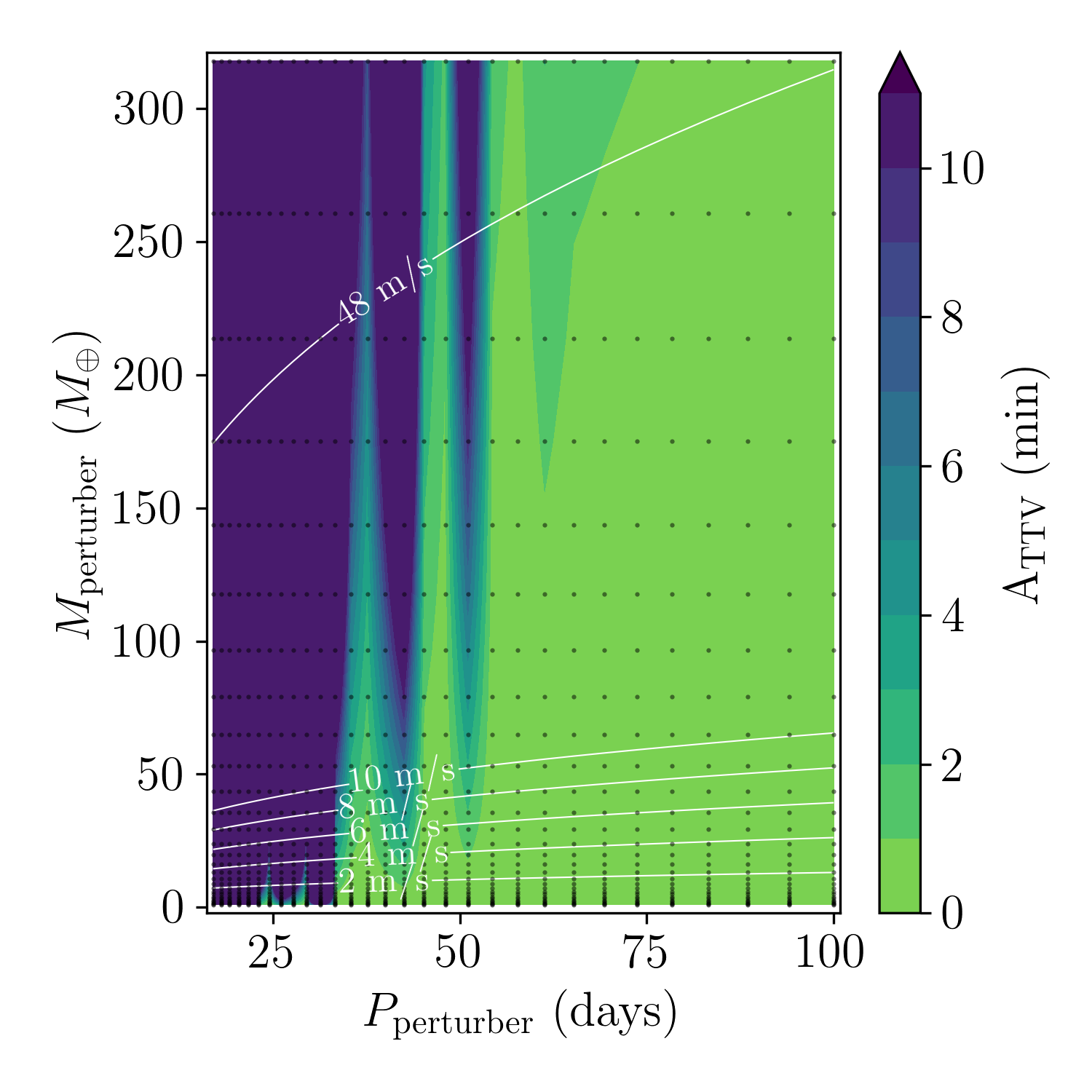}
    
    \caption{TTV amplitude ($A_\mathrm{TTV}$) map from the 900 numerical integration of a possible perturber with
    30 log-values of mass and period for HAT-P-17.
    The gray dots are the mass-period combinations used for each simulation.
    The white lines are the RV semi-amplitude ($K_\mathrm{RV}$) of the perturber. 
    The three plots have different initial values of eccentricity ($e_\mathrm{perturber}$) and mutual inclination ($\Delta i$),
    and same argument of pericenter ($\omega_\mathrm{perturber}=90\degr$).
    \textit{Left:} $e_\mathrm{perturber}=0.0,\ \Delta i=60\degr$;
    \textit{center:} $e_\mathrm{perturber}=0.1,\ \Delta i=60^\circ$;
    \textit{right:} $e_\mathrm{perturber}=0.0,\ \Delta i=0^\circ$.
    }
    \label{fig:grid_hatp17}
\end{figure*}

\begin{figure*}
    \centering
	\includegraphics[width=0.33\textwidth]{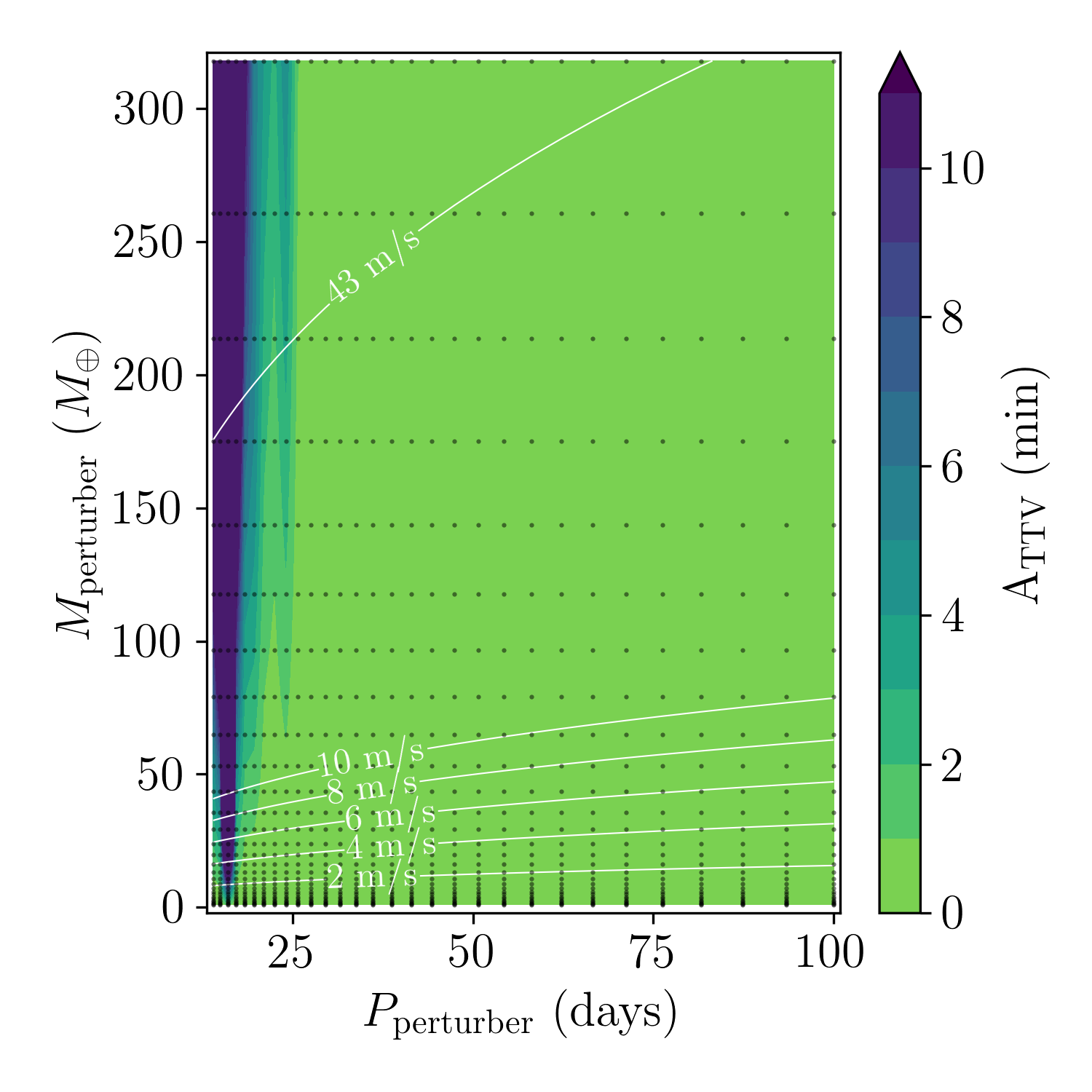}
	\includegraphics[width=0.33\textwidth]{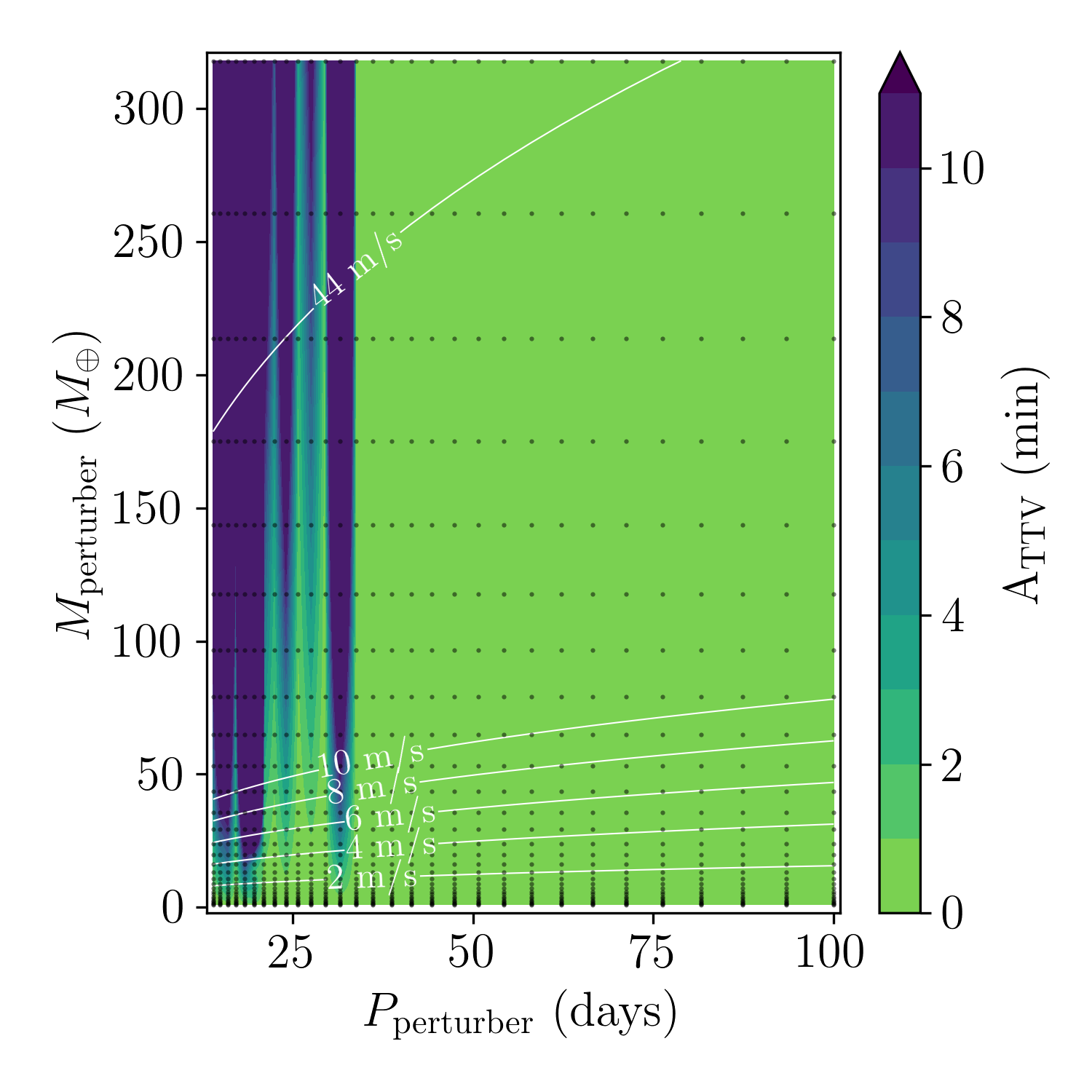}
	\includegraphics[width=0.33\textwidth]{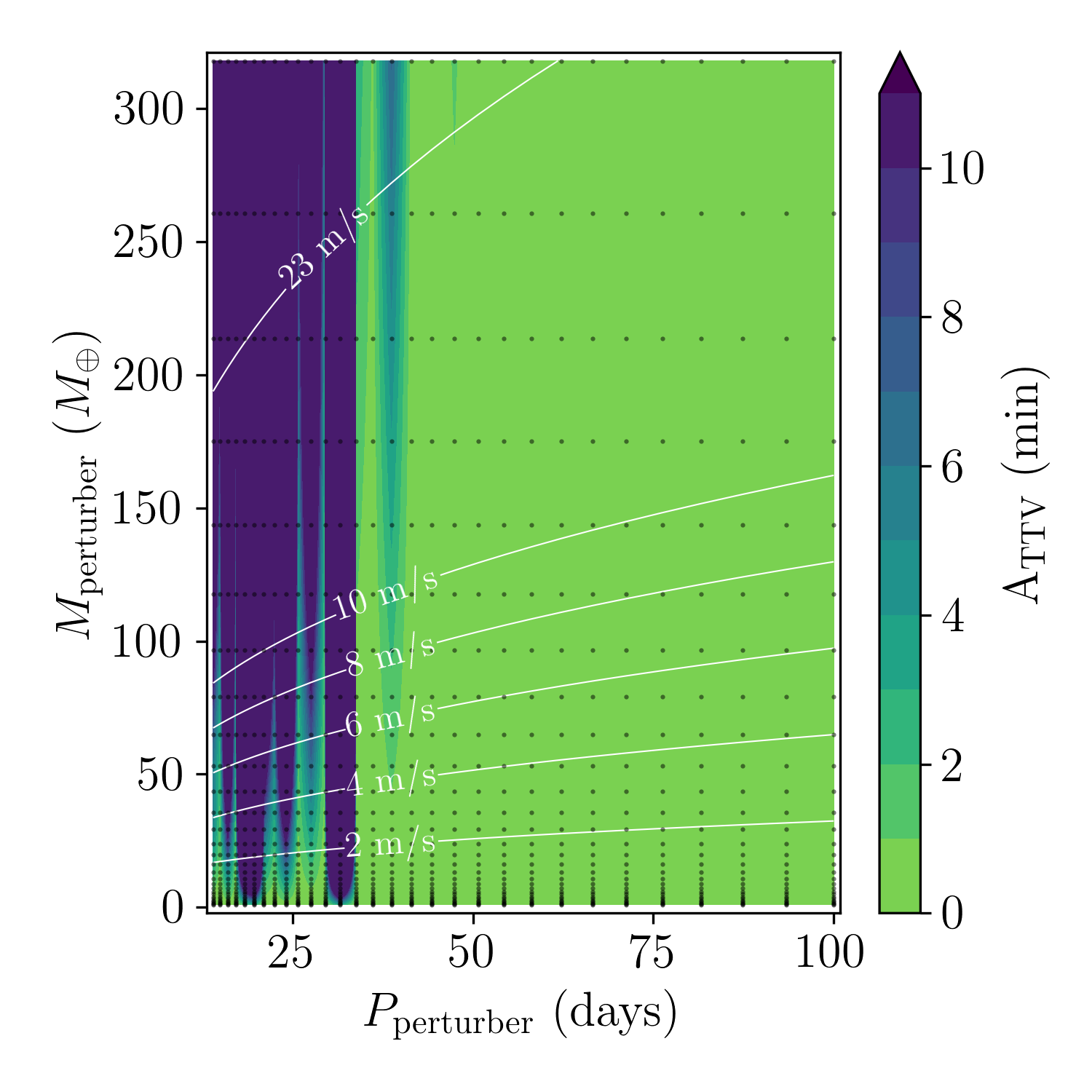}

    \caption{As Fig.~\ref{fig:grid_hatp17} for KELT-6.
    \textit{Left:} $e_\mathrm{perturber}=0.0,\ \Delta i=0\degr$;
    \textit{center:} $e_\mathrm{perturber}=0.1,\ \Delta i=0^\circ$;
    \textit{right:} $e_\mathrm{perturber}=0.1,\ \Delta i=60^\circ$.
    }
    \label{fig:grid_kelt6}
\end{figure*}

\begin{figure*}
    \centering
	\includegraphics[width=0.33\textwidth]{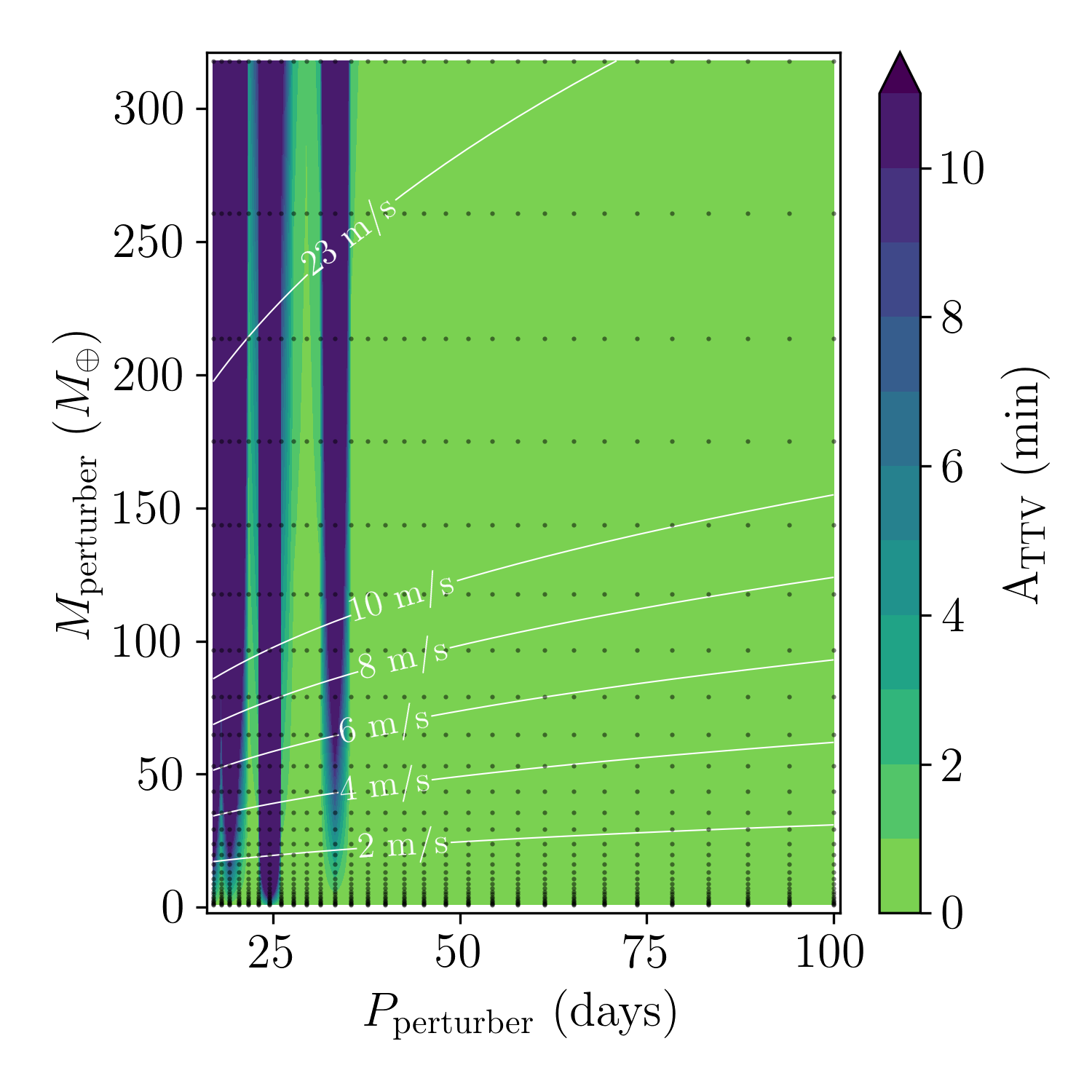}
	\includegraphics[width=0.33\textwidth]{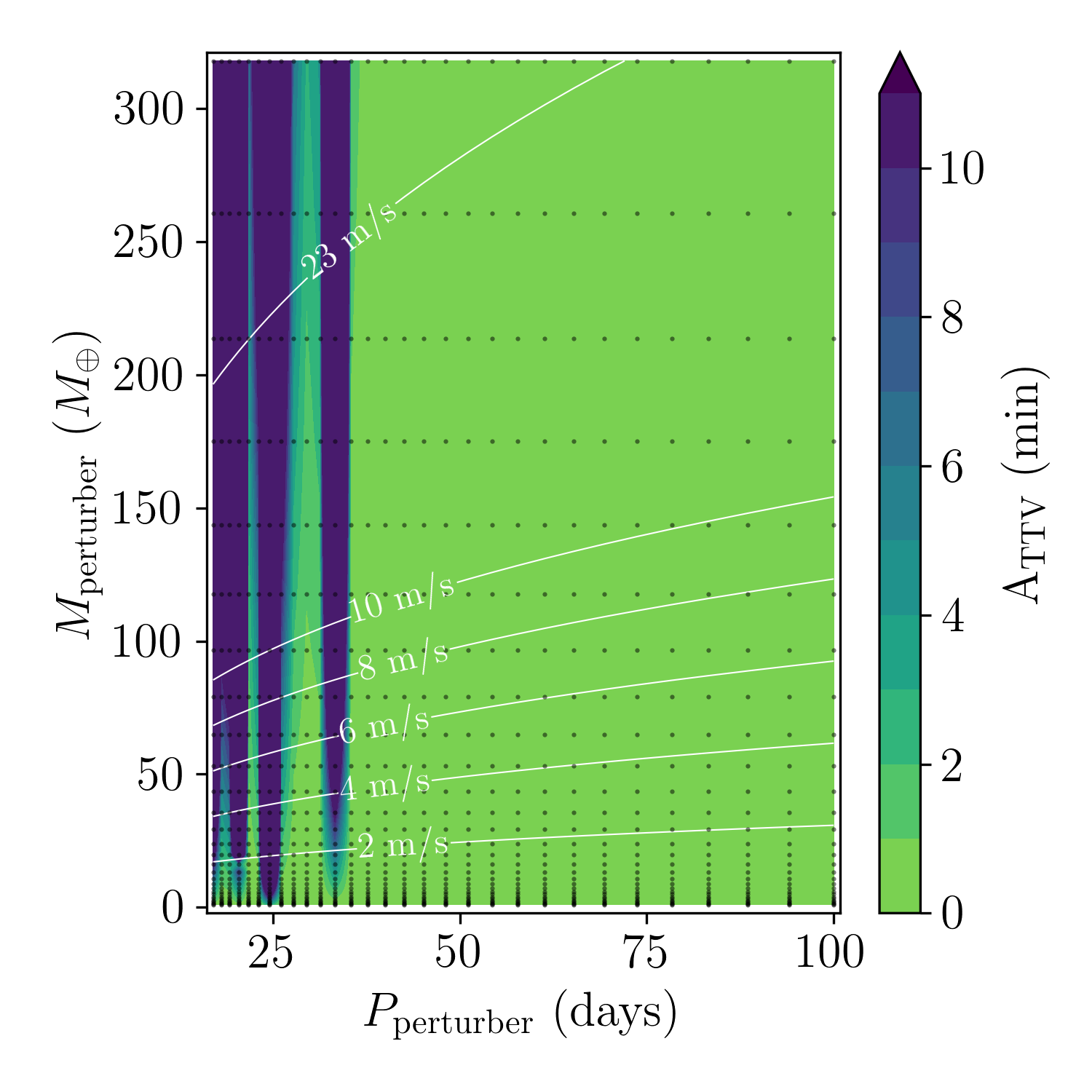}
	\includegraphics[width=0.33\textwidth]{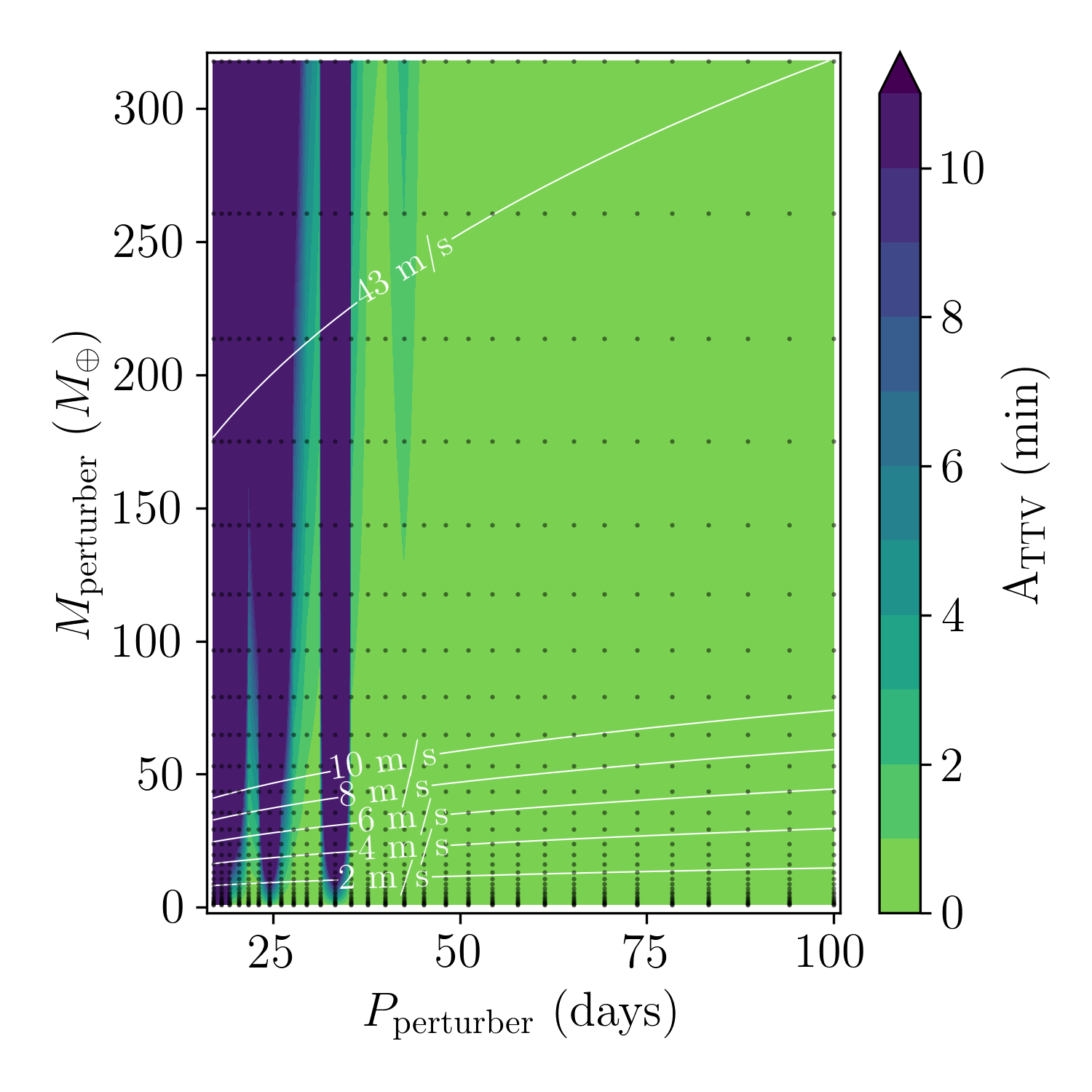}

    \caption{As Fig.~\ref{fig:grid_hatp17} for WASP-8.
    }
    \label{fig:grid_wasp8}
\end{figure*}

\begin{figure*}
    \centering
	\includegraphics[width=0.33\textwidth]{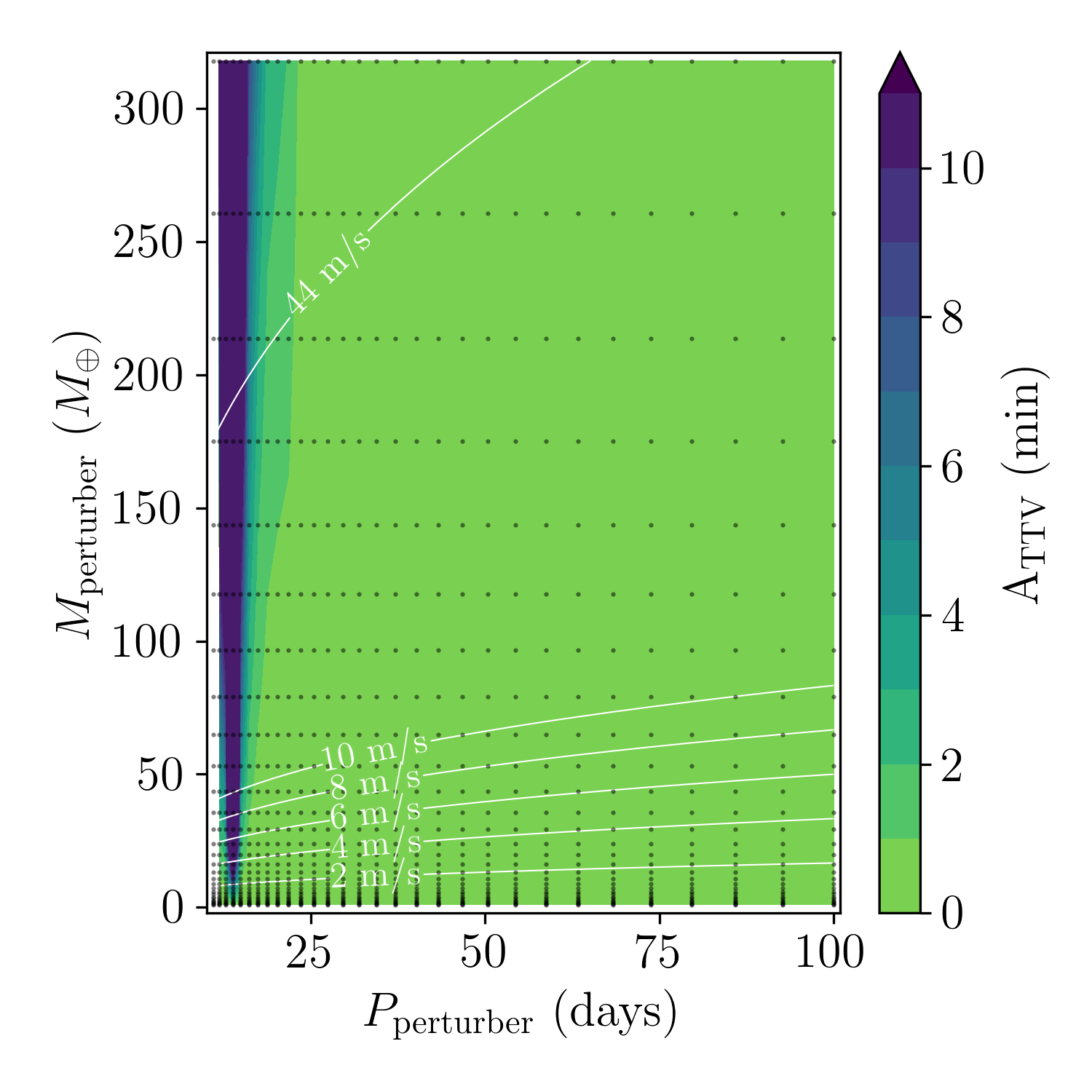}
	\includegraphics[width=0.33\textwidth]{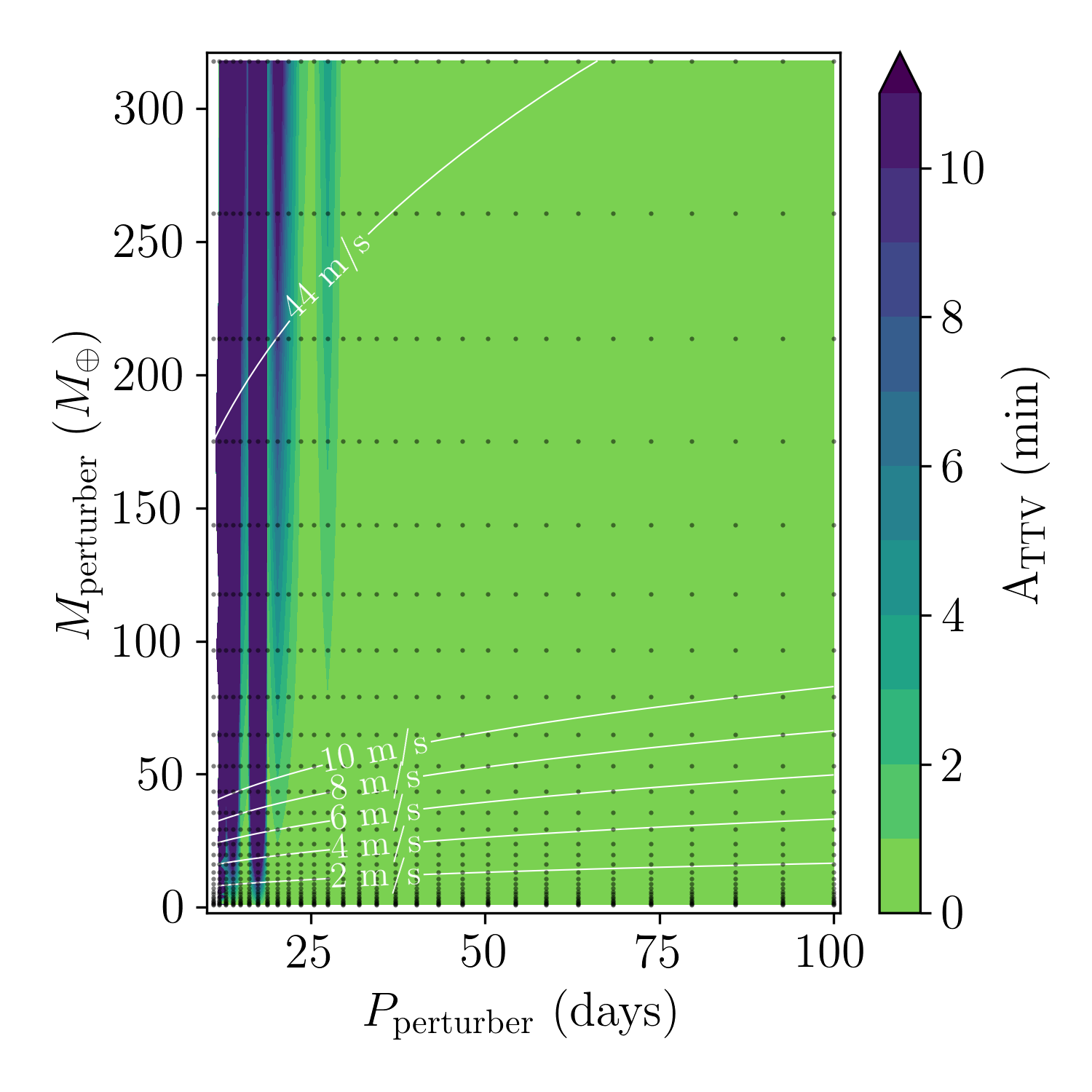}
	\includegraphics[width=0.33\textwidth]{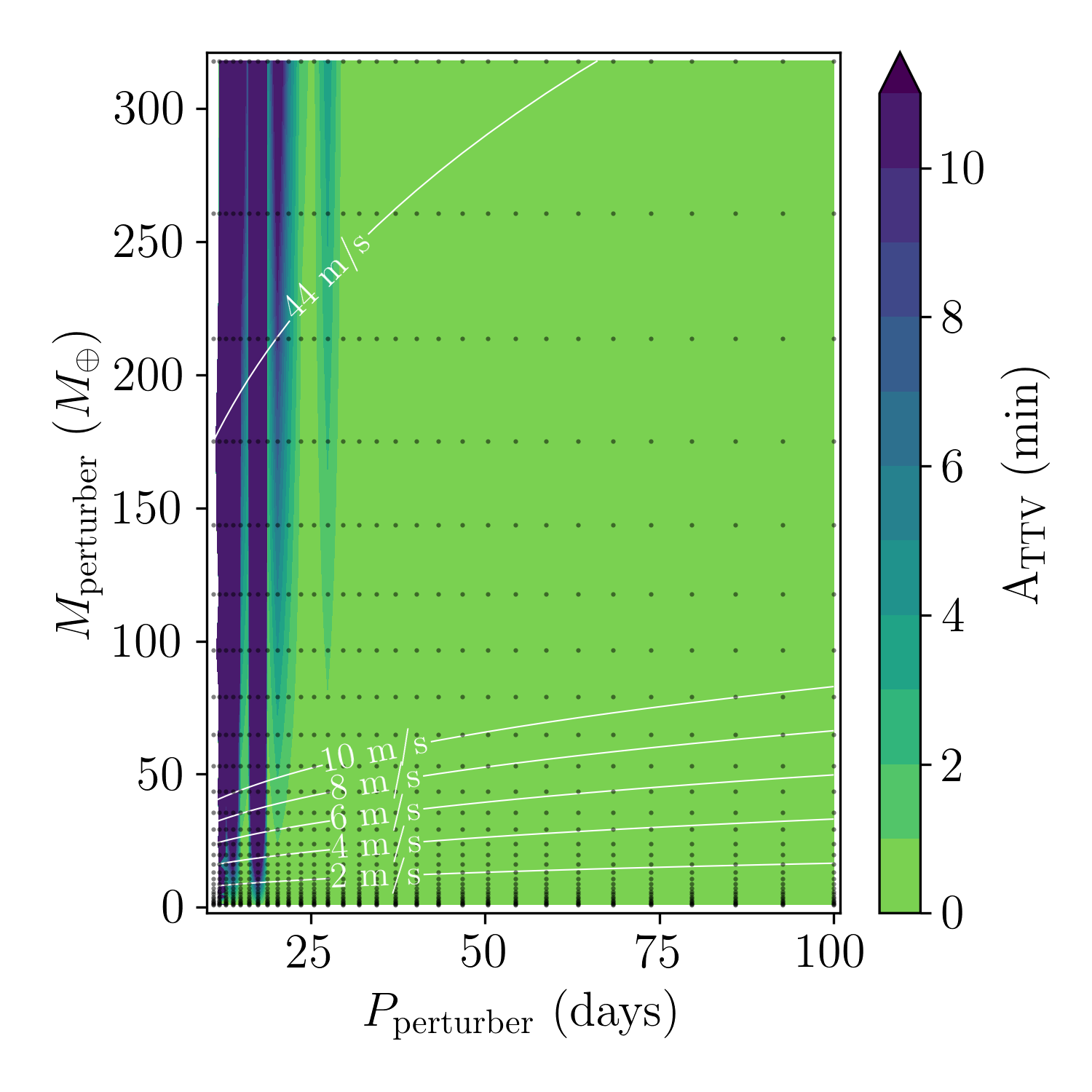}

    \caption{As Fig.~\ref{fig:grid_kelt6} for WASP-38.
    }
    \label{fig:grid_wasp38}
\end{figure*}

\begin{figure*}
    \centering
	\includegraphics[width=0.33\textwidth]{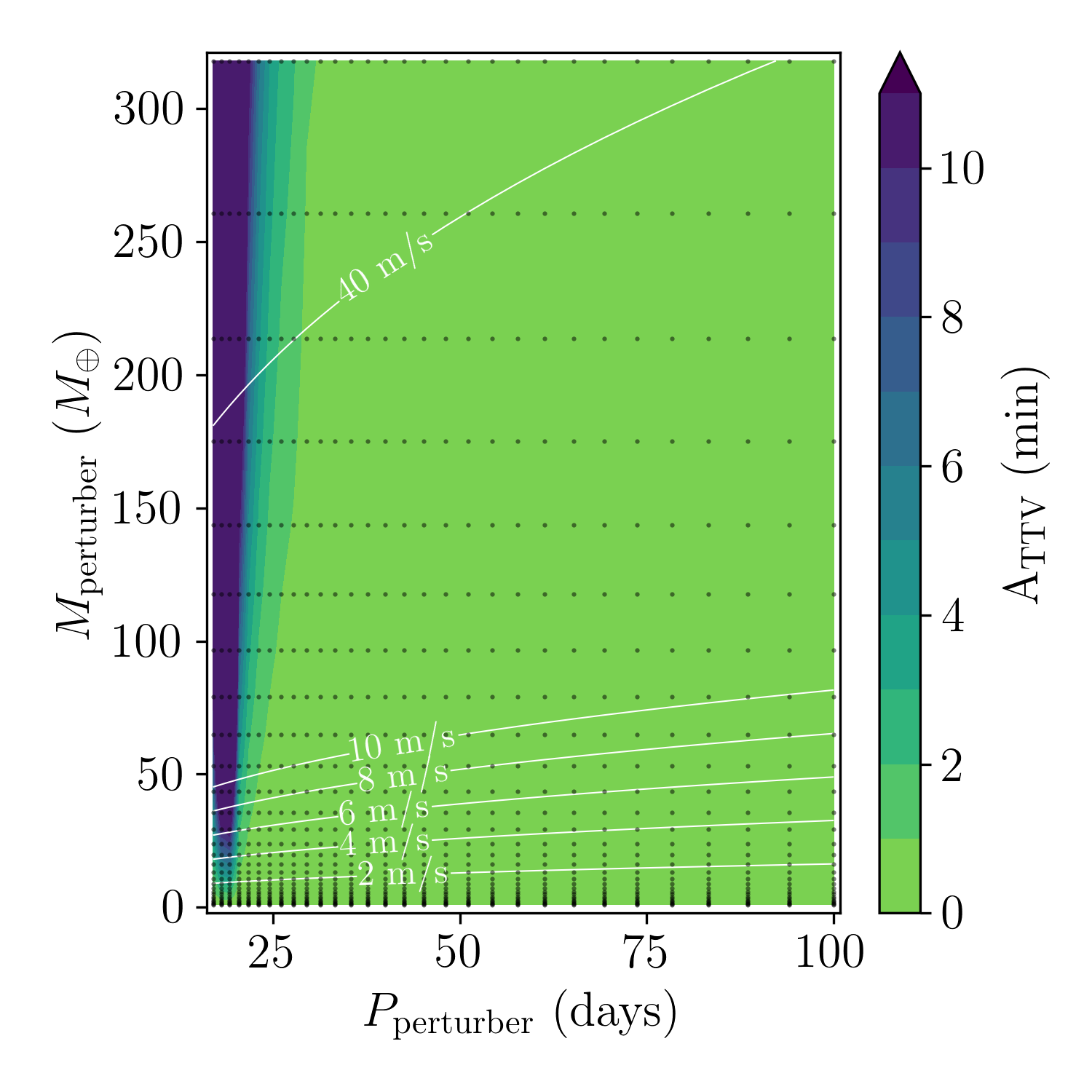}
	\includegraphics[width=0.33\textwidth]{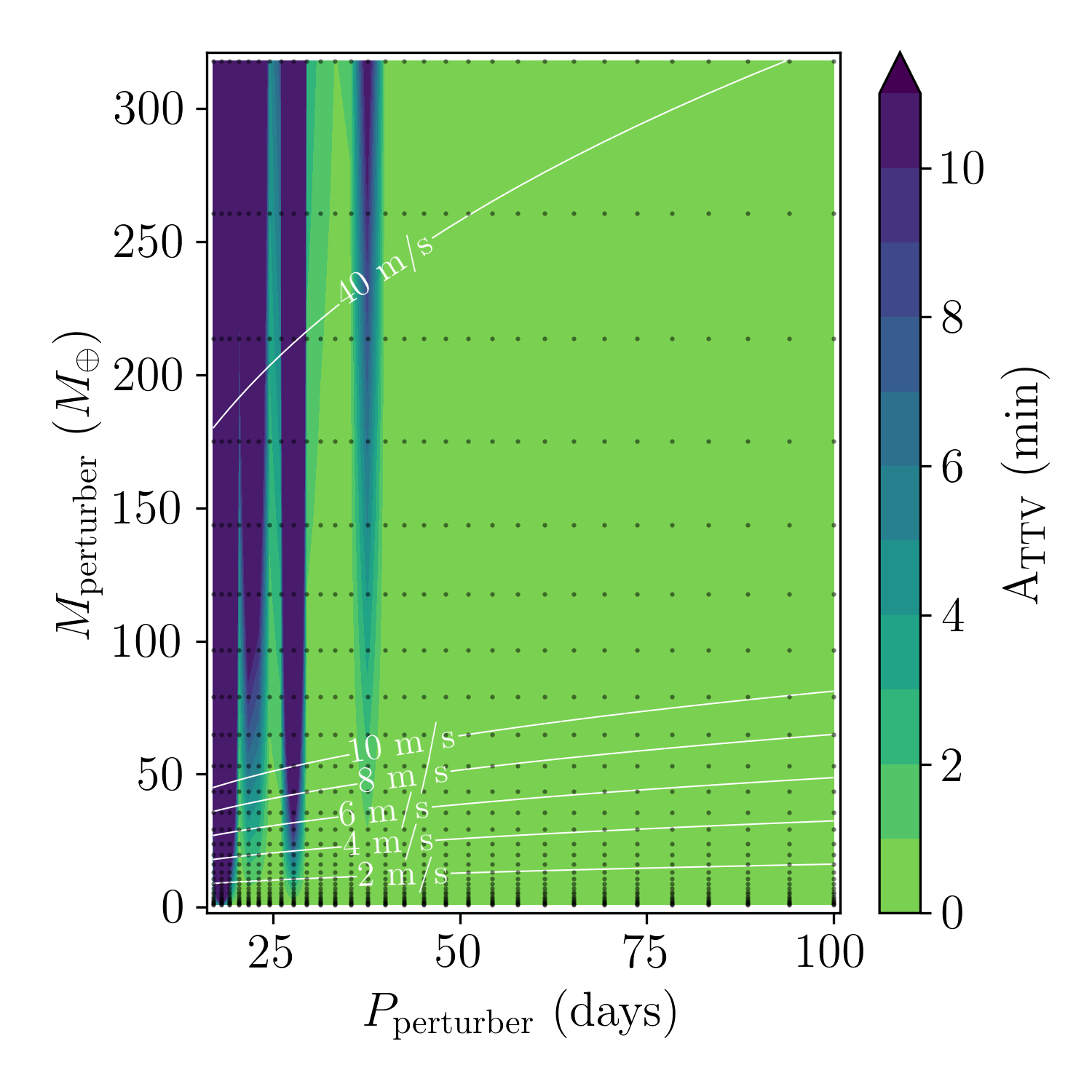}
	\includegraphics[width=0.33\textwidth]{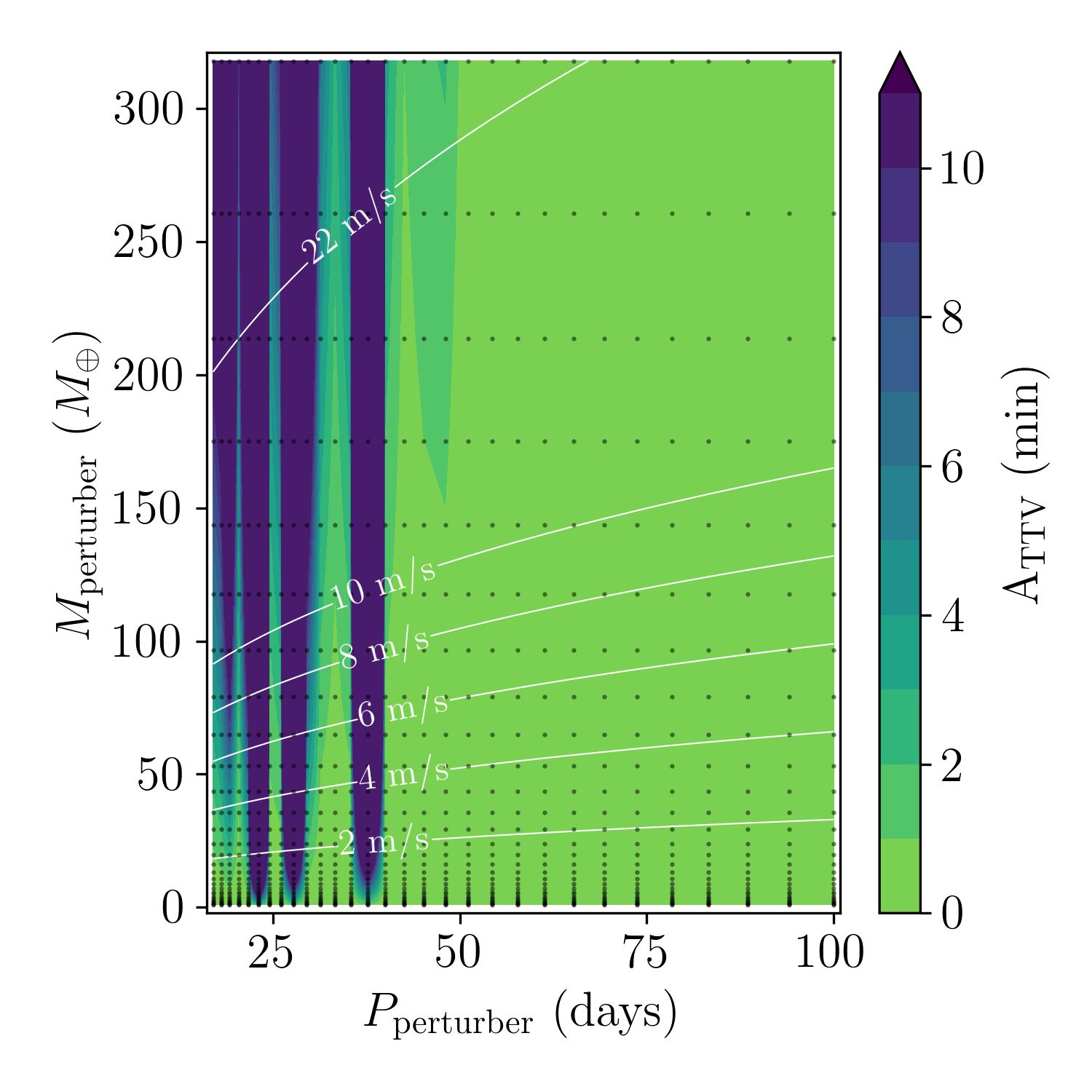}

    \caption{As Fig.~\ref{fig:grid_kelt6} for WASP-106.
    }
    \label{fig:grid_wasp106}
\end{figure*}

\begin{figure*}
    \centering
	\includegraphics[width=0.33\textwidth]{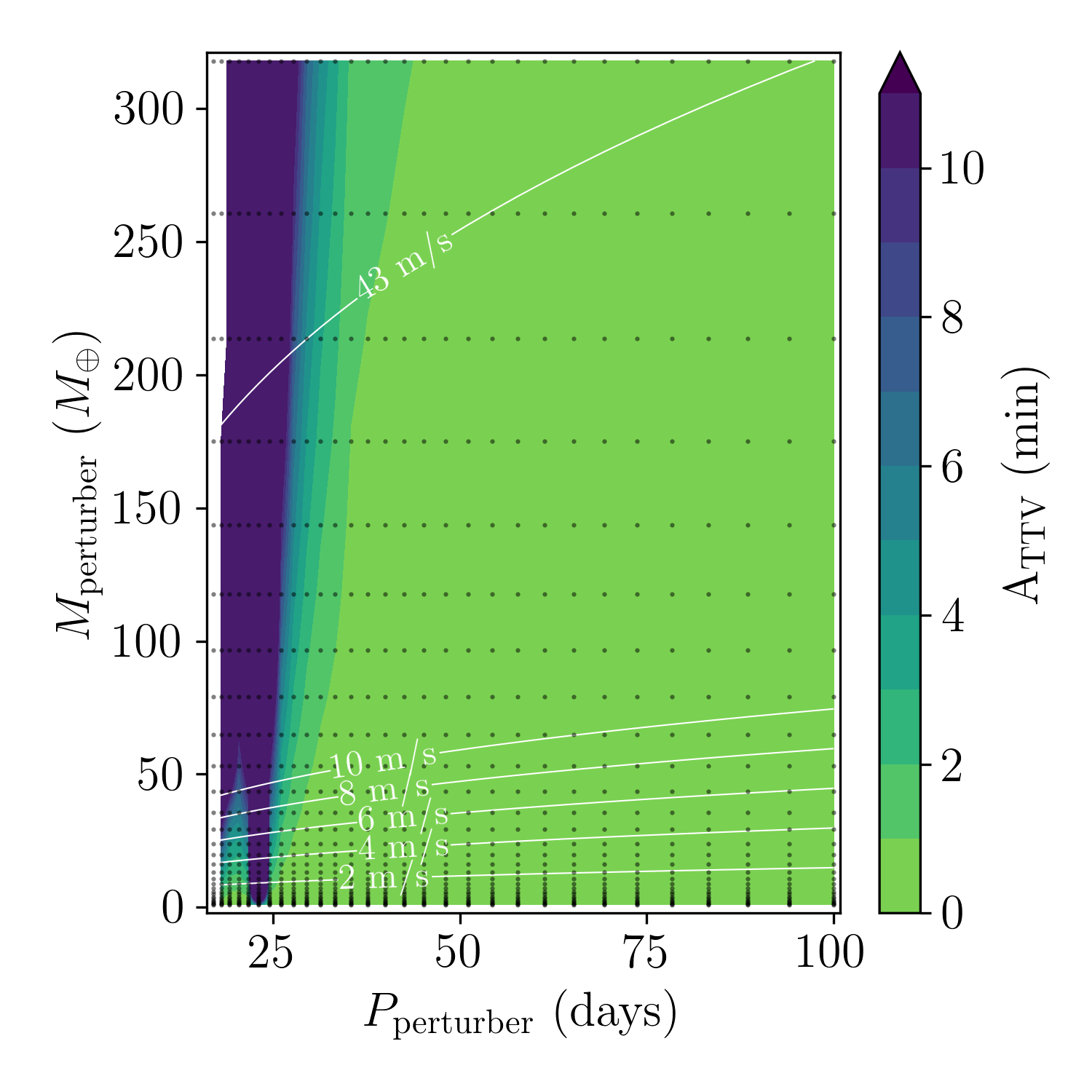}
	\includegraphics[width=0.33\textwidth]{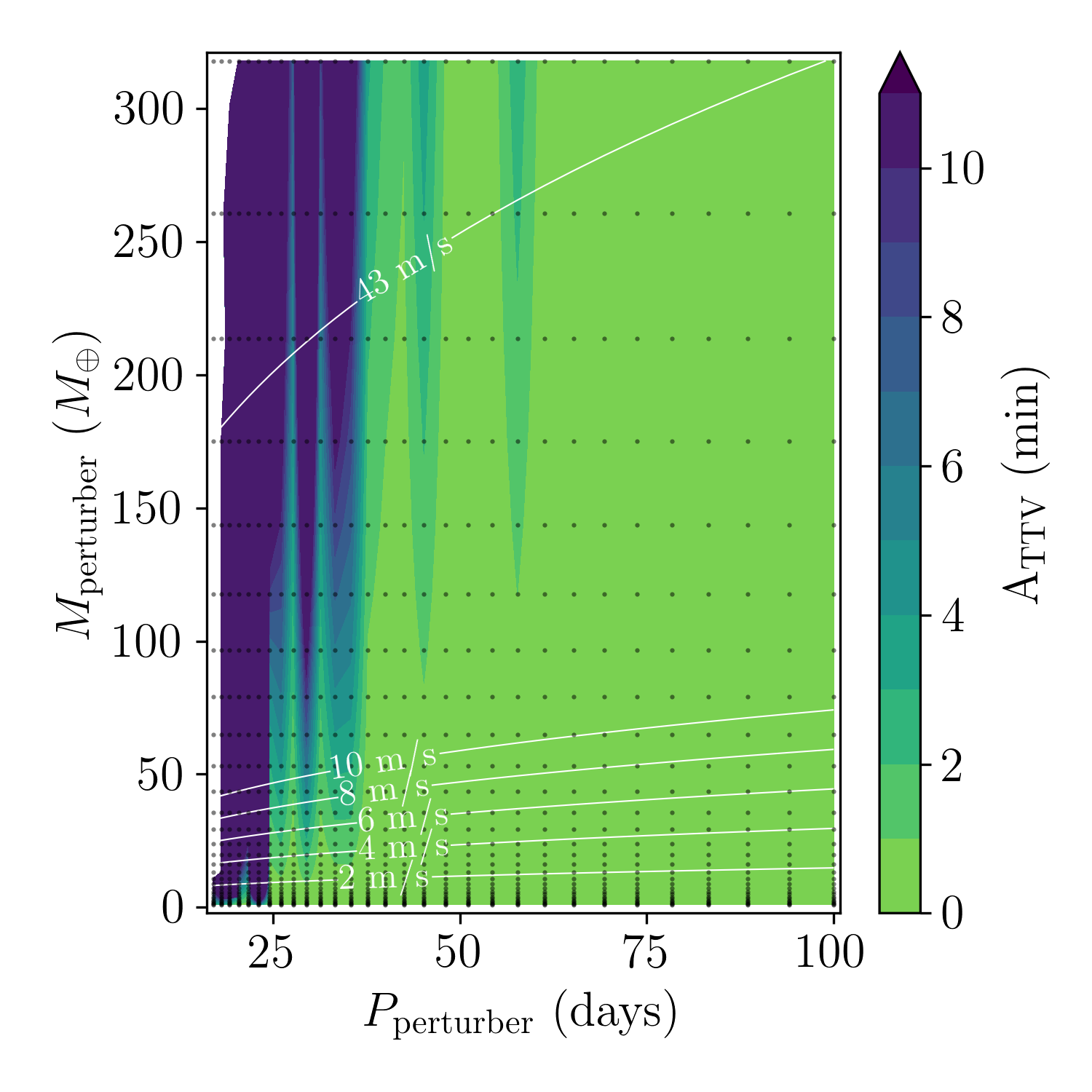}
	\includegraphics[width=0.33\textwidth]{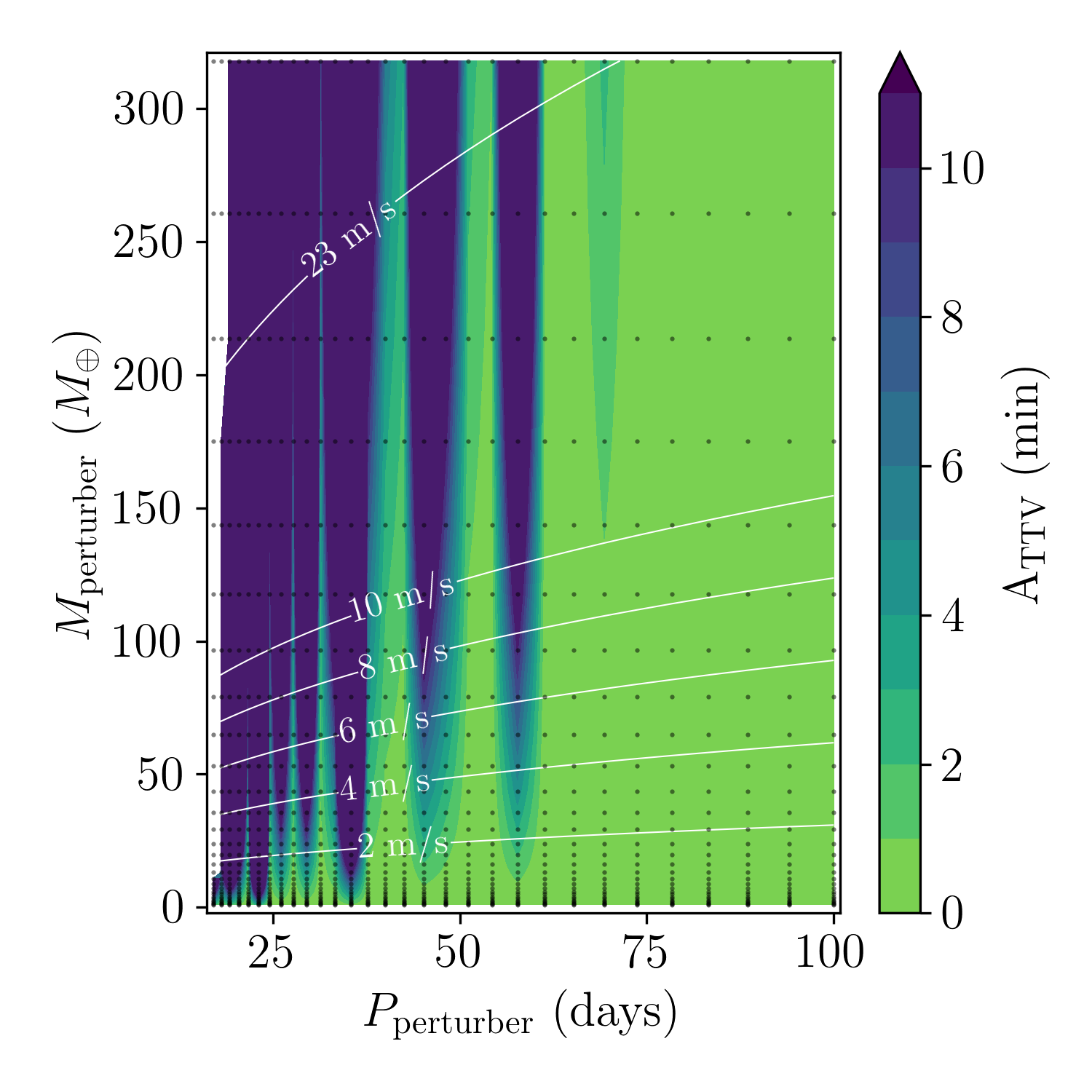}

    \caption{As Fig.~\ref{fig:grid_kelt6} for WASP-130.
    }
    \label{fig:grid_wasp130}
\end{figure*}

\begin{figure*}
    \centering
	\includegraphics[width=0.33\textwidth]{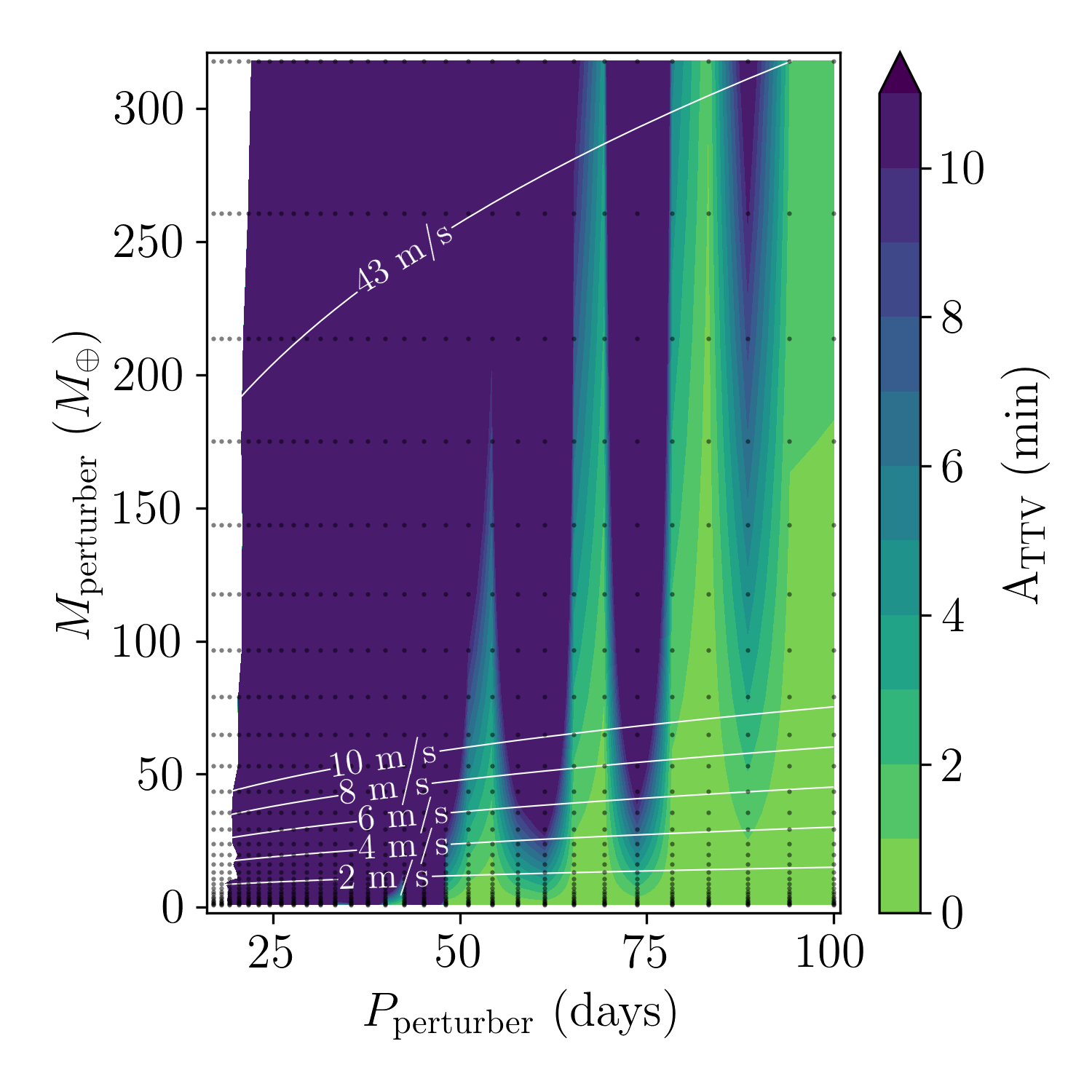}
	\includegraphics[width=0.33\textwidth]{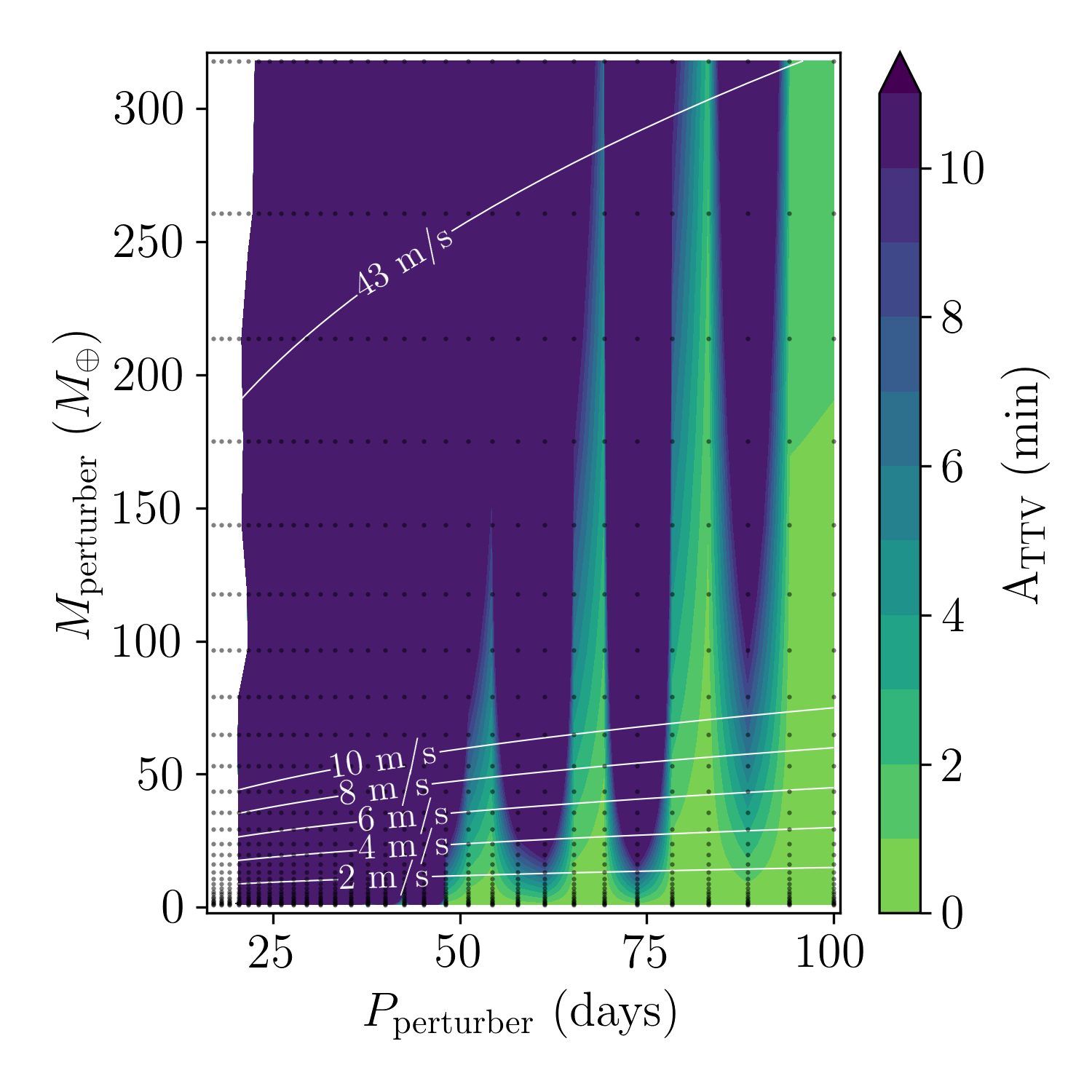}
	\includegraphics[width=0.33\textwidth]{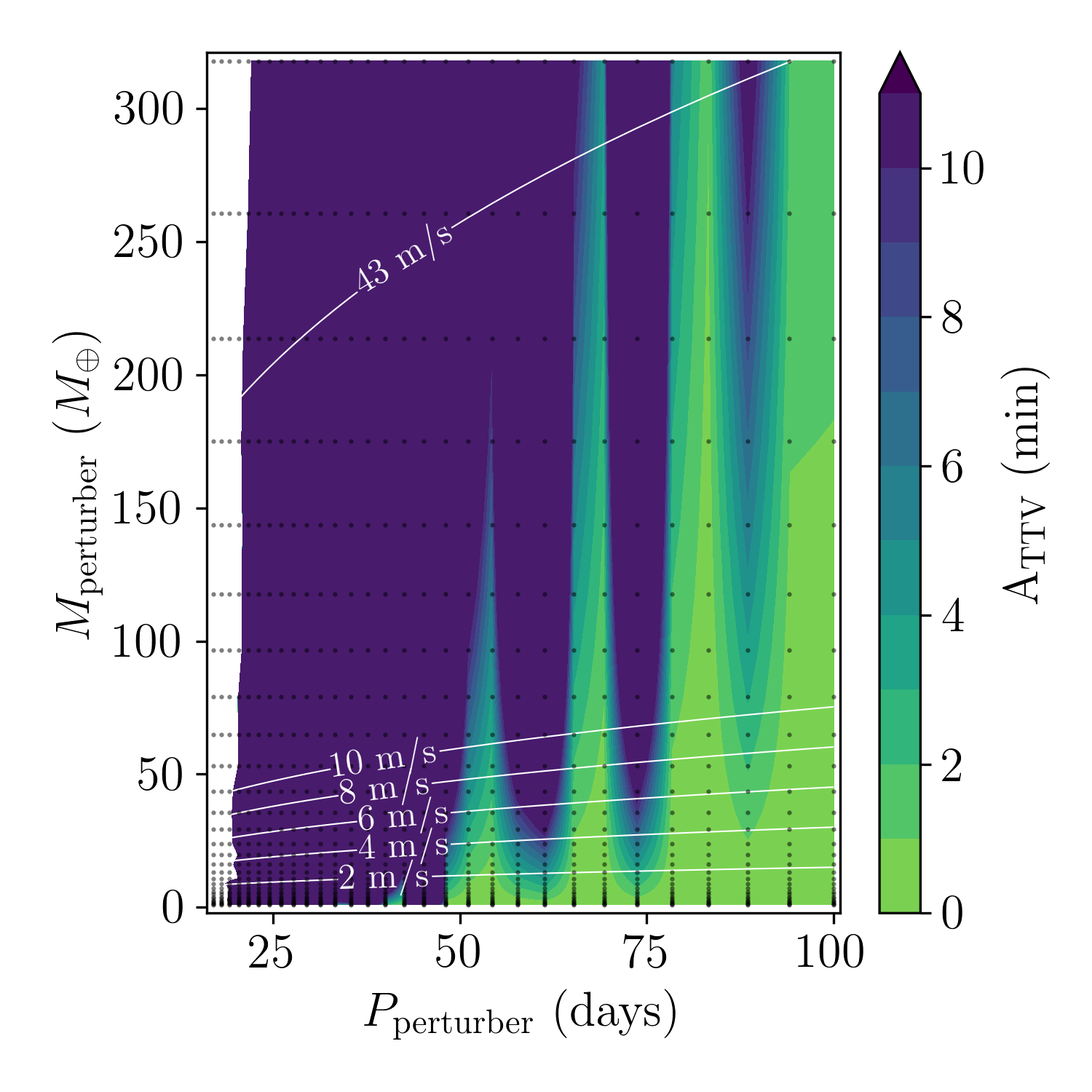}

    \caption{As Fig.~\ref{fig:grid_hatp17} for K2-287.
    }
    \label{fig:grid_k2-287}
\end{figure*}


\bsp	
\label{lastpage}
\end{document}